\documentclass[12pt,letterpaper]{article}
\usepackage[margin=1in]{geometry}
\usepackage{graphicx}
\graphicspath{{figures/}}  

\usepackage{xcolor}
\usepackage{longtable}  
\usepackage{booktabs}
\usepackage{array}
\usepackage{amsmath, amssymb, amsthm}
\usepackage[mathlines]{lineno}
\usepackage{newtx}
\usepackage{setspace}

\usepackage{fvextra}  
\usepackage[skins,breakable]{tcolorbox}
\newtcolorbox{promptbox}[1]{%
  breakable, enhanced,
  colback=gray!3, colframe=gray!50, boxrule=0.5pt, arc=1.5pt,
  left=5pt, right=5pt, top=4pt, bottom=4pt,
  fonttitle=\bfseries\footnotesize, coltitle=black, colbacktitle=gray!15,
  title={#1},
}

\usepackage[american]{babel}
\usepackage{csquotes}
\usepackage[citestyle=numeric-comp,bibstyle=nature,sorting=none,backend=biber]{biblatex}
\usepackage{hyperref}
\usepackage{nameref}
\usepackage{cleveref}
\usepackage{authblk}

\hypersetup{
    colorlinks=true,
    linkcolor=purple,
    filecolor=purple,      
    urlcolor=purple,
    citecolor=purple,
    pdfpagemode=FullScreen,
}

\title{Inverse planning of social interactions in relationships}
\author[1,*]{Alicia M. Chen}
\author[2]{Ashley J. Thomas}
\author[1]{Joshua B. Tenenbaum}
\author[1]{Rebecca Saxe}
\date{\vspace{-5ex}}

\affil[1]{Department of Brain and Cognitive Sciences, MIT}
\affil[2]{Department of Psychology, Harvard University}
\affil[*]{Corresponding author: aliciach@mit.edu}

\newcommand{\nRecruitedOneA}{481}                                       
\newcommand{\nRetainedOneA}{481}
\newcommand{\nExcludedOneA}{0}
\newcommand{\ageMinOneA}{\ensuremath{18}}
\newcommand{\ageMaxOneA}{\ensuremath{83}}
\newcommand{\ageMeanOneA}{\ensuremath{39.5}}
\newcommand{\ageSdOneA}{\ensuremath{12.6}}
\newcommand{\nFemaleOneA}{231}
\newcommand{\nMaleOneA}{232}
\newcommand{\nNonconformingOneA}{15}
\newcommand{\nAbstainOneA}{3}
\newcommand{\llFullOneA}{\ensuremath{-0.83}}                            
\newcommand{\llPreregOneA}{\ensuremath{-0.83}}                          
\newcommand{\dllReweightOneA}{\ensuremath{0.00}}                        
\newcommand{\ciReweightOneA}{\ensuremath{[0.00,\ 0.00]}}
\newcommand{\statReweightOneA}{\dllReweightOneA~\ciReweightOneA}
\newcommand{\dllBaseOneA}{\ensuremath{0.08}}                            
\newcommand{\ciBaseOneA}{\ensuremath{[0.03,\ 0.13]}}
\newcommand{\statBaseOneA}{\dllBaseOneA~\ciBaseOneA}
\newcommand{\dllDiscOneA}{\ensuremath{3.30}}                            
\newcommand{\ciDiscOneA}{\ensuremath{[2.96,\ 3.62]}}
\newcommand{\statDiscOneA}{\dllDiscOneA~\ciDiscOneA}
\newcommand{\pWvOneA}{\ensuremath{9.01}}
\newcommand{\pWeOneA}{\ensuremath{3.63}}
\newcommand{\pWdOneA}{\ensuremath{4.24}}
\newcommand{\pGammaOneA}{\ensuremath{0.68}}
\newcommand{\pAlphaObsOneA}{\ensuremath{0.84}}
\newcommand{\pSigmaOneA}{\ensuremath{0.248}}
\newcommand{\pEtaOneA}{\ensuremath{0.00}}
\newcommand{\nRecruitedOneB}{241}                                       
\newcommand{\nRetainedOneB}{237}
\newcommand{\nExcludedOneB}{4}
\newcommand{\ageMinOneB}{\ensuremath{18}}
\newcommand{\ageMaxOneB}{\ensuremath{82}}
\newcommand{\ageMeanOneB}{\ensuremath{40.8}}
\newcommand{\ageSdOneB}{\ensuremath{13.4}}
\newcommand{\nFemaleOneB}{117}
\newcommand{\nMaleOneB}{119}
\newcommand{\nNonconformingOneB}{4}
\newcommand{\nAbstainOneB}{1}
\newcommand{\llFullOneB}{\ensuremath{-5.87}}                            
\newcommand{\llPreregOneB}{\ensuremath{-6.30}}                          
\newcommand{\dllReweightOneB}{\ensuremath{0.44}}                        
\newcommand{\ciReweightOneB}{\ensuremath{[0.23,\ 0.65]}}
\newcommand{\statReweightOneB}{\dllReweightOneB~\ciReweightOneB}
\newcommand{\dllBaseOneB}{\ensuremath{0.34}}                            
\newcommand{\ciBaseOneB}{\ensuremath{[0.23,\ 0.47]}}
\newcommand{\statBaseOneB}{\dllBaseOneB~\ciBaseOneB}
\newcommand{\dllDiscOneB}{\ensuremath{5.74}}                            
\newcommand{\ciDiscOneB}{\ensuremath{[5.11,\ 6.28]}}
\newcommand{\statDiscOneB}{\dllDiscOneB~\ciDiscOneB}
\newcommand{\pWvOneB}{\ensuremath{2.93}}
\newcommand{\pWeOneB}{\ensuremath{3.30}}
\newcommand{\pWdOneB}{\ensuremath{2.55}}
\newcommand{\pGammaOneB}{\ensuremath{1.51}}
\newcommand{\pAlphaObsOneB}{\ensuremath{2.35}}
\newcommand{\pSigmaOneB}{\ensuremath{0.262}}
\newcommand{\pEtaOneB}{\ensuremath{3.07}}
\newcommand{\nRecruitedTwoA}{240}                                       
\newcommand{\nRetainedTwoA}{239}
\newcommand{\nExcludedTwoA}{1}
\newcommand{\ageMinTwoA}{\ensuremath{18}}
\newcommand{\ageMaxTwoA}{\ensuremath{83}}
\newcommand{\ageMeanTwoA}{\ensuremath{38.6}}
\newcommand{\ageSdTwoA}{\ensuremath{12.9}}
\newcommand{\nFemaleTwoA}{117}
\newcommand{\nMaleTwoA}{121}
\newcommand{\nNonconformingTwoA}{2}
\newcommand{\nAbstainTwoA}{0}
\newcommand{\llFullTwoA}{\ensuremath{0.97}}                             
\newcommand{\llPreregTwoA}{\ensuremath{0.82}}                           
\newcommand{\dllReweightTwoA}{\ensuremath{0.15}}                        
\newcommand{\ciReweightTwoA}{\ensuremath{[0.07,\ 0.23]}}
\newcommand{\statReweightTwoA}{\dllReweightTwoA~\ciReweightTwoA}
\newcommand{\dllBaseTwoA}{\ensuremath{4.48}}                            
\newcommand{\ciBaseTwoA}{\ensuremath{[4.09,\ 4.87]}}
\newcommand{\statBaseTwoA}{\dllBaseTwoA~\ciBaseTwoA}
\newcommand{\dllDiscTwoA}{\ensuremath{0.08}}                            
\newcommand{\ciDiscTwoA}{\ensuremath{[0.02,\ 0.13]}}
\newcommand{\statDiscTwoA}{\dllDiscTwoA~\ciDiscTwoA}
\newcommand{\pWvTwoA}{\ensuremath{6.05}}
\newcommand{\pWeTwoA}{\ensuremath{0.94}}
\newcommand{\pWdTwoA}{\ensuremath{0.84}}
\newcommand{\pGammaTwoA}{\ensuremath{0.32}}
\newcommand{\pAlphaObsTwoA}{\ensuremath{5.83}}
\newcommand{\pSigmaTwoA}{\ensuremath{0.218}}
\newcommand{\pEtaTwoA}{\ensuremath{8.85}}
\newcommand{\nRecruitedTwoB}{122}                                       
\newcommand{\nRetainedTwoB}{120}
\newcommand{\nExcludedTwoB}{2}
\newcommand{\ageMinTwoB}{\ensuremath{19}}
\newcommand{\ageMaxTwoB}{\ensuremath{80}}
\newcommand{\ageMeanTwoB}{\ensuremath{40.2}}
\newcommand{\ageSdTwoB}{\ensuremath{14.5}}
\newcommand{\nFemaleTwoB}{59}
\newcommand{\nMaleTwoB}{60}
\newcommand{\nNonconformingTwoB}{2}
\newcommand{\nAbstainTwoB}{1}
\newcommand{\llFullTwoB}{\ensuremath{-4.61}}                            
\newcommand{\llPreregTwoB}{\ensuremath{-4.62}}                          
\newcommand{\dllReweightTwoB}{\ensuremath{0.01}}                        
\newcommand{\ciReweightTwoB}{\ensuremath{[-0.05,\ 0.07]}}
\newcommand{\statReweightTwoB}{\dllReweightTwoB~\ciReweightTwoB}
\newcommand{\dllBaseTwoB}{\ensuremath{2.57}}                            
\newcommand{\ciBaseTwoB}{\ensuremath{[2.08,\ 3.06]}}
\newcommand{\statBaseTwoB}{\dllBaseTwoB~\ciBaseTwoB}
\newcommand{\dllDiscTwoB}{\ensuremath{3.11}}                            
\newcommand{\ciDiscTwoB}{\ensuremath{[2.50,\ 3.75]}}
\newcommand{\statDiscTwoB}{\dllDiscTwoB~\ciDiscTwoB}
\newcommand{\pWvTwoB}{\ensuremath{3.95}}
\newcommand{\pWeTwoB}{\ensuremath{1.93}}
\newcommand{\pWdTwoB}{\ensuremath{1.46}}
\newcommand{\pGammaTwoB}{\ensuremath{0.91}}
\newcommand{\pAlphaObsTwoB}{\ensuremath{2.96}}
\newcommand{\pSigmaTwoB}{\ensuremath{0.254}}
\newcommand{\pEtaTwoB}{\ensuremath{0.42}}
\newcommand{\nRecruitedThreeA}{241}                                     
\newcommand{\nRetainedThreeA}{239}
\newcommand{\nExcludedThreeA}{2}
\newcommand{\ageMinThreeA}{\ensuremath{18}}
\newcommand{\ageMaxThreeA}{\ensuremath{81}}
\newcommand{\ageMeanThreeA}{\ensuremath{39.1}}
\newcommand{\ageSdThreeA}{\ensuremath{13.2}}
\newcommand{\nFemaleThreeA}{119}
\newcommand{\nMaleThreeA}{119}
\newcommand{\nNonconformingThreeA}{3}
\newcommand{\nAbstainThreeA}{0}
\newcommand{\llFullThreeA}{\ensuremath{-10.62}}                         
\newcommand{\llPreregThreeA}{\ensuremath{-10.86}}                       
\newcommand{\dllReweightThreeA}{\ensuremath{0.24}}                      
\newcommand{\ciReweightThreeA}{\ensuremath{[0.08,\ 0.40]}}
\newcommand{\statReweightThreeA}{\dllReweightThreeA~\ciReweightThreeA}
\newcommand{\dllBaseThreeA}{\ensuremath{0.15}}                          
\newcommand{\ciBaseThreeA}{\ensuremath{[0.08,\ 0.22]}}
\newcommand{\statBaseThreeA}{\dllBaseThreeA~\ciBaseThreeA}
\newcommand{\dllDiscThreeA}{\ensuremath{4.30}}                          
\newcommand{\ciDiscThreeA}{\ensuremath{[3.84,\ 4.77]}}
\newcommand{\statDiscThreeA}{\dllDiscThreeA~\ciDiscThreeA}
\newcommand{\pWvThreeA}{\ensuremath{3.44}}
\newcommand{\pWeThreeA}{\ensuremath{2.63}}
\newcommand{\pWdThreeA}{\ensuremath{1.98}}
\newcommand{\pGammaThreeA}{\ensuremath{3.10}}
\newcommand{\pAlphaObsThreeA}{\ensuremath{3.10}}
\newcommand{\pSigmaThreeA}{\ensuremath{0.308}}
\newcommand{\pEtaThreeA}{\ensuremath{2.88}}
\newcommand{\nRecruitedThreeB}{239}                                     
\newcommand{\nRetainedThreeB}{238}
\newcommand{\nExcludedThreeB}{1}
\newcommand{\ageMinThreeB}{\ensuremath{18}}
\newcommand{\ageMaxThreeB}{\ensuremath{73}}
\newcommand{\ageMeanThreeB}{\ensuremath{39.1}}
\newcommand{\ageSdThreeB}{\ensuremath{13.1}}
\newcommand{\nFemaleThreeB}{122}
\newcommand{\nMaleThreeB}{117}
\newcommand{\nNonconformingThreeB}{0}
\newcommand{\nAbstainThreeB}{0}
\newcommand{\llFullThreeB}{\ensuremath{-7.82}}                          
\newcommand{\llPreregThreeB}{\ensuremath{-7.72}}                        
\newcommand{\dllReweightThreeB}{\ensuremath{-0.10}}                     
\newcommand{\ciReweightThreeB}{\ensuremath{[-0.15,\ -0.05]}}
\newcommand{\statReweightThreeB}{\dllReweightThreeB~\ciReweightThreeB}
\newcommand{\dllBaseThreeB}{\ensuremath{2.21}}                          
\newcommand{\ciBaseThreeB}{\ensuremath{[1.90,\ 2.51]}}
\newcommand{\statBaseThreeB}{\dllBaseThreeB~\ciBaseThreeB}
\newcommand{\dllDiscThreeB}{\ensuremath{1.77}}                          
\newcommand{\ciDiscThreeB}{\ensuremath{[1.33,\ 2.19]}}
\newcommand{\statDiscThreeB}{\dllDiscThreeB~\ciDiscThreeB}
\newcommand{\pWvThreeB}{\ensuremath{3.09}}
\newcommand{\pWeThreeB}{\ensuremath{0.21}}
\newcommand{\pWdThreeB}{\ensuremath{0.14}}
\newcommand{\pGammaThreeB}{\ensuremath{1.23}}
\newcommand{\pAlphaObsThreeB}{\ensuremath{27.03}}
\newcommand{\pSigmaThreeB}{\ensuremath{0.282}}
\newcommand{\pEtaThreeB}{\ensuremath{0.58}}
\newcommand{\rStudyOne}{\ensuremath{0.991}}                             
\newcommand{\ciStudyOne}{\ensuremath{[0.987,\ 0.995]}}
\newcommand{\ceilStudyOne}{\ensuremath{0.997}}                          
\newcommand{\rStudyTwo}{\ensuremath{0.987}}                             
\newcommand{\ciStudyTwo}{\ensuremath{[0.978,\ 0.994]}}
\newcommand{\ceilStudyTwo}{\ensuremath{0.997}}                          
\newcommand{\rStudyThree}{\ensuremath{0.988}}                           
\newcommand{\ciStudyThree}{\ensuremath{[0.980,\ 0.993]}}
\newcommand{\ceilStudyThree}{\ensuremath{0.996}}                        
\newcommand{\nAltsMedianRange}{\mbox{\ensuremath{2}--\ensuremath{4}}}   
\newcommand{\rNonfoodOwnFit}{\ensuremath{0.988}}

\newcommand{\rNonfoodFoodFit}{\ensuremath{0.977}}
\newcommand{\ciNonfoodFoodFit}{\ensuremath{[0.965,\ 0.988]}}
\newcommand{\rNonfoodPooledFit}{\ensuremath{0.981}}

\newcommand{\dllPoolOneA}{\ensuremath{-0.07}}                           
\newcommand{\ciPoolOneA}{\ensuremath{[-0.12,\ -0.02]}}
\newcommand{\statPoolOneA}{\dllPoolOneA~\ciPoolOneA}
\newcommand{\dllPoolOneB}{\ensuremath{-0.27}}                           
\newcommand{\ciPoolOneB}{\ensuremath{[-0.40,\ -0.15]}}
\newcommand{\statPoolOneB}{\dllPoolOneB~\ciPoolOneB}
\newcommand{\dllPoolTwoA}{\ensuremath{-0.07}}                           
\newcommand{\ciPoolTwoA}{\ensuremath{[-0.17,\ 0.05]}}
\newcommand{\statPoolTwoA}{\dllPoolTwoA~\ciPoolTwoA}
\newcommand{\dllPoolTwoB}{\ensuremath{0.11}}                            
\newcommand{\ciPoolTwoB}{\ensuremath{[0.05,\ 0.17]}}
\newcommand{\statPoolTwoB}{\dllPoolTwoB~\ciPoolTwoB}
\newcommand{\dllPoolThreeA}{\ensuremath{0.06}}                          
\newcommand{\ciPoolThreeA}{\ensuremath{[0.01,\ 0.10]}}
\newcommand{\statPoolThreeA}{\dllPoolThreeA~\ciPoolThreeA}
\newcommand{\dllPoolThreeB}{\ensuremath{-0.02}}                         
\newcommand{\ciPoolThreeB}{\ensuremath{[-0.11,\ 0.06]}}
\newcommand{\statPoolThreeB}{\dllPoolThreeB~\ciPoolThreeB}
\newcommand{\dllPoolAll}{\ensuremath{-0.06}}                            
\newcommand{\ciPoolAll}{\ensuremath{[-0.09,\ -0.03]}}
\newcommand{\statPoolAll}{\dllPoolAll~\ciPoolAll}
\newcommand{\nRecruitedTotal}{1{,}564}                                  
\newcommand{\nRetainedTotal}{1{,}554}
\newcommand{\nScenarios}{16}                                            
\newcommand{\nRuns}{20}                                                 
\newcommand{\nFolds}{16}                                                
\newcommand{\nFoldsTrain}{15}                                           
\newcommand{\altTemperature}{\ensuremath{0.7}}                          
\newcommand{\scoreTemperature}{\ensuremath{0.2}}                        
\newcommand{\lmName}{Llama-3.3-70B-Instruct-Turbo}                      

\begin{document}

\maketitle

\setstretch{1.15}

\begin{abstract}
We propose a formal account of how structured, shared knowledge about social relationships shapes action interpretation. 
The model represents relationships as constraints in a social environment, analogous to boundaries or obstacles in a physical environment and operating within the same generative model, but exerting distinct constraints on action.
As an initial test of this framework, we draw on research across the social sciences to capture in the models how one dimension of relationships -- formality versus intimacy -- shapes how people interpret interpersonally vulnerable behavior. 
We test this account in stories of naturalistic everyday situations, extending structured models of action understanding to open-ended contexts. 
Across six preregistered experiments ($N = \nRetainedTotal$), the model captures people's inferences about desires, physical environments, and social relationships. 
This work formalizes how relationships can constrain -- and be revealed through -- everyday action.
\end{abstract}

\section*{Introduction}

People make rich inferences from sparse information by reasoning about the causes and constraints that could have generated what they observed \autocite{tenenbaum2011grow,anderson2013adaptive,gopnik2004theory}. 
One instance of this capacity is when people observe other people's actions and then infer their beliefs, intentions, and knowledge -- that is, use a `theory of mind' \autocite{dennett1989intentional,gopnik1997words,wellman2014making}. 
These inferences allow people to navigate complex social environments by predicting, explaining, and intervening on others' actions \autocite{ho2022planning}. 

A standard cognitive model of theory of mind is ``inverse planning.'' 
Inverse-planning models formalize the idea that observers recover unobservable mental states by inverting a generative model of rational action \autocite{baker2009action,baker2017rational,jara2016naive}. 
For example, if Alice walks past Korean and Lebanese food trucks to reach a Mexican food truck, observers can invert a model of Alice's action planning (how she trades off the value of different foods against the physical effort required to reach them) to infer that she \textit{knows} a Mexican food truck is available, and \textit{prefers} it to the other options \autocite{baker2017rational}. 
This influential family of models captures people's reasoning across many kinds of social behaviors \autocite{wu2021too,xiang2023collaborative,goodman2016pragmatic,radkani2025people,houlihan2023emotion}, and has served as a foundation for robust and interpretable social reasoning in AI systems \autocite{zhang2025autotom,ying2025language}.

Yet many everyday actions involve people acting with others, based on mutual beliefs and knowledge \autocite{sebanz2006joint,tomasello2005understanding,lewis1969convention}. 
These joint actions introduce utilities that depend on the constraints imposed by the relationship between the people involved. 
For example, imagine that Alice walked \textit{with Bob} to the Mexican food truck. 
Suppose that when they arrive, they order several different items and share them, taking bites from the same dishes and trading items back and forth. 
A standard inverse-planning model would explain this action in terms of their preferences: Alice and Bob each wanted to eat several dishes, and sharing was an efficient way to satisfy those preferences. 
By contrast, people observing this situation would likely also infer something else: that Alice and Bob have an existing relationship, likely as either romantic partners or close friends, and almost certainly not as coworkers or people meeting for the first time. 
This inference is possible because the observers know that the way people share food depends on the nature of the dyad's relationship. 
Between romantic partners or close friends, eating from the same dish may feel easy and ordinary; between coworkers or acquaintances, the same action may feel uncomfortable and inappropriate \autocite{tybur2020behavioral,miller1998food}. 

Across social psychology, developmental psychology, anthropology, and other areas of social science, social relationships have been studied as central features of the social environment that shape both behavior and action interpretation \autocite{thomas2024cognitive,basyouni2022mapping,fiske1992four,clark1988interpersonal,clark2014understanding,earp2021social,powell2022adopted,sahlins2013kinship,rai2011moral,chuey2026young,pepe2026infants,liberman2017origins,brown1987politeness,descola2013beyond,levi1971elementary}. 
People eat, travel, converse, disclose information, share space, and solve practical problems with friends, partners, relatives, coworkers, teachers, and strangers; the same action can have different meanings depending on the relationship in which it occurs \autocite{fiske1992four,clark1988interpersonal}. 
In the current research, we expand the capacity of inverse-planning models to capture human inferences from social interactions, by incorporating social relationships into the generative model of dyadic action selection. 

Here we focus on an aspect of relationships that has been extensively characterized across the social sciences, but has not been formalized computationally: how they function as \textit{social constructs}, with properties that are part of common knowledge in terms of the way they influence action understanding \autocite{gilbert1989social}. 
This knowledge may be organized around a small set of dimensions or categories that recur across places and times and are each tied to distinct behavioral signatures \autocite{cheng2025conceptual,fiske1992four,thomas2024cognitive,brown1987politeness}. 
Construed this way, relationships constrain joint action in the way that the physical world constrains action in standard inverse-planning models: 
as structural features of a situation that influence context-sensitive judgments about the costs of actions, in ways that are part of shared intuitive knowledge\footnote{For example, `the height of a barrier determines how effortful it is to cross it' vs. `the closeness of a relationship determines how uncomfortable it is to exchange body fluids'} \autocite{baker2009action,baker2017rational,jara2020naive,liu2017ten,liu2026physical}.
Treating relationships as social constructs is distinct from two ways that computational models have previously characterized social relationships, which have primarily been based on (1) whether or not a social connection exists, for the purpose of statistical learning about networks and groups \autocite{basyouni2022mapping,davis2026inferring,gershman2020social,son2023abstract,son2021cognitive,aslarus2025early}, and (2) the aspects of relationships that can be located within an individual agent's decision making, such as how much one person values another person's outcomes \autocite{kleiman2017learning,ullman2009help,powell2022adopted,shum2019theory,jern2014reasoning}. 

As a first test of this framework, we focus on one dimension of relationships: a relationship's position on an axis from \textit{formal} to \textit{intimate}. 
Intimacy -- often described in terms of `closeness' or `communality' -- is a central construct in theories of reasoning about social relationships, with widely documented behavioral signatures \autocite{fiske1992four,thomas2024cognitive,prager1997psychology,clark1988interpersonal,brown1987politeness}.  
Research on intimacy and communality motivates a theoretically grounded, model-specifiable hypothesis about how this dimension enters action planning: intimacy changes the utility of actions involving interpersonal vulnerability, including bodily contact \autocite{miller1998food,thomas2022early,clark1988interpersonal,shaver1988intimacy}. 
Crucially, intimacy is not reducible to how often people interact or how much they value another person's welfare: coworkers or members of the same political party, for example, may interact often and be strongly invested in one another's success while being in a relatively formal relationship \autocite{fiske1992four,thomas2024cognitive}.
By formalizing intimacy in this way, we can specify how one central dimension of relationships enters inverse planning and test the account across actions and domains.
Establishing this inferential logic provides a foundation for extending the approach to other dimensions of relationships in future work.

To test this account, we first construct an experimental and modeling framework in which humans and models reason about short vignettes describing naturalistic social situations. Drawing on approaches in computational linguistics that use language models (LMs) to generate and evaluate alternative utterances for pragmatic language understanding \autocite{tsvilodub2025integrating,tsvilodub2026computational,qiu2025same}, we use LMs to identify and represent the information (the alternative actions and their features) relevant for understanding behavior in open-ended social situations. 
This approach situates theory-of-mind reasoning within a broader computational perspective in which distributional representations identify what is relevant in an open-ended situation and symbolic models perform structured reasoning over those representations \autocite{wong2025modeling}.
The framework enables systematic comparisons among theory-of-mind models in everyday social situations where relationships matter, without relying on the highly constrained settings and predetermined choice sets typical of previous tests of such models \autocite{gelpi2025towards,jin2024mmtom}.

Next, we treat relationships as part of a \textit{sociological} environment. 
Like distance, barriers, or other physical constraints \autocite{lewin1939field}, the relationship between two people is a latent feature of the situation that observers may not know directly, but can infer from action because it changes the utilities of acting. 
We capture how relationships enter the utility function by how they introduce a sociological cost of \textit{discomfort}, so that a single action can be explained jointly in terms of desires, the physical environment (which determines the physical costs of actions), and relationships (which determine the sociological costs of actions).
The same action can therefore be more or less diagnostic of a relationship depending on the other goals and constraints in the situation. 
This allows us to move beyond treating actions as direct qualitative cues to relationships \autocite{miller1998food,thomas2022early}, and toward an account of how observers infer relationships from action in graded, quantitative, and context-sensitive ways.

Combining these two components, we build and test models that take sparse information about a pair of people engaged in a social action, retrieve the contextual knowledge needed for inverse planning, and use a probabilistic model to jointly update beliefs about people's desires and goals, the physical environment, and the sociological environment (here, the relationship between the people). 
The resulting models capture three key facets of human inference about social interactions: (1) how sparse observations of the social world can support rich inferences about unobservable variables, (2) how action choices reveal and thereby may sustain different kinds of relationships, and (3) how joint inference over this causal model captures the graded strength of these inferences.  

By integrating theoretical work on relationships from across the social sciences with the formal machinery of inverse planning, this work characterizes the cognitive processes that support reasoning about actions in relationships, while extending inverse-planning models to capture a central dimension of human social life. 

\begin{figure}[hp]
  \centering
  \includegraphics[width=\textwidth]{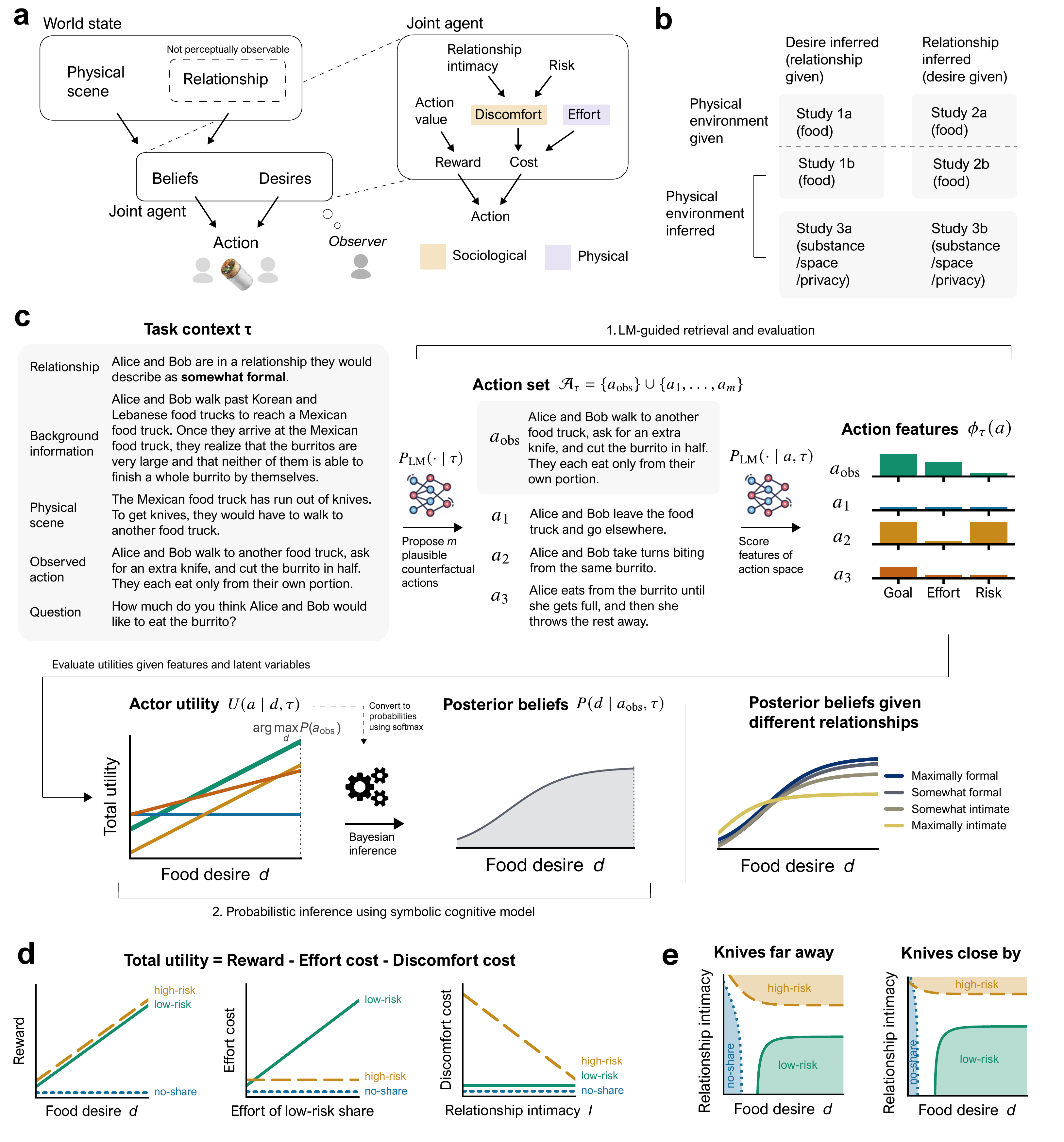}
  \caption{\textbf{Model and experimental design.}
  \textbf{(a)} Model schematic. 
  \textbf{(b)} Across the experiments, either relationship or desire is given; the physical environment is given in the single-inference experiments and inferred in the joint-inference experiments. 
  \textbf{(c)} Given a vignette and observed action, an LM generates plausible unchosen actions and scores every action on goal satisfaction, effort and interpersonal risk. A probabilistic inverse-planning model uses these features to infer the latent variable(s) queried in the experiment.
  The example illustrates how relationship context shifts the posterior over desire.
  \textbf{(d)} How the utility components vary with desire, relationship intimacy and the physical environment. 
  \textbf{(e)} How posterior judgments vary based on the physical scene. 
  The utility and posterior curves use the LM-elicited features from the example scenario with illustrative weights.}
  \label{fig:model-schematic}
\end{figure}

\section*{A model of social inverse planning}

Inverse-planning models assume that agents choose actions in proportion to their expected utility (reward minus cost), and that observers invert this intuitive theory of an actor to recover the latent variables that made the observed action worth taking \parencite{baker2009action,baker2017rational,jara2016naive}.
In standard applications, rewards are determined by the agent's desires and motivations, and costs arise from properties of the physical environment -- distance, barriers, or terrain -- that determine how much physical effort an action requires \parencite{baker2017rational,zhi2020online,jara2020naive,liu2017ten}.

We abstract joint action under shared goals and mutual knowledge into a joint ``we-agent,'' or collective actor, which captures the dyad-level mutual knowledge characteristic of stable social relationships that enables joint action without explicit bargaining or negotiation \parencite{tomasello2005understanding,chater2022paradox,shteynberg2023theory}.
We propose that given the environment $\mathbf{e} = [e_\mathrm{physical}, e_\mathrm{sociological}]$, the cost $C(a \mid \mathbf{e})$ of a joint action $a$ factors into two separable components: the physical effort of executing the action, which is based on the physical environment, and the \textit{discomfort} of taking it, which is based on the sociological environment (here, the relationship between the people involved) (\autoref{fig:model-schematic}a): 

\begin{align}
  C(a \mid \mathbf{e}) = \operatorname{effort}(a \mid e_\mathrm{physical}) + \operatorname{discomfort}(a \mid e_\mathrm{sociological}).
\end{align}

This factorization makes explicit that physical and sociological costs can exert distinct, and sometimes opposing, influences on action utilities. 
An action that is instrumentally rewarding or practically efficient may nevertheless be socially costly: for example, in a crowded conference hall, the shortest path to the water station may require walking too close to others, making a longer route preferable despite its greater physical cost. 
Our experiments exploit cases in which these costs come apart to test whether action interpretation requires reasoning about both standard physical costs and relationship-modulated discomfort.

\subsection*{Intimacy as one feature of the sociological environment}

We conduct an initial test of the framework by focusing on one aspect of interpersonal discomfort: how it depends on \textit{intimacy} (\autoref{fig:model-schematic}). 
Across cultures, times, and ages, intimacy reduces the discomfort of actions involving bodily \autocite{suvilehto2015topography,carsten2004after,sahlins2013kinship,fischler2011commensality,peng2013physiological}, spatial \autocite{hall1966hidden}, and emotional \autocite{prager1997psychology,clark1988interpersonal,shaver1988intimacy,taylor1975self} access.
This reflects a loosening of interpersonal boundaries \autocite{altman1973social,aron1992inclusion} and, in some domains, \textit{consubstantiation} -- the partial merging of selves through shared substance \parencite{carsten1995substance,thomas2022early,miller1998food}. 
As a result, actions that include close contact and transfer of bodily fluids would be costly in a formal relationship but relatively low-cost in an intimate one. 

Intimacy is therefore an ideal test case for our models: it generates salient costs along a quantifiable action dimension that influences choice in the same setting as the standard utilities in many previous tests of inverse-planning models (food preferences).
We formalize intimacy $I$ (a part of the sociological environment $e_\mathrm{sociological}$) as a graded state\footnote{Although relationships are often described as categorical types \autocite{clark1979interpersonal,fiske1992four}, both closeness and relationship strength are also represented as graded constructs \autocite{aron1992inclusion,thomas2024cognitive}.} of a dyadic relationship that systematically lowers the discomfort of risky interpersonal actions -- those that expose each person's vulnerabilities to the other:

\begin{align}
    \operatorname{discomfort}(a \mid I) = \operatorname{risk}(a) \cdot (1 - I)^{\gamma}
\end{align}

Here, $\operatorname{risk}(a)$ captures the extent to which an action creates interpersonal access, exposure, contact, dependency, or vulnerability within a broader sociocultural context, independently of the particular relationship. 
The parameter $\gamma$ determines the shape of how intimacy attenuates the discomfort of close contact.

Taken together, the we-agent abstraction and the above discomfort term make explicit two levels of shared knowledge, following \textcite{clark1996using}'s distinction between `communal' and `personal' common ground. Communal common ground -- the shared understanding that intimacy reduces the discomfort of vulnerable action --  is encoded as fixed model structure, analogous to how standard models encode shared intuitive physics \autocite{liu2026physical,baker2017rational,jara2016naive}. 
Personal common ground -- the mutual knowledge specific to a dyad, represented here by their degree of intimacy -- is represented here as a scalar variable.  
The observer can therefore use \textit{communal} common ground to recover a dyad's \textit{personal} common ground from action -- much as observers use shared knowledge of intuitive physics to recover physical structure from action -- without explicitly modeling the processes by which that mutual knowledge is formed in the first place (which concerns what \textit{constitutes} a relationship rather than how it functions in dyadic action planning; see Discussion). 

\subsection*{Using large statistical models for context-specific retrieval}

What we infer when we observe Alice and Bob sharing one burrito at the Mexican food truck depends on what other action options we believe they were choosing between.
For example, if the food truck was about to close, was out of food, or Alice and Bob were out of money, then sharing may simply have been the best available way for both people to eat; in that context, it is less diagnostic of an intimate relationship. 
The composition of this option set can strongly shape people's inferences \autocite{jern2017people}. 
Therefore, one important part of the social inference problem is figuring out what options an actor was selecting between and evaluating the features of those options.

One possibility is that observers solve this problem by considering actions relevant to the current situation and the problem they are trying to solve \autocite{phillips2019how}.
Language models (LMs) have been proposed as a class of models well suited for approximating this process, since they capture large-scale regularities of human-described situations, which can be used to approximate aspects of retrieval from associative memory \autocite{wong2025modeling,tsvilodub2026computational}. 
Related work uses large statistical models to translate continuous visual or linguistic input into symbolic representations for theory-of-mind reasoning, often with the goal of building more interpretable AI systems \autocite{zhang2025autotom,ying2023neuro,ying2025language,gelpi2025towards,ying2024pragmatic,tsvilodub2025integrating,tsvilodub2026computational}. 
Here, we apply a similar approach, for the different purpose of comparing \textit{cognitive} models in the kinds of open-ended situations that have been difficult to expand them into. 

We use separate LM components to `propose' and `evaluate' alternative actions, similar to the architecture described in \textcite{tsvilodub2026computational}. 
First, an LM component generates the alternative actions an observer might use to interpret an action, conditioned on the scenario context, the observed action, and the question the observer is answering. 
This follows how alternatives are handled in computational pragmatics -- where the alternatives comprehenders consider are shaped by the observed utterance \autocite{degen2015processing,hu2023expectations} and the question under discussion\footnote{The reported results also include a post-hoc model extension where observers give question-relevant alternatives more weight, see Methods.} \autocite{roberts2012information,kao2014nonliteral,degen2023rational} -- but combines these conditioning sources in the generation step itself using language models. 
Once the alternative actions are generated, a separate LM component maps these open-ended linguistic descriptions into the elements required by the cognitive model (here desire, risk, and physical effort). 
The resulting combined architecture means that the model can operate directly over the kinds of open-ended input that humans see in everyday life \autocite[it is ``stimulus-computable'';][]{frank2025cognitive,berke2026toward} (\autoref{fig:model-schematic}c; see Methods for detail). 

This approach addresses several limitations of standard tests of inverse-planning models.
Identifying a set of alternatives and computing their features is something that people need to do before any model-based inverse reasoning, but standard tests of inverse-planning models typically bypass this step.
They either give the agent an experimenter-specified set of possible actions whose features are manually coded \autocite{jern2017people,radkani2025people,houlihan2023emotion}, or embed the agent in a discretized physical environment, such as a ``grid world,'' where the structure of the task determines the action space and the costs associated with acting are specified by the experimenter and assumed to be common knowledge \autocite{baker2009action,jara2020naive,baker2017rational}.
In everyday settings, however, neither the relevant alternatives nor their features are specified in advance; observers must identify them from the particular context and broader world knowledge \autocite{wong2025modeling,dennett1984cognitive}.
Moreover, computing over the full set of options -- even in discretized settings -- makes model-based theory of mind computationally expensive, motivating proposals that observers rely on alternative, cheaper strategies for social inference \autocite{ho2022planning,harootonian2025mentalizing,wang2025modeling,zhi2020online}.
By contrast, our approach assumes that structured theory-of-mind reasoning happens within a restricted scope: the symbolic model evaluates only a small, context-specific set of alternatives and features supplied by separate associative components \autocite{wong2025modeling}.
On this view, expensive model-based reasoning is more plausible as an account of human inference if it operates over a small, contextually limited comparison set \autocite{vul2014one,sanborn2016bayesian,lieder2020resource}.
Together, this approach enables tests of structured cognitive models in the kinds of continuous, open-world settings that resemble the problems human observers need to solve in everyday social life.

\subsection*{Overview of experiments}

Across all studies, observers read stories covering a range of everyday social situations, and see a dyad choose to either forgo some or all of an outcome (e.g., leave the food truck), expend effort to attain it with less interpersonal vulnerability (e.g., get a knife to cut a burrito), or accept greater vulnerability to achieve the outcome and avoid that effort (e.g., bite from the same burrito).
The model explains an action choice in terms of three variables: the dyad's \textit{desire} for the outcome, which determines its reward; the \textit{physical environment}, which determines the effort required; and the relationship's \textit{intimacy}, which determines the discomfort of interpersonal vulnerability.
Because these variables provide competing explanations for the same choice, changing or learning one should change what the action implies about the others.

Manipulating or measuring the three variables defines a design space that 
the experiments traverse systematically (Study 1, Study 2) (\autoref
{fig:model-schematic}b). 
In Study~1, observers infer the dyad's desire from its action and relationship; in Study~2, they infer the relationship from the action and the dyad's desire.
Within each study, the physical environment is either specified in the vignette or inferred jointly with the primary target (the `a' and `b' experiments), testing whether a single action can coherently influence beliefs about multiple causes.
Study~3 tests whether the same model generalizes beyond the primary domain we test (food) to other kinds of interpersonal vulnerability. 
If the model is correct, then it should be able to capture people's 
judgments in all of these cases. 

We compare the full social model with two simpler ablated versions, each inspired by part of the prior literature.
The \textit{vanilla model} includes the standard utility terms from prior inverse-planning work, omitting the discomfort term \autocite{baker2017rational,jara2016naive,jara2020naive,baker2009action}.
The \textit{discomfort-only} model formalizes a direct-cue account based on documented associations between vulnerable actions and intimacy, omitting reward and physical effort \autocite{miller1998food,suvilehto2015topography,thomas2022early}.
For each experiment, we fit the full social model and its ablations by maximum likelihood to participants' trial-level belief updates, carrying variation across repeated LM elicitations into the likelihood.
Within an experiment, the predictions were generated by fitting the model on all but one scenario and then predicting the held-out scenario. 
To assess the qualitative fits between the models, we report Pearson correlation between humans and model predictions in the main text; the model comparisons under trial-level log-likelihood are in the Supplementary Information. 
The experiments and procedures were preregistered at \url{https://osf.io/47cyq/}.
The elicitation, fitting, and evaluation approach is detailed in the Methods; deviations from the preregistered plan and results under the preregistered procedure are reported in SI Section~\ref{si:preregistration-deviations}.

\section*{Results}

\subsection*{Relationship context changes inferences about reward and cost}

\begin{figure}[tp]
  \centering
  \includegraphics[width=\textwidth]{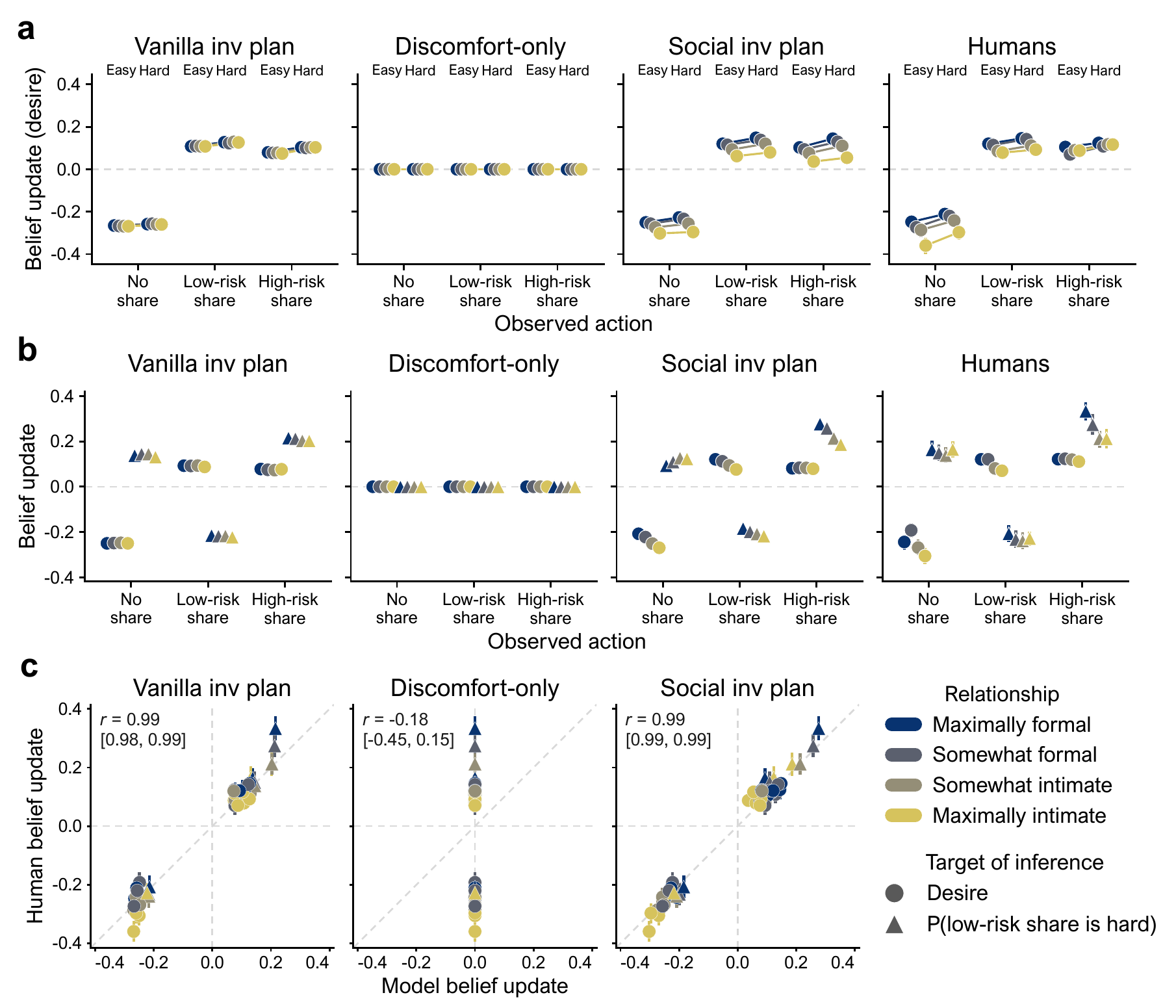}
  \caption{\textbf{Relationship context changes inferences about desire and physical constraints.}
  \textbf{(a)} Study~1a results (desire judgments, given relationship and information about the physical environment that determined whether low-risk sharing was `easy' or `hard').
  \textbf{(b)} Study~1b results (desire and physical environment judgments, given the relationship). 
  \textbf{(c)} Correlation across scenario-averaged condition means between model predictions and human judgments.
  Points are condition means and error bars are 95\% bootstrapped confidence intervals resampling participants.}
  \label{fig:study1-results}
\end{figure}

In standard inverse planning, observed behavior provides evidence about what an agent wants and what physical costs they were willing to incur: taking a difficult route suggests that the goal justified the effort, whereas avoiding it suggests that the route was costly \autocite{jara2020naive,baker2017rational}. 
Study 1 asks how these inferences change when actions occur within a social relationship. 
We use stories set in U.S. cultural contexts about food, a domain widely used for testing inverse planning models \autocite{jern2014reasoning,baker2017rational,
jern2017people}. 
The difference in our experimental framework is that it concerns the common situation in which two people eat food \textit{together} and might plausibly share it, rather than choosing food alone.
Sharing utensils, cups, or food can provide access to something desirable while creating uncomfortable bodily contact related to the likelihood and quantity of saliva transferred; additional effort, such as retrieving another utensil, can be exerted to avoid that contact \autocite{miller1998food,thomas2022early,fischler2011commensality}.

Study 1 manipulated the relationship between two people and asked how it shaped observers' inferences about desire for food (Study 1a, $N = \nRetainedOneA$) and their joint inferences about desire and physical cost (Study 1b, $N = \nRetainedOneB$). 
Participants read \nScenarios{} everyday food-sharing scenarios (see \autoref{fig:model-schematic} for an example and SI Section~\ref{si:food-scenarios} for all scenarios). 
Each scenario described the relationship between the two characters as maximally formal, somewhat formal, somewhat intimate, or maximally intimate. 
In Study 1a, participants were given information determining how effortful low-risk sharing would be -- for example, whether utensils were nearby or at another food truck. In Study 1b, this information was omitted, and participants inferred it along with desire for the food. 
Participants then observed one of three actions: no sharing (e.g., they walk away from the food truck), low-risk sharing (e.g., they retrieve a knife to divide the burrito into separate portions), or high-risk sharing (e.g., they take bites directly from the burrito) (see \autoref{fig:model-schematic}d for a visualization of the utility components under each of these actions). 
We confirmed that the scenario manipulations had their intended effects on the LM-derived features (\autoref{fig:si-feature-structure}; \autoref{fig:si-manipulation-checks}).
Before and after observing the action, participants estimated how much the two people would like the food on a continuous scale from ``not at all'' to ``extremely''; the main dependent variable was belief update (posterior minus prior).
The model captures how, after reading a scenario, an observer generates the alternative actions relevant for interpreting the observed action, assesses the features that determine their utilities\footnote{Note that we don't make commitments to the \textit{mechanism} by which these first two steps occur; see Discussion.}, and works backward from the observed choice to infer the latent variables that made it worth taking (\autoref{fig:model-schematic}c). 

First, across Studies 1a and 1b, participants showed standard qualitative signatures of inverse planning (\autoref{fig:study1-results}; scenario-level results in SI Section~\ref{si:scenario-level}), serving as a check that our paradigm works as intended: eating a food increased inferred desire for it, especially when doing so required more effort; when the physical world was unknown, participants inferred that the chosen action required relatively little effort \autocite{baker2009action,baker2017rational,jara2020naive}.

The relationship context modulated both inferences.  
Sharing was stronger evidence for high desire in more formal relationships, where greater interpersonal discomfort meant that more desire was needed to explain the choice. Conversely, declining to share was stronger evidence for low desire in intimate relationships, where sharing would have carried little sociological cost (\autoref{fig:study1-results}a). 
The relationship also changed what actions implied about the physical world. After high-risk sharing (e.g., biting from the same burrito), observers were more likely to infer that low-risk sharing would have been effortful (e.g., that utensils were far away) in more formal relationships, where the chosen action already carried greater discomfort (\autoref{fig:study1-results}b). 
Only the full social model reproduced this relationship dependence.
It captured the overall pattern of belief updates at $r = \rStudyOne$ (95\% CI \ciStudyOne), against a split-half noise ceiling of \ceilStudyOne{} (\autoref{fig:study1-results}c).
The vanilla model reproduced the standard desire inferences and their dependence on physical effort but was nearly flat across relationship levels, which enter its predictions only through the composition of the LM-generated comparison sets.
The discomfort-only model lacked reward and physical-cost terms, so it could not produce belief updates about desire; we include it here for completeness. 
Together, Studies~1a and 1b show that relationship intimacy changes ordinary inverse-planning inferences about desires and costs.

\subsection*{Actions reveal sociological and physical structure}

In Study 1, the relationship was a known feature of the sociological environment. 
We next asked whether people can use inverse planning to infer the relationship itself. 
Here we test whether people use the same generative model to make graded inferences about relationships as \textit{jointly held} beliefs about the \textit{social} environment that guide dyadic action selection. 

The design of Study 2 paralleled Study 1, but reversed the direction of inference: participants were given information about desire and inferred the relationship.
To manipulate desire, the scenarios varied details that informed the value of the food to the characters, such as whether the characters were hungry with no other options or full and preferred something else.
Participants observed no sharing, low-risk sharing, or high-risk sharing, and rated their prior and posterior beliefs about how the characters would describe their relationship, on a continuous scale from maximally formal to maximally intimate.
As in Study 1a, Study 2a ($N = \nRetainedTwoA$) specified information about the environment that determined the physical effort required for the low-risk action, whereas in Study 2b ($N = \nRetainedTwoB$) participants also inferred that physical context from the scenario.

\begin{figure}[tp]
  \centering
  \includegraphics[width=\linewidth]{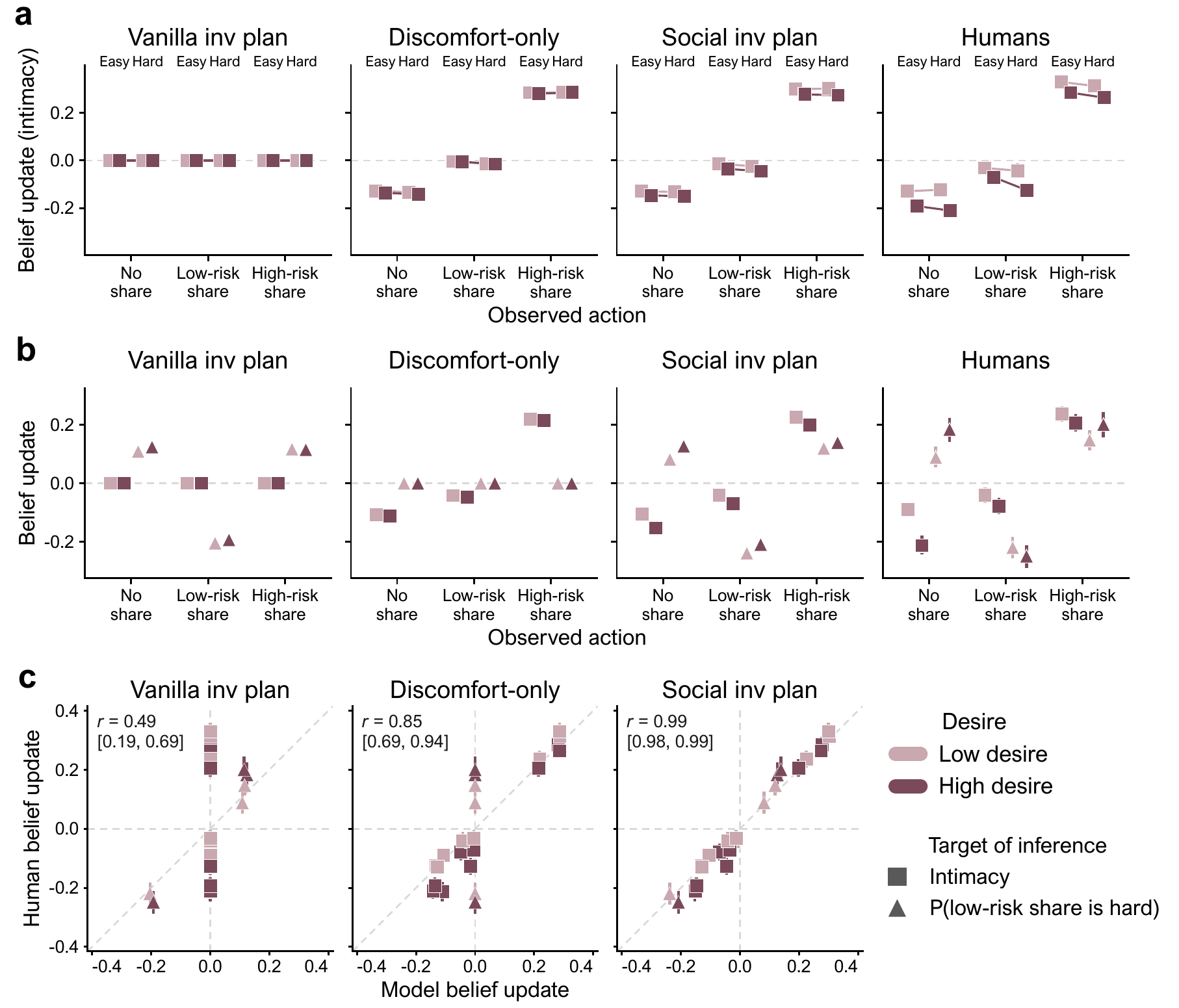}
  \caption{\textbf{Actions reveal sociological and physical structure.}
  \textbf{(a)} Study~2a results (relationship judgments, given desire and information about the physical environment that determined whether low-risk sharing was `easy' or `hard').
  \textbf{(b)} Study~2b results (relationship and physical environment judgments, given desire).
  \textbf{(c)} Correlation across scenario-averaged condition means between model predictions and human judgments.
  Points are condition means and error bars are 95\% confidence intervals from resampling participants.}
  \label{fig:study2-results}
\end{figure}

Across Studies 2a and 2b, the predicted qualitative pattern appeared: greater interpersonal vulnerability shifted observers' beliefs toward more intimate relationships, consistent with prior research on food and saliva sharing \autocite{miller1998food,thomas2022early}.
Crucially, the same action's diagnosticity also depended on the other causes that could explain it. 
High-risk sharing produced weaker intimacy inferences when the characters strongly desired the food or when the lower-risk alternative required greater effort (\autoref{fig:study2-results}a,b). 
For example, biting from the same burrito was weaker evidence for intimacy when the characters strongly wanted the burrito or when obtaining a knife was difficult, because either factor provided an alternative explanation for choosing the socially riskier action.
Study 2b further showed that observers could jointly infer sociological and physical structure: when both intimacy and the physical environment were uncertain, high-risk sharing simultaneously increased inferred intimacy and the inferred effort required for low-risk sharing (\autoref{fig:study2-results}b). 

Only the full social model captured this dependence on competing explanations.
It reproduced the overall pattern of updates at $r = \rStudyTwo$ (95\% CI \ciStudyTwo), against a noise ceiling of \ceilStudyTwo{} (\autoref{fig:study2-results}c).
The vanilla model could not infer relationship intimacy from action because it had no representation of relationship structure.
The discomfort-only direct-cue model reproduced the direction of the intimacy updates but not their dependence on desire and effort or the joint recovery of social and physical structure.
Together, Studies~2a and 2b show that observers treat relationship intimacy -- an aspect of sociological structure -- as a latent feature of the action-generating environment, making inferences about relationships by considering how it integrates with desires and physical constraints within the same generative model.

\subsection*{The model generalizes beyond food to other ways of being vulnerable}

\begin{figure}[tp]
  \centering
  \includegraphics[width=\linewidth]{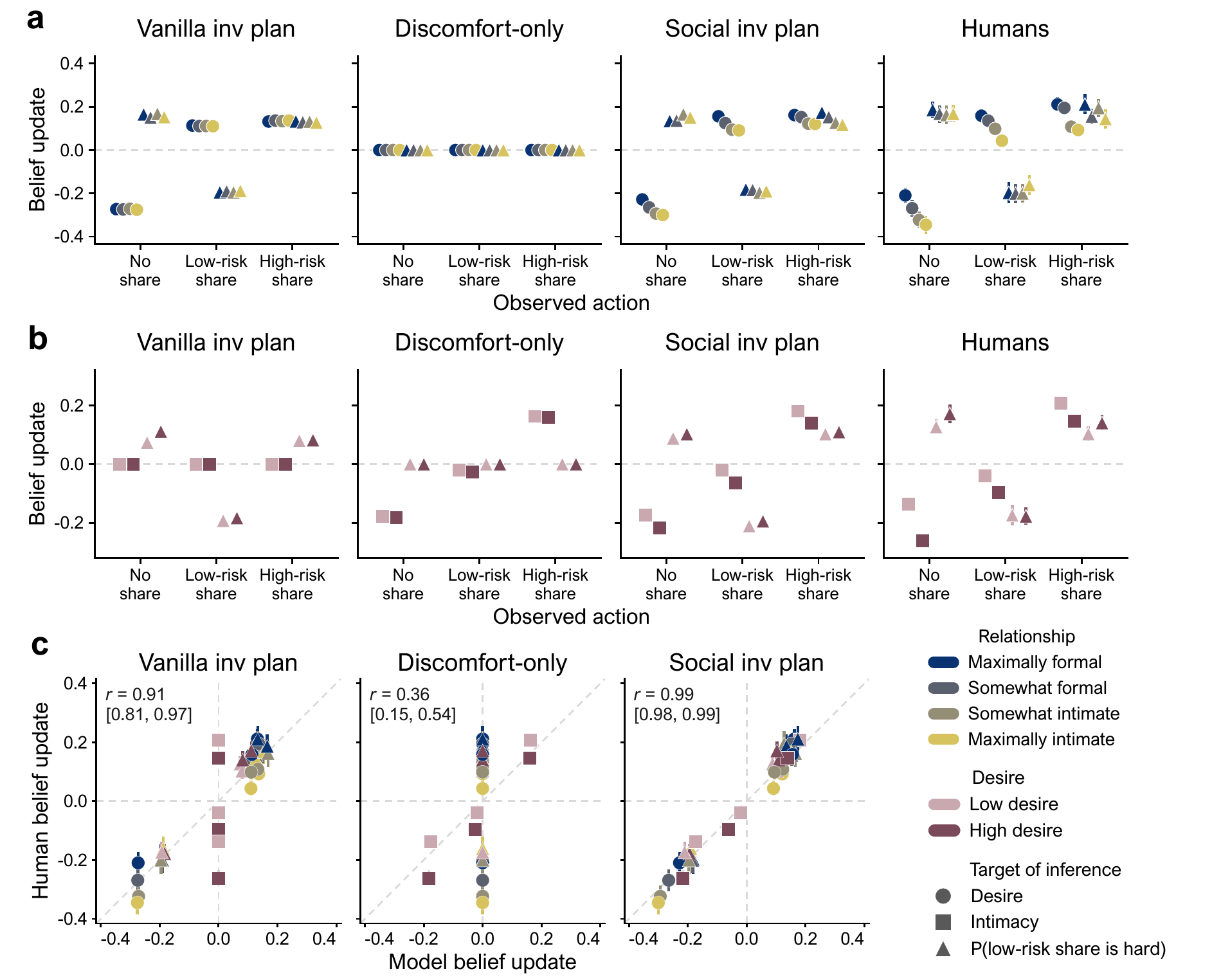}
  \caption{\textbf{The model generalizes beyond food sharing.}
  \textbf{(a)} Study~3a results (desire and physical environment judgments, given relationship).
  \textbf{(b)} Study~3b results (relationship and physical environment judgments, given desire).
  \textbf{(c)} Correlation across scenario-averaged condition means between model predictions and human judgments.
  Points are condition means and error bars are 95\% bootstrapped confidence intervals (resampling participants).}
  \label{fig:study3-results}
\end{figure}

The studies so far focus on food sharing, in which the relevant interpersonal risk comes from the likelihood and quantity of saliva transfer. 
However, many social actions expose people to one another in other ways. 
In Study 3, we test whether the model generalizes to other forms of interpersonal vulnerability. 
We consider three such forms, each corresponding to part of the prior literature. 
The first is \textit{substance}: allowing another person to touch one's body or to come into contact with bodily substances.
The set of bodily regions that people allow each other to contact expands with the closeness of the relationship, so that incidental or formal contact such as a handshake or fist bump is acceptable even between strangers, whereas closer or more sustained contact -- a hug, holding hands, sitting pressed side to side -- is reserved for more intimate relationships \autocite{jourard1966body,suvilehto2015topography}.
The second is \textit{space}: being physically close to another while in a vulnerable state such as being asleep or undressed.
Here the interpersonal risk comes from letting one's guard down in another's presence, which is normally reserved for close relationships \autocite{hall1966hidden,kennedy2009personal,tompkins2026development}. 
The third is \textit{privacy}: disclosing personal information, or granting entry to one's private spaces or possessions.
Relationships deepen through progressively more vulnerable disclosure and access, from public facts toward private inner layers, so the depth of access a person grants to another tracks how close the relationship is \autocite{altman1973social,clark1988interpersonal,prager1997psychology,shaver1988intimacy,petronio2002boundaries,richardson2025children}.

These forms of interpersonal vulnerability have largely been studied in separate research traditions, yet they share a common structure: in each, the same behavior can be comfortable and expected in one relationship yet awkward, intrusive, or inappropriate in another, based on how formal or intimate the relationship is.
If our model captures people's inferences across all of these domains, it would suggest that these phenomena are expressions of a common inferential structure, in which one dimension of relationships -- intimacy -- reduces the cost of actions that expose one person to another, and can be recovered from those actions based on the same causal model.

We wrote \nScenarios{} scenarios, structurally matched to the food scenarios, spanning the three above categories (see SI Section~\ref{si:nonfood-scenarios} for all scenarios).
The substance scenarios used non-food objects whose use transfers bodily substances -- saliva, sweat, or skin oils -- or requires direct skin contact, such as a chapstick, a sweat-damp towel, a shared pair of earbuds, a hairbrush, or rubbing sunscreen onto someone's back.
The space scenarios involved being together while physically unguarded, such as sleeping under one blanket, sharing a sleeping bag or bed, or changing in a locker room near each other.
The privacy scenarios involved disclosing sensitive information or granting access to a private sphere, such as confiding the painful details of a breakup, trading sensitive gossip, comparing salary details, hosting someone in an untidied home, or letting someone use an unlocked phone. 
Using these scenarios, Study~3 parallels the two joint inference studies from Studies~1 and~2 (the `b' variants in \autoref{fig:model-schematic}b).
In Study~3a ($N = \nRetainedThreeA$), as in Study~1b, the relationship is given and observers infer desire and the physical environment. 
In Study~3b ($N = \nRetainedThreeB$), as in Study~2b, desire is given and observers infer the relationship and the physical environment. 

\begin{figure}[tp]
  \centering
  \includegraphics[width=\linewidth]{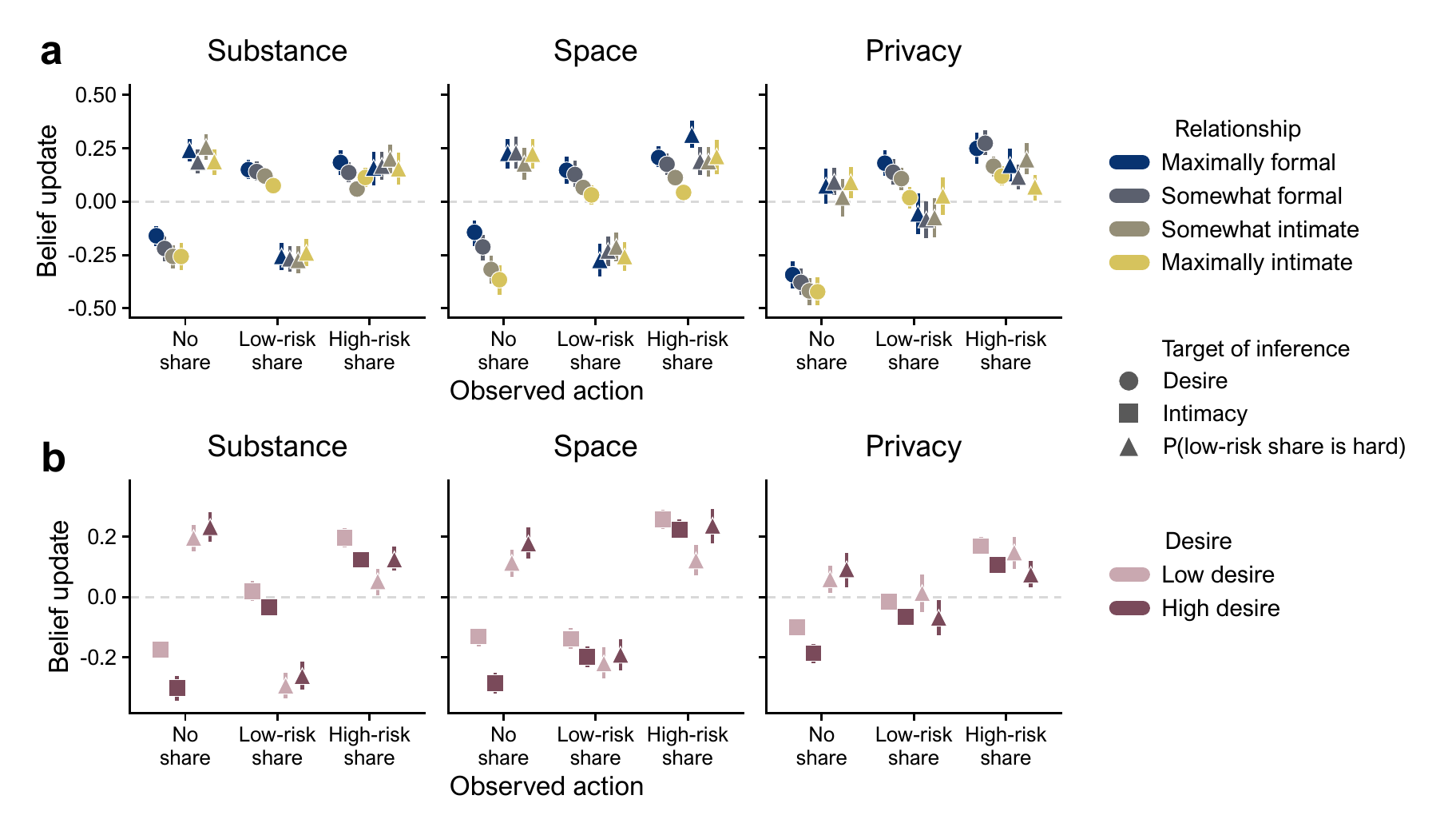}
  \caption{\textbf{Human belief updates by non-food domain.}
  Study~3a (\textbf{a}) and Study~3b (\textbf{b}) condition means split by substance, space and privacy.
  The same qualitative action signatures appeared in each domain.
  Error bars are 95\% bootstrapped confidence intervals (resampling participants).}
  \label{fig:study3-domains}
\end{figure}

Studies~3a and 3b reproduced these qualitative patterns across non-food forms of interpersonal vulnerability.
When the relationship was given, its intimacy changed what vulnerable actions implied about desire and physical environment (Study~3a).
When desire was given, more vulnerable actions shifted beliefs toward greater intimacy, and observers jointly inferred intimacy and the physical constraints on lower-risk alternatives (Study~3b; \autoref{fig:study3-results}a,b).
The broad action patterns appeared in each of the substance, space and privacy domains (\autoref{fig:study3-domains}).

The full social model captured the pattern across the non-food scenarios ($r = \rStudyThree$, 95\% CI \ciStudyThree, against a noise ceiling of \ceilStudyThree; \autoref{fig:study3-results}c).
As in the food studies, the vanilla model could not capture relationship-dependent inference, whereas the discomfort-only model could not capture how desire and physical effort changed the meaning of the same action.
Moreover, utility weights estimated from the food experiments alone predicted non-food judgments about as well as weights estimated directly from the non-food data ($r = \rNonfoodFoodFit$, 95\% CI \ciNonfoodFoodFit, against $r = \rNonfoodOwnFit$ for weights fit to the non-food data; SI Section~\ref{si:generalization}).
Together, these findings suggest that the inferential structure identified in Studies~1 and~2 is not specific to eating or saliva transfer -- such as a contamination response specific to eating \autocite{rozin1987perspective} -- but extends across distinct forms of interpersonal vulnerability. 

\section*{Discussion}

We developed an experimental and computational framework for incorporating social relationships into inverse planning.
Across experiments, people read brief descriptions of naturalistic social interactions resembling the sparse evidence available in everyday social observation and drew rich, systematic inferences from them.
Their judgments were best captured by a model in which relationship-dependent costs are integrated with goal value and physical effort in a single generative model of action selection.
The same model structure supported inferences about desires, beliefs about the physical environment, and beliefs about the sociological environment. 

Standard tests of inverse planning typically abstract away from social context, even when they examine actions, such as navigating a physical environment to retrieve food, that commonly occur around other people \autocite{baker2009action,baker2017rational,jara2016naive,jara2020naive}.
Our results show that when actions are taken together and embedded in everyday social interactions with other people, the sociological environment must be represented explicitly for the models to explain inferences about goals and costs. 
By representing actions as taken jointly by dyads, our model specifies a level of abstraction at which aspects of relationships that are `social constructs' enter inverse planning. 

This research extends inverse planning toward more naturalistic and everyday social situations.
We used a hybrid `neurosymbolic' approach that combines language-model and probabilistic components.
Our approach can be viewed as an instance of on-demand probabilistic model synthesis \autocite{wong2025modeling}: a language model supplies context-specific model components, and a structured probabilistic model performs coherent inference over them.
Rather than synthesizing the entire model from scratch, however, we held the structure of the cognitive model fixed and used the language model to generate and evaluate the components that enter it.
This addresses a central challenge in open-world action understanding of how observers generate alternative actions, given that an observed action is informative only relative to the actions the agent could have taken but did not \autocite{gergely2003teleological,jara2016naive,baker2017rational}. 
This alternative-generation problem has been characterized extensively in pragmatic language understanding, formalized using the same class of theory-of-mind models \autocite{goodman2016pragmatic}, where an utterance's meaning depends on what a speaker could have said but did not \autocite{degen2015processing,degen2023rational}. 
Approaches in computational linguistics and cognitive AI have approximated alternative generation and evaluation using neural language models \autocite{hu2023expectations,qiu2025same,tsvilodub2025integrating,tsvilodub2026computational}, without committing to the cognitive processes by which human observers generate those alternatives \autocite{buccola2022conceptual,phillips2019how,bear2020comes,morris2021generating}. Here, we extend this strategy beyond pragmatics to broader action understanding, making structured cognitive models testable in open-world contexts and bringing inverse-planning models closer to the problem humans face in everyday social reasoning.

Notably, the same model structure supports inferences not only about the standard targets of inverse planning, but also about relationships themselves.
Unlike walls, distances, or other features of the physical environment, relationships are less likely to be directly perceived, so observers must learn about them by interpreting everyday behavior \autocite{miller1998food,thomas2022early,liberman2018secret,fiske1992four}.
Our account complements two existing approaches to social relationship reasoning: (1) `learning' models that describe how people integrate information about who is connected into representations of networks and groups \autocite{son2023abstract,lau2018discovering,davis2026inferring,basyouni2022mapping}, and (2) vicarious-utility models that describe how much one agent values another's welfare \autocite{kleiman2017learning,ullman2009help}.
Here, we characterize a third feature of relationships -- the formality-intimacy axis -- that cannot be reduced to either social connectedness or concern for another's welfare, but nevertheless plays an important role in social behavior.
How formal or intimate a relationship is reflects a dyad's mutual knowledge that functions as a stable feature of their shared social environment \autocite{searle1995construction,gilbert1989social}.
Our model explains how a single observation can provide graded evidence about this shared structure.

A complete formal account of reasoning about social relationships should therefore incorporate several ways that relationships are represented.
To understand relationships, observers need to reason about whether a social connection exists at all, how much one person values another's welfare, and what shared expectations govern joint action. 
The present research focuses on this third component, and how it may be inferred as an environmental variable in an integrated planning model.
Together, these components span cognitive mechanisms that operate at multiple levels: interpreting the beliefs, desires, and utilities of individual minds; representing joint action and shared knowledge within a dyad; and accumulating evidence across observations into broader representations of social groups and networks.
A unified formal account of intuitive sociology would need all of these components and mechanisms. 

The shared expectations that people often have in stable relationships enable joint action to proceed without explicit negotiation \autocite{lewis1969convention,clark1996using,tomasello2005understanding,chater2022paradox}.
Consistent with this idea, our model represents the dyad as a single planner \autocite{gilbert1989social,sugden2003logic}.
This abstraction captures the dyad-level mutual knowledge that is most relevant for the third-party observer tasks in our experiments, where judgments concern the dyad (``their relationship'') rather than either individual, and where observers do not have direct access to either person's mental states. 
At the same time, this leaves out the processes by which mutual knowledge is formed, negotiated, or repaired.
Risky or vulnerable actions may not only express an existing relationship, but also create and affirm the shared knowledge that makes a relationship close or intimate in the first place \autocite{altman1973social,rossignac2026relationships,rossignac2021merged}. 
In this respect, expectations about relationships may develop in a way analogous to shared meaning in language: across repeated interactions, shared expectations guide the interpretation of actions, while actions in turn update those expectations \autocite{lewis1969convention,clark1996using,hawkins2023partners}. 
Separating within-dyad beliefs may be especially important for understanding cases where expectations diverge, such as cases where an action is performed reluctantly or nonconsensually \autocite{sommers2020commonsense,demaree2022autonomy}. 

The domains we studied -- commensality, social touch, personal space, and self-disclosure -- each have their own research traditions and proposed mechanisms \autocite{fischler2011commensality,suvilehto2015topography,hall1966hidden,altman1973social}, and the sociological cost in each plausibly draws on a different source -- contamination, threat, or reputation.
Yet one computational structure -- a sociological cost, attenuated by intimacy and traded off against reward and effort within a planning model -- captured inference in our setting across all of them, without committing to the proximal sources of that cost.
Across scenarios, the lower-vulnerability route required additional effort to preserve an interpersonal boundary: retrieving a clean utensil to avoid saliva transfer and producing an indirect or roundabout account to avoid explicit informational disclosure play the same role in the model \autocite{brown1987politeness,pinker2008logic}.
In the context of pragmatic language, utility-theoretic models have already shown how social goals and communicative efficiency can be traded off in action selection \autocite{yoon2020polite,goodman2016pragmatic}.
Our results suggest a parallel structure for social action more broadly: people choose among alternatives that vary in instrumental value, effort, and sociological cost, and observers invert those choices to infer what the dyad wanted, what physical constraints they faced, and what relationship they had.
Because a single observed choice can locate the dyad on this trade-off, it can carry graded information about desire, effort, and intimacy all at once.

One benefit of formalizing this reasoning with a cognitive model is that it can help distinguish mechanisms that are difficult to separate using verbal theories alone.
For example, people report more comfort with infection-risky contact -- sharing utensils or drinking from the same cup -- with partners they value more, including honest or agreeable strangers \autocite{tybur2020behavioral}, and people are less likely to share food or act in other vulnerable ways across social hierarchies or groups \autocite{miller1998food,reicher2016core}.
These associations could arise in at least two ways.
One possibility, proposed by \textcite{tybur2020behavioral}, is that the expected benefits of associating with a valuable or cooperative partner directly offset the costs of possible infection.
However, another possibility is that partners that value each other's welfare are more likely to be in intimate relationships, and that intimacy is instead the key variable that directly makes risky contact feel less costly.
By making relationship-modulated discomfort explicit and separable from vicarious value, reward, and physical cost, the model provides a way to test these possibilities quantitatively.

Several important future directions are outside the scope of the present studies.
First, we focused on intimacy, but other aspects of sociological structure may affect action planning by shaping the utility of actions in different ways.
Asymmetries in social rank, for instance, may change the utility of exchanges that differ in \textit{sequence} and \textit{directionality} -- who initiates, and whether people alternate or follow a precedent -- whereas closeness may matter more for the \textit{content} of an exchange, such as food versus formal gifts \autocite{chen2026expectations,graeber2012debt,brown1987politeness}.
Other kinds of social interactions are defined less by dyadic expectations than by institutional roles and obligations (e.g., doctor and patient), which could involve a distinct role-based stance \autocite{jara2024institutional,baker2026people}.
Mapping dimensions of relationships and social structure onto the action features they affect could help both models and human observers distinguish among, reason about, and communicate different features of relationships in the context of all of these features of social structures. 

Second, we tested how U.S. participants evaluated familiar U.S. contexts. 
An important question is which aspects of these findings generalize across cultures. 
One source of variation is which actions are considered interpersonally risky in the first place.
Sharing from a communal dish, greeting a new acquaintance by kissing their cheek, or disclosing personal information such as salary are common in some cultures or communities, even between people in relatively formal relationships \autocite{curtis2011disgust,wang2021cross,tybur2018people}.
The way closeness decreases discomfort, however, may be more general \autocite{curtis2004evidence,kamilouglunorms}. 
Expressed using our model, one hypothesis therefore is that the presence and the functional form of the discomfort term is general, the LM-elicited components are culturally specific, reflecting the statistics of humans' and LMs' training distribution\footnote{Note that we used a U.S.-developed LLM, prompted to act as a ``regular adult from the United States.'' LM outputs disproportionately reflect the Western norms captured in their training data and procedures, and prompt-based strategies don't fully correct this \autocite{tao2024cultural}. Cross-cultural extensions would require retrieval components validated on data from the community being modeled.}; and the degree to which intimacy attenuates discomfort (the $\gamma$ term in the model) varies across individuals. 
The model can be used to generate quantitative predictions that distinguish these sources of variation. 

Together, our results show that social relationships are part of the causal structure through which action is interpreted.
By integrating intimacy into a generative model of action choice, this work extends inverse planning beyond mental states and physical constraints to a broader class of sociological utilities that are relevant for action selection, action interpretation, and the maintenance of relationships.

\newpage

\section*{Methods}

\subsection*{Participants}

We recruited adults through Prolific who were fluent in English and lived in the United States.
Before starting the experiment, participants had three attempts to pass a quiz about the instructions; during the experiment, they completed one attention check and three memory-checks questions.
We retained participants who passed the attention check and answered at least one memory-check question correctly.
Participants who completed one experiment could not participate in a later one.
Each experiment took approximately 20 minutes, and participants received \$5.
Participants gave informed consent, and all procedures were approved by the MIT Committee for the Use of Humans as Experimental Subjects. 

We preregistered the sample size for each experiment to provide approximately 20 observations for each condition $\times$ scenario combination.
Across the six experiments, we recruited \nRecruitedTotal{} participants and retained \nRetainedTotal{}; participant characteristics and exclusions by experiment are reported in SI Section~\ref{si:supplementary-methods}.

\subsection*{Materials and procedure}

At the beginning of the experiment, participants were told that some relationships are formal, such as relationships with an employee, religious leader, shopkeeper, or new acquaintance, while others are close and intimate, such as relationships with a romantic partner, sibling, or best friend.
They were then told that they would read short descriptions of two people interacting in everyday situations and use their best judgment to evaluate information that the descriptions left out.

Studies~1 and 2 used \nScenarios{} food-sharing scenarios, and Study~3 used \nScenarios{} structurally matched non-food scenarios.
For each scenario, we wrote three possible observed actions (not sharing, lower-vulnerability sharing, and higher-vulnerability sharing), two world-state descriptions that determined how effortful the lower-vulnerability action would be, and two descriptions that varied how much the characters valued the outcome (e.g., how much they wanted to eat a food or retrieve an object).
The relationship was described as maximally formal, somewhat formal, somewhat intimate, or maximally intimate.
The complete scenarios are provided in SI Section~\ref{si:scenarios}.

Participants evaluated their beliefs before and after observing the action.
Depending on the experiment, they rated how much the characters desired the outcome, from ``Not at all'' to ``Extremely''; how the characters would describe their relationship, from ``Maximally formal'' to ``Maximally intimate''; and/or the relative likelihood of two physical-world states, with ``Equally likely'' at the midpoint.
For each judgment, the dependent variable was the posterior rating minus the prior rating.

Every experiment manipulated the observed action.
Study~1a also manipulated the relationship and physical world and measured desire; Studies~1b and 3a manipulated the relationship and measured both desire and the physical world; Study~2a manipulated desire and the physical world and measured the relationship; and Studies~2b and 3b manipulated desire and measured both the relationship and the physical world. 
Participants saw every scenario once, with conditions balanced across scenarios and participants, and the trial order randomized.
The experiments were built in jsPsych \autocite{de2023jspsych}, and the data were collected using DataPipe \autocite{de2024datapipe}.

\subsection*{LM elicitation}

We used \texttt{meta-llama/\lmName}, queried through the Together AI API, to construct the inputs to the cognitive model.
For each scenario and experimental condition, the LM was first prompted to generate plausible alternative actions that an observer might use to interpret the observed action (the ``comparison set'').
A separate set of queries then evaluated the actions on the features that enter the utility function.

Formally, if $\tau$ contains the information in the scenario, the observed action, and the question participants answered,
\begin{equation}
  \begin{aligned}
    \{a_1, \ldots, a_m\} &\sim P_{\mathrm{LM}}(\cdot \mid \tau), \qquad
    \mathcal{A}_{\tau} = \{a_{\mathrm{obs}}\} \cup \{a_1, \ldots, a_m\}, \\
    \phi_{\tau}(a) &= \big(g_{\tau}(a),\, \operatorname{risk}_{\tau}(a),\, \operatorname{effort}_{\tau}(a)\big).
  \end{aligned}
  \label{eq:lm-elicitation}
\end{equation}
The resulting action set $\mathcal{A}_\tau$ contains the observed action and the alternatives generated for that scenario.
Note that because $\tau$ includes the observed action and the question, $\mathcal{A}_\tau$ is an observer-constructed \textit{local} action set rather than an assertion about the actor's complete action set before the action was observed in the first place.

The feature map $\phi_\tau$ assigns each action a measure of goal satisfaction $g_\tau(a)$, interpersonal risk $\operatorname{risk}_\tau(a)$, and the total effort required across the dyad $\operatorname{effort}_\tau(a)$ (see below for how these features factor into the model).
For the disclosure scenarios, note that effort includes the work required to produce an account, such as the length of the explanation, consistent with how cost is treated in models of pragmatics \autocite{goodman2016pragmatic}. 
These features were rated on a $0$--$6$ scale and rescaled to $[0,1]$.
When desire or intimacy was given in the vignette rather than inferred, the LM also rated the magnitude implied by the corresponding description. 
As a check, we verified that the elicited features preserved the designed structure of the manipulations (Figure \ref{fig:si-feature-structure}-\ref{fig:si-manipulation-checks}).

The alternative-generation prompt templates were the same across experiments and included the condition information available to participants, the observed action, and the question they answered.
Thus, the experimental conditions could change which actions were generated as alternatives.
For feature scoring, goal satisfaction and risk for each action were scored independent of condition, and effort was scored under each physical-world state. 
We checked the properties of the generated alternatives: \nAltsMedianRange{} alternatives were typically generated (\autoref{fig:si-action-sets}a), and the sets predominantly featured action types other than the one observed, with composition changing modestly with relationship and desire (\autoref{fig:si-action-sets}c,d).
Note that comparing the generated alternatives or feature scores with human alternative generation or human feature ratings is not of interest here -- all model variants used the same LM-generated comparison sets and feature scores, so the analysis compares the cognitive models conditional on a common elicited representation -- but component-level validation has been done in prior work \autocite{tsvilodub2025integrating,tsvilodub2026computational,hu2023expectations}. 

We repeated alternative generation $\nRuns$ times for each scenario $\times$ condition cell; for each run, we then pooled and scored the actions at the scenario level as described above.
We carried the variation across these complete elicitation runs into the model likelihood.
The action-generation step was sampled at temperature \altTemperature{} and the feature-scoring step (and the given-magnitude desire and intimacy ratings) was sampled at temperature \scoreTemperature{}.
Further implementation details and complete prompts are in SI Sections~\ref{si:supplementary-methods} and~\ref{si:prompts}.

\subsection*{Probabilistic model}

In the full social model, observers have an intuitive theory that an actor chooses actions proportional to their relative total utilities, where the utility of an action is

\begin{align}
    U_\mathrm{total}(a \mid d, I, e_\mathrm{physical}, \tau) &= R(a \mid d) - \operatorname{effort}(a; e_\mathrm{physical}) - \operatorname{discomfort}(a; I)\\
    &= w_v \cdot d \cdot g_\tau(a) - w_e \cdot \operatorname{effort}_\tau(a; e_\mathrm{physical}) - w_d \cdot \operatorname{risk}_\tau(a) \cdot (1 - I)^\gamma,
    \label{eq:utility}
\end{align}

Here, $d \in [0,1]$ is how much the characters desire the outcome, $I \in [0,1]$ is the intimacy of their relationship, and $e_\mathrm{physical} \in \{0,1\}$ indexes the experimentally manipulated physical environment, which determines the effort of the lower-vulnerability action.
The observer treats $\mathcal{A}_\tau$ as a local approximation to the actions against which the observed choice is evaluated and uses a softmax to compute the local action probability $p_\tau$.
The observer then inverts this model to recover the latent variables $\mathbf{z}$ that were not given in that experiment.
\begin{align}
    p_\tau(a \mid \mathbf{z}, \mathcal{A}_\tau)
    &=
    \frac{\exp \left(U_\mathrm{total}(a \mid \mathbf{z}, \tau)\right)}
    {\sum_{b \in \mathcal{A}_\tau}\exp \left(U_\mathrm{total}(b \mid \mathbf{z}, \tau)\right)},
    \qquad a \in \mathcal{A}_\tau,\\
    q_{\alpha_{\mathrm{obs}}}(\mathbf{z} \mid a_{\mathrm{obs}}, \mathcal{A}_\tau, \tau)
    &\propto p_\tau(a_{\mathrm{obs}} \mid \mathbf{z}, \mathcal{A}_\tau)^{\alpha_{\mathrm{obs}}}P(\mathbf{z}).
    \label{eq:inverse-planning}
\end{align}

In the model, we set uniform priors over the inferred variables, so that its predictions directly reflect the diagnosticity of the observed action. 
In principle, the LM could also estimate priors, but we chose not to, to keep the predictions focused on action diagnosticity. 
Note that because of these experimental and modeling choices (we elicit point estimates rather than distributions, and our model does not estimate the prior distributions), the approach cannot capture how the \textit{concentration} of the prior might modulate the size of the belief update.
However, this limitation applies equally to all model variants and therefore should not affect comparisons between them.
Participants' prior and posterior ratings are shown in \autoref{fig:si-prior-posterior-levels}.

The vanilla model omits the discomfort term, and the discomfort-only model omits the reward and effort terms.
To compute the posteriors, we discretized the continuous variables in increments of $0.01$ and used exact enumeration by normalizing JAX arrays. We used the semantics of memo, a probabilistic programming language for expressing theory of mind models, to verify the model outputs \autocite{chandra2025domain}.

\subsection*{Question-relevant reweighting}

In the preregistered specification, the observer equally weighs every action in the generated comparison set.
After examining the residuals, we added a post-hoc extension that gives more weight to question-relevant actions in the comparison set when the observed action is surprising.
We applied the reweighting to physical-state and intimacy judgments, whose evidence mostly depends on forgone actions, but not to desire judgments, for which the observed action is usually directly informative and the preregistered procedure already captured well (this follows the representational hierarchy between `value inference' and `belief inference' detailed in \textcite{wu2022representational}).
The full specification, scope, and comparison with the preregistered specification are reported in SI Section~\ref{si:reweighting-detail}.

\subsection*{Fitting and model comparison}

Because each LM elicitation produces a different comparison set and set of feature ratings, the model can make a slightly different prediction $\delta_k$ on each run.
We retained this variation by scoring each participant's belief update $u$ under an equally weighted mixture of the $K$ individual runs:
\[
    p(u) = \frac{1}{K}\sum_{k=1}^{K}\mathcal{N}(u \mid \delta_k,\,\sigma^2).
\]
The spread of the $\delta_k$ values captures variation introduced by the LM elicitation, while $\sigma$ captures the remaining variation in participants' responses.
For each experiment and model variant, we jointly fit the utility parameters, the observer inverse temperature $\alpha_{\mathrm{obs}}$, the response-noise scale $\sigma>0$, and, where applicable, the reweighting gain $\eta$ by maximum likelihood.
The actor inverse temperature was fixed at one for identifiability, and the joint-inference studies used the corresponding bivariate mixture.
Further details about the likelihood are reported in SI Section~\ref{si:supplementary-methods}, with the parameter estimates in \autoref{tab:fitted-params} and predictive checks in \autoref{fig:si-variability-checks}a,b.

We generated all reported predictions out of sample, using cross-validation that held out one scenario at a time.
For each of the \nFolds{} folds, the parameters were fit to \nFoldsTrain{} scenarios and used to predict the remaining one. 
The utility weights and intimacy exponent were constrained to be positive. 
Following previous work, our primary measure of model performance was the correlation across scenario-averaged condition means, comparing held-out model predictions with human belief updates \autocite{baker2017rational,jara2020naive,houlihan2023emotion}.
We report this Pearson correlation relative to a participant split-half noise ceiling. The reported 95\% confidence intervals are from resampling the condition-mean pairs.

\paragraph{Cross-experiment generalization}

Because the weights were fit separately in each experiment, we also tested whether the same utility could generalize between experiments.
We fit one set of utility weights jointly to the four food experiments and applied those weights unchanged to the two non-food experiments, re-estimating only the observer parameters ($\alpha_{\mathrm{obs}}$, $\sigma$, and $\eta$) and cross-validating as above.
An additional analysis in SI Section~\ref{si:generalization} fits one utility across all six experiments.

\newpage

\section*{Code and data availability} 

All de-identified code and data are available at \url{https://github.com/aliciamchen/sip} and deposited on Zenodo at \url{https://doi.org/10.5281/zenodo.22088285}.

\section*{Acknowledgments}

We thank Robert Hawkins, Jae-Young Son, Marlene Berke, Lio Wong, and Pablo Rodriguez Osztreicher for helpful discussions and feedback. 
This research was supported by the Simons Foundation Autism Research Initiative through the Simons Center for the Social Brain at MIT (award \#6951949 to R.S.), as well as by AFOSR (FA9550-22-1-0387), the ONR Science of AI program (N00014-23-1-2355), and a Schmidt AI2050 Fellowship to J.B.T.

\section*{AI disclosure}

We disclose the use of AI tools in the preparation of this work.
AI coding agents (Claude Code, Cursor Grok/Composer, and Codex) were used for code generation after December 2025.
Coding agents were also used to generate initial drafts of the `non-food' stimuli in Study 3 using the `food' stimuli as a basis, and of portions of the Methods and Supplementary Information. 
AI tools were used for language assistance and consistency checks across the manuscript, preregistrations, and code base.
All AI-assisted outputs were reviewed and edited by the authors. 
The authors take full responsibility for the content of the manuscript.
The instruction files for the coding agents are available in the project repository.

\section*{Competing interests}

The authors declare no competing interests.

\newpage

\printbibliography

\newpage

\appendix

\setcounter{figure}{0}
\renewcommand{\thefigure}{S\arabic{figure}}
\renewcommand{\theHfigure}{S\arabic{figure}}
\setcounter{table}{0}
\renewcommand{\thetable}{S\arabic{table}}
\renewcommand{\theHtable}{S\arabic{table}}
\setcounter{section}{0}
\renewcommand{\thesection}{\arabic{section}}
\renewcommand{\theHsection}{S\arabic{section}}
\renewcommand{\thesubsection}{\arabic{section}.\arabic{subsection}}
\renewcommand{\theHsubsection}{S\arabic{section}.\arabic{subsection}}

{\centering\section*{Supplementary information}\par}

\makeatletter
\renewcommand\l@paragraph[2]{}
\renewcommand\l@subparagraph[2]{}
\makeatother
\tableofcontents
\newpage

\section{Supplementary methods}
\label{si:supplementary-methods}

\subsection{Participant characteristics and exclusions}

We retained participants who passed the attention check and answered at least one of these three questions correctly.
In Study~1a, the preregistered rule excluded a participant only if they failed the attention check and answered all three memory-check questions incorrectly; no participant met this criterion, so we changed it for the following studies.

\begin{table}[h!]
  \centering
  \small
  \setlength{\tabcolsep}{4pt}
  \begin{tabular}{lrrrrl}
    \toprule
    Study & Recruited & Retained & Excluded & Age, M (SD) & Gender, F/M/NC/NR \\
    \midrule
    1a & \nRecruitedOneA & \nRetainedOneA & \nExcludedOneA & \ageMeanOneA{} (\ageSdOneA) & \nFemaleOneA/\nMaleOneA/\nNonconformingOneA/\nAbstainOneA \\
    1b & \nRecruitedOneB & \nRetainedOneB & \nExcludedOneB & \ageMeanOneB{} (\ageSdOneB) & \nFemaleOneB/\nMaleOneB/\nNonconformingOneB/\nAbstainOneB \\
    2a & \nRecruitedTwoA & \nRetainedTwoA & \nExcludedTwoA & \ageMeanTwoA{} (\ageSdTwoA) & \nFemaleTwoA/\nMaleTwoA/\nNonconformingTwoA/\nAbstainTwoA \\
    2b & \nRecruitedTwoB & \nRetainedTwoB & \nExcludedTwoB & \ageMeanTwoB{} (\ageSdTwoB) & \nFemaleTwoB/\nMaleTwoB/\nNonconformingTwoB/\nAbstainTwoB \\
    3a & \nRecruitedThreeA & \nRetainedThreeA & \nExcludedThreeA & \ageMeanThreeA{} (\ageSdThreeA) & \nFemaleThreeA/\nMaleThreeA/\nNonconformingThreeA/\nAbstainThreeA \\
    3b & \nRecruitedThreeB & \nRetainedThreeB & \nExcludedThreeB & \ageMeanThreeB{} (\ageSdThreeB) & \nFemaleThreeB/\nMaleThreeB/\nNonconformingThreeB/\nAbstainThreeB \\
    \bottomrule
  \end{tabular}
  \caption{\textbf{Participant characteristics and exclusions.} Gender counts describe the recruited sample; NC denotes non-conforming and NR denotes declined to report. Age ranges for Studies~1a--3b were \ageMinOneA{}--\ageMaxOneA{}, \ageMinOneB{}--\ageMaxOneB{}, \ageMinTwoA{}--\ageMaxTwoA{}, \ageMinTwoB{}--\ageMaxTwoB{}, \ageMinThreeA{}--\ageMaxThreeA{}, and \ageMinThreeB{}--\ageMaxThreeB{}, respectively.}
  \label{tab:participants}
\end{table}

\subsection{LM elicitation details}

In \autoref{eq:lm-elicitation}, $P_{\mathrm{LM}}(\cdot \mid \tau)$ refers to the distribution of action sets produced by the entire elicitation procedure, including the prompt, sampling temperature, removal of duplicates within a run, and any retries after a parsing failure (rather than the LM's token-level probability distribution). 
We sampled the alternative actions at temperature \altTemperature{} and scored their features at temperature \scoreTemperature{}; the latter temperature was also used to rate desire and intimacy when those variables were given in the vignette.
The generation prompt asked the LM to explain its response before returning the action array, which we retained for auditing the generated sets.

The generation and scoring calls were handled differently.
For feature scoring, each scenario and run used one pooled action list containing all three possible observed actions and the alternatives generated across all of that scenario's condition cells.
After case-insensitive removal of duplicate strings, the entire list was randomly ordered and presented without identifying which actions the characters had taken.
(Pooling placed repeated actions in a common scoring frame, while randomizing and removing the observed-action labels prevented the three experimenter-written actions from always receiving a marked status or fixed position. This approach also saved tokens compared to scoring each comparison set separately.) 
The scoring responses followed a fixed JSON format, and we repeated a query if a rating fell outside the requested scale.
For action generation, we instead allowed free-form responses: requiring a fixed schema caused the LM to return empty lists in scenarios where it was harder to generate alternatives.

The three feature questions corresponded directly to the terms in the utility function.
Goal satisfaction $g_\tau(a)$ measured how fully the action achieved the characters' goal, independent of how much they desired the outcome.
Thus, reward is based on the value of the outcome, $d$, and how fully the action delivers it, $g_\tau(a)$, consistent with previous utility-based models of action understanding \autocite{jara2016naive,jara2020naive,baker2009action}.
Effort included the total work required across both characters -- physical work, time, equipment, preparation, and, for the disclosure scenarios, the work required to produce the account.
Risk measured the interpersonal access, exposure, contact, dependency, or vulnerability created by the action.
The LM rated each feature from $0$ to $6$, and we rescaled the ratings to $[0,1]$.
The given desire and intimacy descriptions were rated from $0$ to $100$ and rescaled in the same way, allowing given and inferred values to use a common scale.
Within each study, all reported model variants used the same LM-generated comparison sets and feature scores.
Note that we preregistered omitting the relationship description during alternative generation for the vanilla model, but we decided to change this so that any difference between the models in the model comparison could be attributed to the cognitive model. 
We retained these separate relationship-free sets and compare them with the relationship-conditioned sets in \autoref{fig:si-action-sets}b.

\subsection{Likelihood details}

In the joint-inference studies (1b, 2b, 3a, and 3b), each run produces a two-dimensional update, so the mixture uses an isotropic bivariate Gaussian with covariance $\sigma^2\mathbf{I}$.
We did not fit a within-component correlation.
The mixture can still predict correlated judgments: if the two predicted updates vary together across elicitation runs, the component means are themselves correlated, and the mixture concentrates probability on response pairs that covary in the same way.
Note that the Gaussian places a small amount of probability outside the bounded response scale, which we accepted for simplicity.  
The likelihood also treats a participant's trials as independent given the model -- the confidence intervals for model comparison account for the repeated observations by resampling participants rather than trials.

\subsection{Comparison-set reweighting}
\label{si:reweighting-detail}

In the preregistered procedure, the observer assigns equal weight to every action in the comparison set. 
For the reported model, we added a post-hoc extension that captures how observers gives more weight to question-relevant actions in the comparison set when the observed action is surprising.
Related work has examined relevance to the ``question under discussion'' in pragmatics \autocite{roberts2012information,kao2014nonliteral}. 

Let $\rho_0(a \mid \mathcal{A}_\tau)=1/|\mathcal{A}_\tau|$ be the preregistered uniform comparison weight.
For latent state $\mathbf{z}$ and scenario context $\tau$, the baseline local action probability is
\begin{equation}
    p_0(a \mid \mathbf{z}, \mathcal{A}_\tau, \tau)
    =
    \frac{\rho_0(a \mid \mathcal{A}_\tau)\exp\!\big(U(a \mid \mathbf{z}, \tau)\big)}
    {\sum_{b \in \mathcal{A}_\tau}\rho_0(b \mid \mathcal{A}_\tau)\exp\!\big(U(b \mid \mathbf{z}, \tau)\big)}.
    \label{eq:baseline-action-probability}
\end{equation}
For each scenario cell and LM run, we calculate the surprise of the observed action under this local action probability:
\begin{equation}
    S
    =
    -\log
    \mathbb{E}_{\mathbf{z} \sim P}
    \left[
        p_0(a_{\mathrm{obs}} \mid \mathbf{z}, \mathcal{A}_\tau, \tau)
    \right],
    \label{eq:action-surprise}
\end{equation}

The second quantity, $v(a)$, measures how much an action is relevant for the question. 
We use a simple heuristic where, when observers need to answer a question, actions are more important based on its contrast from the observed action for the `question under discussion'. For physical-world question, it is the action's effort difference between the two possible states:

\begin{equation}
    v(a)
    =
    \left|
        \operatorname{effort}_\tau(a; e_\mathrm{physical} = \mathrm{high})
        -
        \operatorname{effort}_\tau(a; e_\mathrm{physical} = \mathrm{low})
    \right|.
    \label{eq:world-sensitivity}
\end{equation}
For an intimacy judgment, $v(a)=|\operatorname{risk}_\tau(a)-\operatorname{risk}_\tau(a_\mathrm{obs})|$.

Taken together, the observer reweighs the comparison set based on $S$ and $v$: 

\begin{equation}
    \rho_\eta(a \mid a_\mathrm{obs}, \mathcal{A}_\tau, \tau)
    =
    \frac{
        \rho_0(a \mid \mathcal{A}_\tau)\exp\!\big(\eta S v(a)\big)
    }{
        \sum_{b \in \mathcal{A}_\tau}
        \rho_0(b \mid \mathcal{A}_\tau)\exp\!\big(\eta S v(b)\big)
    },
    \qquad \eta \geq 0.
    \label{eq:surprise-gate}
\end{equation}

When $\eta=0$, this recovers the preregistered specification.

\begin{table}[h!]
\centering
\small
\begin{tabular}{lllll}
\toprule
Study & Observer infers & Reweighted target & Sensitivity score $v(a)$ & Fitted $\eta$ \\
\midrule
1a & desire & none & not applied & not fitted \\
1b & desire, physical state & physical state & effort difference (\autoref{eq:world-sensitivity}) & \pEtaOneB \\
2a & intimacy & intimacy & $|\operatorname{risk}_\tau(a) - \operatorname{risk}_\tau(a_\mathrm{obs})|$ & \pEtaTwoA \\
2b & intimacy, physical state & both & risk difference $+$ effort difference & \pEtaTwoB \\
3a & desire, physical state & physical state & effort difference (\autoref{eq:world-sensitivity}) & \pEtaThreeA \\
3b & intimacy, physical state & both & risk difference $+$ effort difference & \pEtaThreeB \\
\bottomrule
\end{tabular}
\caption{\textbf{Comparison-set reweighting by study.} When a study asks two questions that depend on the alternatives, the two sensitivity scores are added and the model computes one joint posterior. $\eta$ is fitted separately by study. Reweighting was not applied in Study~1a, so its reported and preregistered versions are identical.}
\label{tab:reweighting-spec}
\end{table}

Overall, this extension captures the intuition that reweighting the action set changes an inference when the observed action provides little direct evidence for the target variable and the alternatives differ on the relevant feature. 
Eating the cake, for example, directly signals desire because desire changes the reward of the chosen action.
By contrast, whether utensils were nearby or far away is typically revealed by the effort of the lower-risk action that the characters \textit{did not} choose.
Similarly, declining to share food has no contamination risk at any level of intimacy, so its implications for intimacy depend on the alternatives that were not chosen.
We therefore applied reweighting to physical-state and intimacy judgments, but not to desire judgments. 
During post-hoc model development, we also tested several alternatives to this reported version, including a surprise term that was constant across actions, as well as a version that increased only the weights of actions sensitive to the physical state.

\begin{table}[h!]
  \centering
  \small
  \setlength{\tabcolsep}{6pt}
\begin{tabular}{lccccccc}
\toprule
Study & $w_v$ & $w_e$ & $w_d$ & $\gamma$ & $\alpha_{\mathrm{obs}}$ & $\sigma$ & $\eta$ \\
\midrule
1a & \pWvOneA & \pWeOneA & \pWdOneA & \pGammaOneA & \pAlphaObsOneA & \pSigmaOneA & \pEtaOneA \\
1b & \pWvOneB & \pWeOneB & \pWdOneB & \pGammaOneB & \pAlphaObsOneB & \pSigmaOneB & \pEtaOneB \\
2a & \pWvTwoA & \pWeTwoA & \pWdTwoA & \pGammaTwoA & \pAlphaObsTwoA & \pSigmaTwoA & \pEtaTwoA \\
2b & \pWvTwoB & \pWeTwoB & \pWdTwoB & \pGammaTwoB & \pAlphaObsTwoB & \pSigmaTwoB & \pEtaTwoB \\
3a & \pWvThreeA & \pWeThreeA & \pWdThreeA & \pGammaThreeA & \pAlphaObsThreeA & \pSigmaThreeA & \pEtaThreeA \\
3b & \pWvThreeB & \pWeThreeB & \pWdThreeB & \pGammaThreeB & \pAlphaObsThreeB & \pSigmaThreeB & \pEtaThreeB \\
\bottomrule
\end{tabular}

  \caption{\textbf{Fitted parameters for the reported model.} Values are maximum-likelihood estimates from the full data for each experiment. The weights were fitted separately by experiment, but the generalization analyses do not require them to differ substantially (SI Section~\ref{si:generalization}), so we do not interpret differences among rows.}
  \label{tab:fitted-params}
\end{table}

\newpage 

\section{Deviations from preregistration}
\label{si:preregistration-deviations}

\begin{figure}[p]
  \centering
  \includegraphics[width=0.75\textwidth]{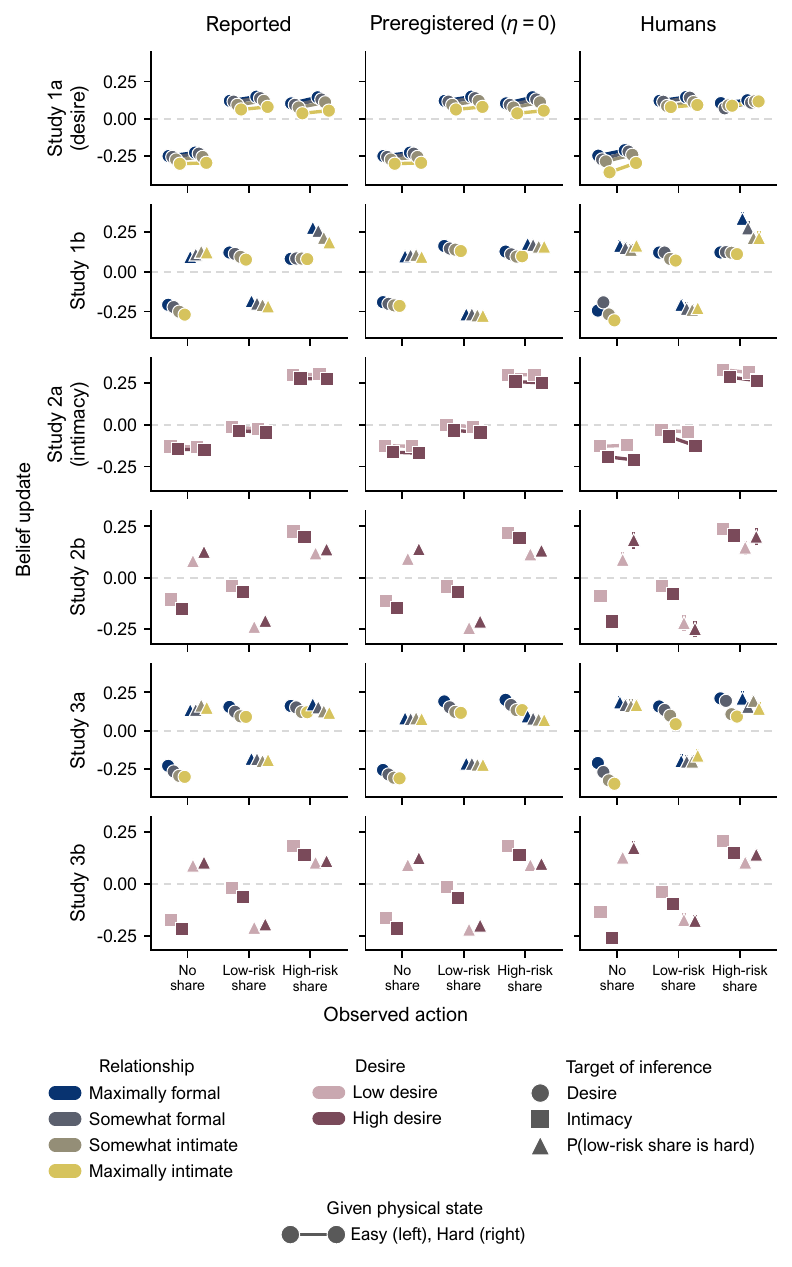}
  \caption{\textbf{Belief updates predicted by the reported and preregistered specifications.} The first two columns show the full social model's out-of-sample predictions with the reported reweighting (left) and with the preregistered uniform weighting ($\eta=0$, middle); the third shows human condition means. Error bars are 95\% confidence intervals from resampling participants.}
  \label{fig:si-prereg-predictions}
\end{figure}

\begin{figure}[p]
  \centering
  \includegraphics[width=\textwidth]{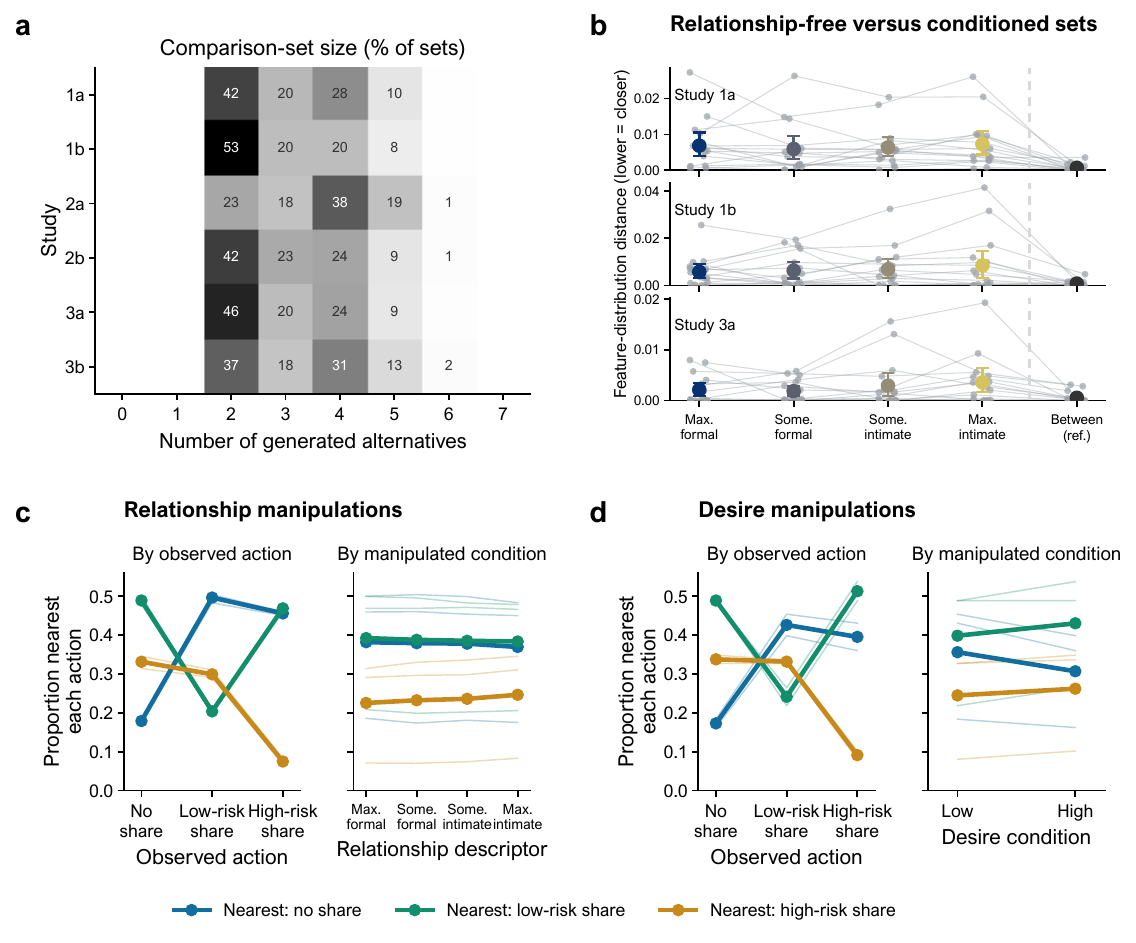}
  \caption{\textbf{Properties of the LM-generated comparison sets.}
  \textbf{(a)} Distribution of the number of generated alternatives by study; cells show percentages.
  \textbf{(b)} Energy distance between the relationship-free sets specified for the preregistered vanilla model and the relationship-conditioned sets used in the reported comparison; lower values indicate more similar feature distributions, and the rightmost category compares sets conditioned on different relationships as a reference. Gray trajectories show scenarios; colored points and error bars show means and bootstrap 95\% confidence intervals.
  \textbf{(c,d)} Composition of generated sets, in studies that manipulate relationship (c) or desire (d). Each alternative is assigned to the nearest observed-action centroid in $(g,\,\operatorname{risk},\,\operatorname{effort})$ space. Bold lines are averages across studies; thin lines show individual relationship, desire, or action conditions.}
  \label{fig:si-action-sets}
\end{figure}

\begin{enumerate}
  \item \textbf{Primary metric.}
  Held-out log-likelihood per trial was preregistered as primary and correlation across scenario-averaged condition means as secondary; we reversed that order (Methods; SI Section~\ref{si:modulation}; \autoref{tab:model-comparison}).

  \item \textbf{Comparison-set reweighting.}
  The reported model adds the reweighting in SI Section~\ref{si:reweighting-detail} (\autoref{eq:surprise-gate}; \autoref{tab:reweighting-spec}).
  \autoref{tab:prereg-deviation} and \autoref{fig:si-prereg-predictions} compare the two specifications on held-out likelihood and predicted belief updates.

  \item \textbf{Shared comparison set for the vanilla model.}
  In Studies~1a, 1b, and 3a, the preregistered vanilla model used relationship-free alternatives; the reported comparison holds the relationship-conditioned elicitations fixed across variants, so that any changes in fit could be localized to the cognitive model (\autoref{fig:si-action-sets}b).
  The ablation comparison under the fully preregistered specification is in \autoref{tab:preregistered-model-comparison}.

  \item \textbf{Exploratory generalization analyses.}
  The cross-experiment analyses in SI Section~\ref{si:generalization} were not preregistered.

  \item \textbf{Study~3b sample size.}
  We preregistered 120 participants for Study~3b but recruited double the sample size through a recruitment error, applying Study~3a's target sample size to both Study~3 experiments.
  All preregistered exclusion criteria were applied unchanged (\autoref{tab:participants}). 
\end{enumerate}

\begin{table}[h]
  \centering \small
\begin{tabular}{lcccc}
\toprule
 & & \multicolumn{2}{c}{Held-out LL / participant} & \\
\cmidrule(lr){3-4}
Study & $\eta$ & Preregistered & Reported & Reported $-$ preregistered \\
\midrule
1a & \pEtaOneA & \llPreregOneA & \llFullOneA & \statReweightOneA \\
1b & \pEtaOneB & \llPreregOneB & \llFullOneB & \statReweightOneB \\
2a & \pEtaTwoA & \llPreregTwoA & \llFullTwoA & \statReweightTwoA \\
2b & \pEtaTwoB & \llPreregTwoB & \llFullTwoB & \statReweightTwoB \\
\midrule
\multicolumn{5}{l}{\itshape Non-food} \\
3a & \pEtaThreeA & \llPreregThreeA & \llFullThreeA & \statReweightThreeA \\
3b & \pEtaThreeB & \llPreregThreeB & \llFullThreeB & \statReweightThreeB \\
\bottomrule
\end{tabular}

  \caption{\textbf{Held-out performance under the reported and preregistered specifications.} The preregistered model sets $\eta=0$ and weights every action in a cell's comparison set uniformly. Values are mean held-out log-likelihoods per participant from the same leave-one-scenario-out protocol on the same trials. The final column is the paired difference with a 95\% confidence interval from resampling participants; positive values favor the reported model.}
  \label{tab:prereg-deviation}
\end{table}

\begin{table}[h]
  \centering
  \small
  \setlength{\tabcolsep}{5pt}
\begin{tabular}{llccc}
\toprule
 & & \multicolumn{3}{c}{Held-out LL / participant} \\
\cmidrule(lr){3-5}
Study & Inferred target & Full & Full $-$ vanilla & Full $-$ discomfort-only \\
\midrule
1a & desire                     & \ensuremath{-0.83} & \ensuremath{0.11}~\ensuremath{[0.05,\ 0.16]} & \ensuremath{3.30}~\ensuremath{[2.96,\ 3.62]} \\
1b & desire $+$ physical        & \ensuremath{-6.30} & \ensuremath{-0.03}~\ensuremath{[-0.16,\ 0.09]} & \ensuremath{5.30}~\ensuremath{[4.71,\ 5.81]} \\
2a & relationship               & \ensuremath{0.82} & \ensuremath{4.33}~\ensuremath{[3.95,\ 4.72]} & \ensuremath{0.08}~\ensuremath{[0.00,\ 0.15]} \\
2b & relationship $+$ physical  & \ensuremath{-4.62} & \ensuremath{2.55}~\ensuremath{[2.06,\ 3.04]} & \ensuremath{3.06}~\ensuremath{[2.44,\ 3.70]} \\
\midrule
\multicolumn{5}{l}{\itshape Non-food} \\
3a & desire $+$ physical        & \ensuremath{-10.86} & \ensuremath{-0.03}~\ensuremath{[-0.14,\ 0.07]} & \ensuremath{4.06}~\ensuremath{[3.62,\ 4.52]} \\
3b & relationship $+$ physical  & \ensuremath{-7.72} & \ensuremath{2.12}~\ensuremath{[1.82,\ 2.43]} & \ensuremath{1.92}~\ensuremath{[1.48,\ 2.34]} \\
\bottomrule
\end{tabular}

  \caption{\textbf{Model comparison under the  preregistered specification.} The preregistered specification sets $\eta=0$ and, in Studies~1a, 1b, and 3a, uses relationship-free comparison sets for the vanilla model. Values are held-out log-likelihoods per participant under leave-one-scenario-out cross-validation; the two columns on the right show the full model's improvement over each ablation, with 95\% confidence intervals from resampling participants. Note that the confidence intervals for Full minus vanilla for Study 1b and 3a cover zero. (Note that using log-likelihood does not capture the condition-level qualitative differences, which are the primary theoretical comparisons of interest; see \autoref{fig:si-prereg-predictions} for the qualitative comparison.)}
  \label{tab:preregistered-model-comparison}
\end{table}

\clearpage
\newpage

\section{Additional results}

\subsection{Scenario-level belief updates}
\label{si:scenario-level}

\begin{figure}[h!]
  \centering
  \includegraphics[width=\textwidth]{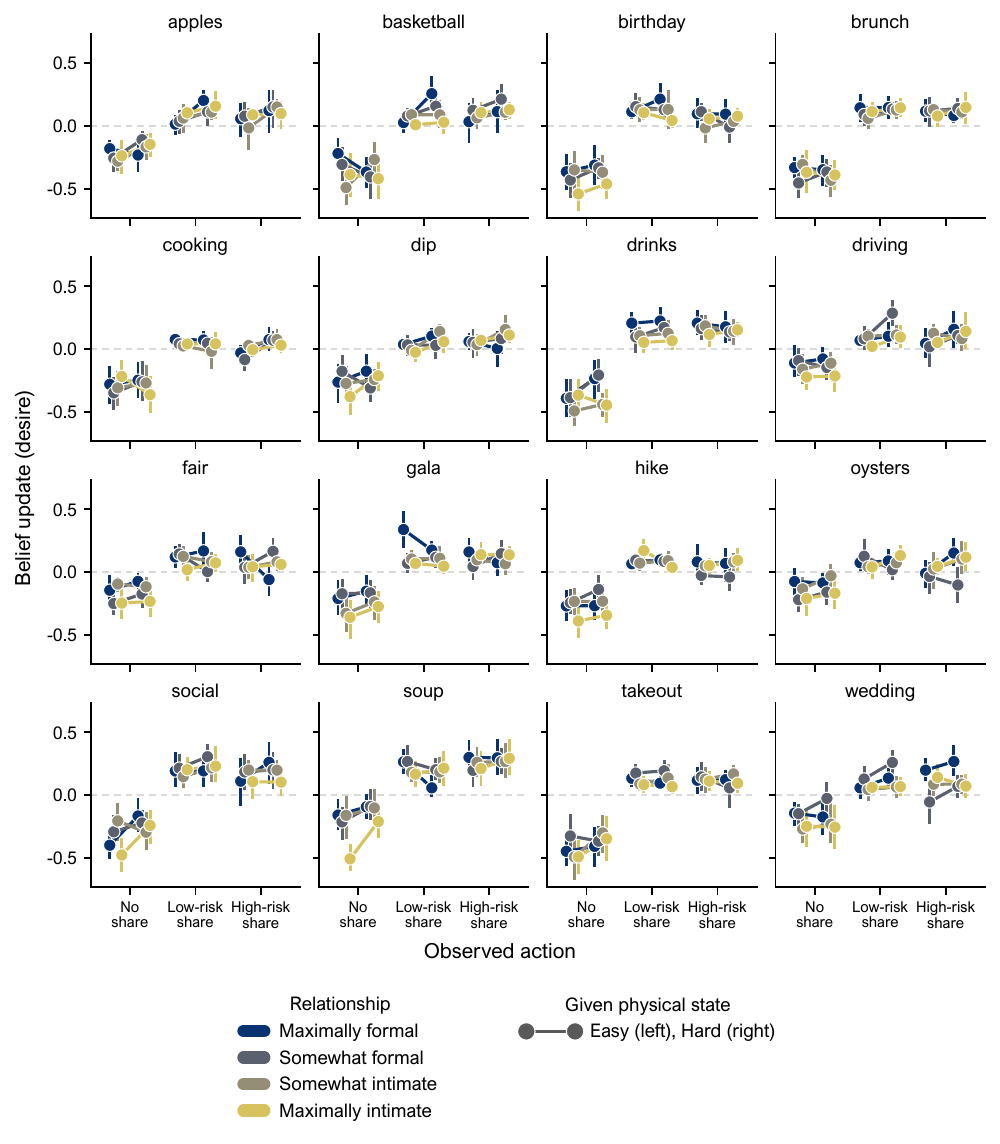}
  \caption{\textbf{Scenario-level results for Study 1a.} Error bars are 95\% confidence intervals from resampling participants.}
  \label{fig:si-scenarios-1a}
\end{figure}

\begin{figure}[p]
  \centering
  \includegraphics[width=\textwidth]{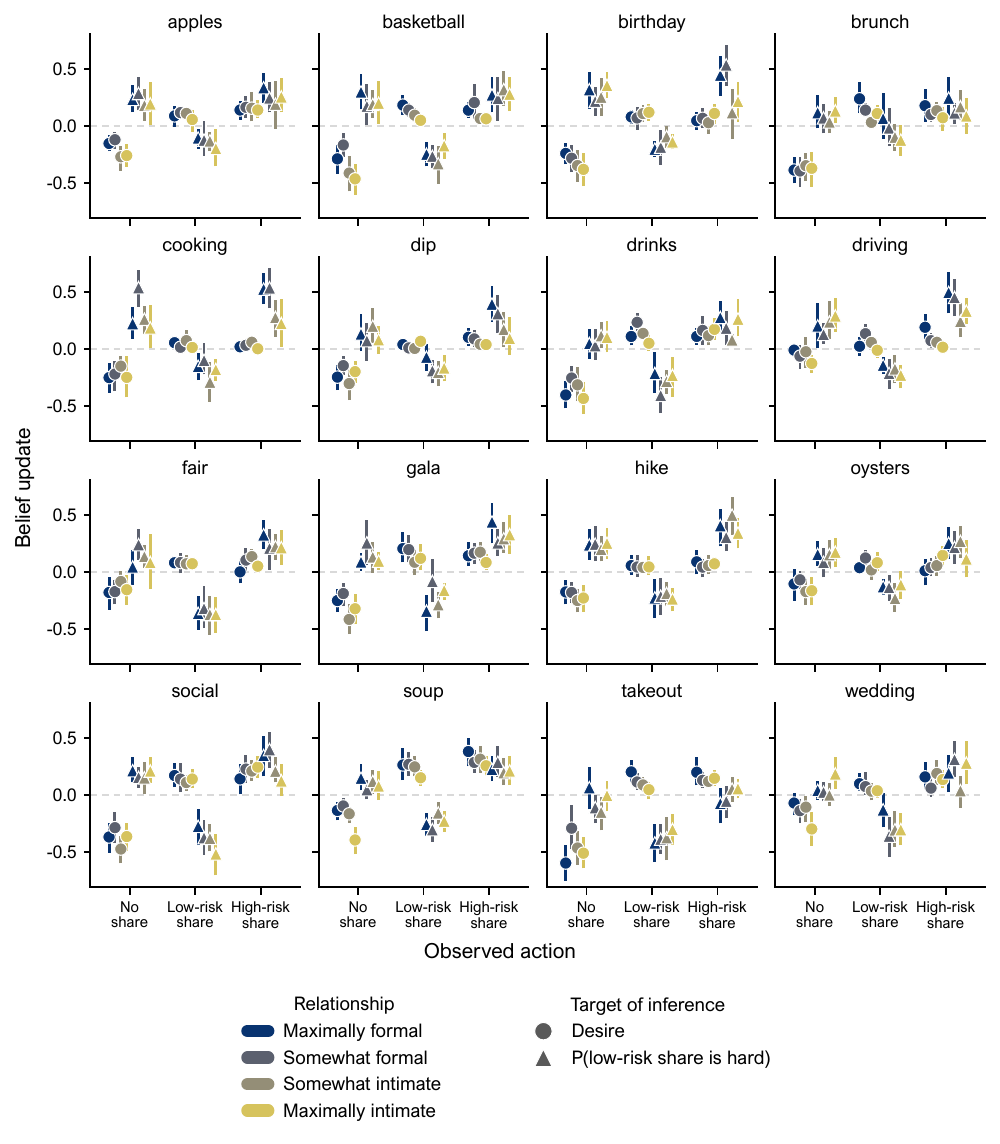}
  \caption{\textbf{Scenario-level results for Study 1b.} Error bars are 95\% confidence intervals from resampling participants.}
  \label{fig:si-scenarios-1b}
\end{figure}

\begin{figure}[p]
  \centering
  \includegraphics[width=\textwidth]{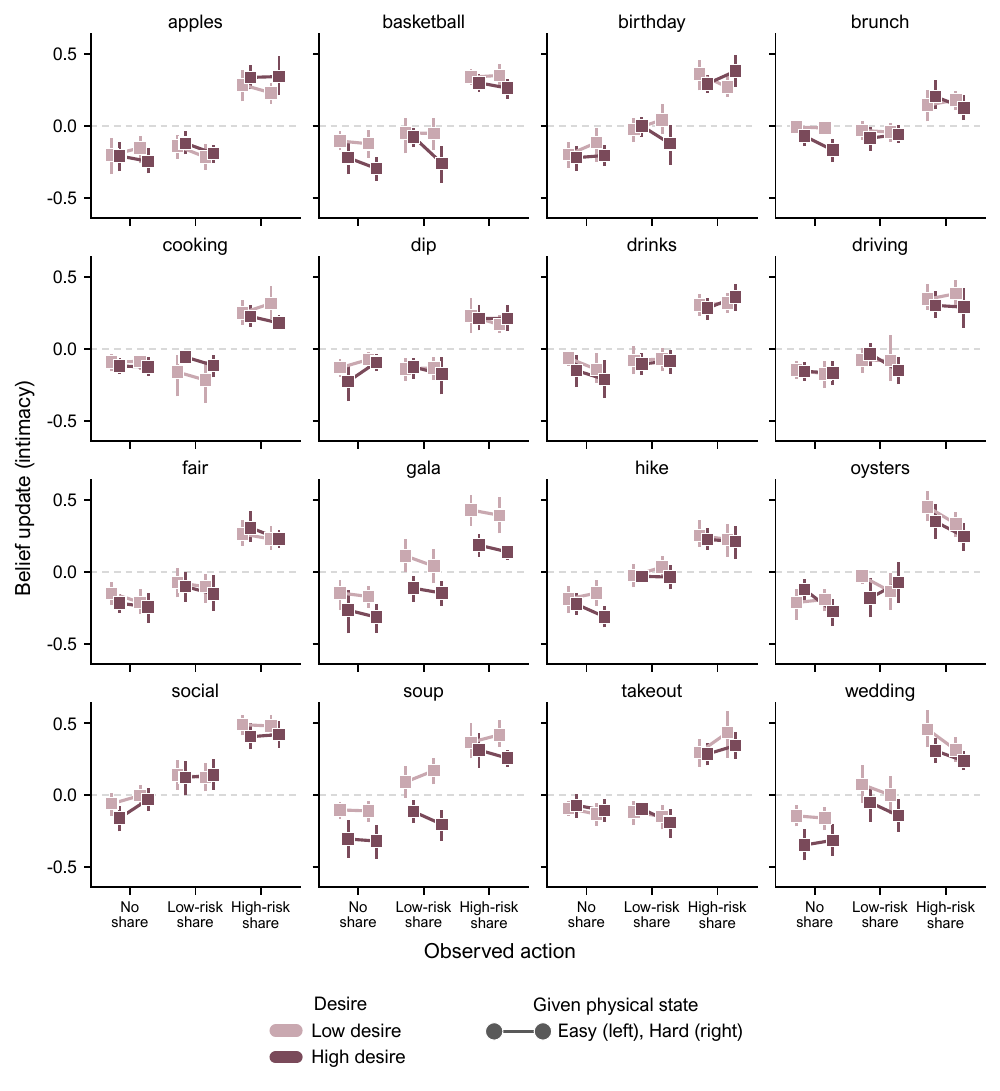}
  \caption{\textbf{Scenario-level results for Study~2a.} Error bars are 95\% confidence intervals from resampling participants.}
  \label{fig:si-scenarios-2a}
\end{figure}

\begin{figure}[p]
  \centering
  \includegraphics[width=\textwidth]{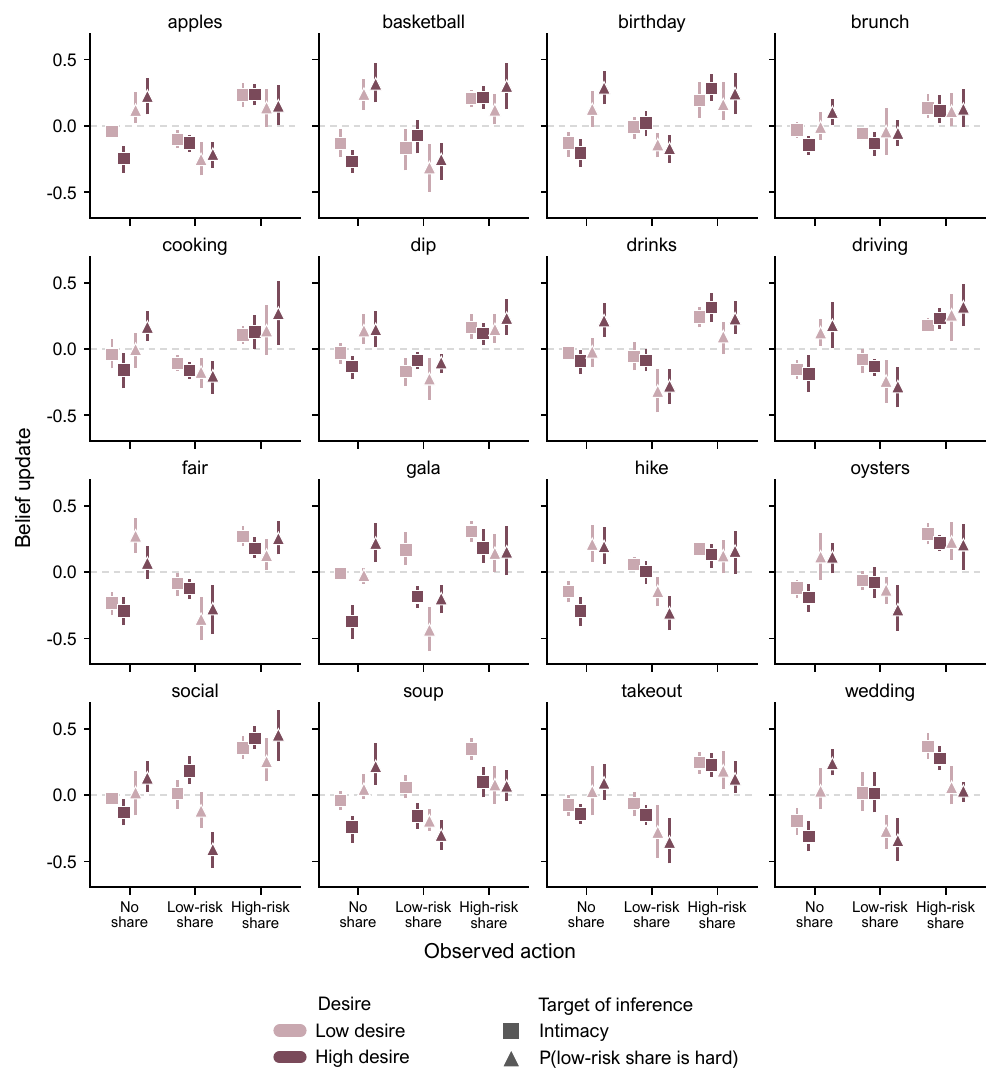}
  \caption{\textbf{Scenario-level results for Study~2b.} Error bars are 95\% confidence intervals from resampling participants.}
  \label{fig:si-scenarios-2b}
\end{figure}

\begin{figure}[p]
  \centering
  \includegraphics[width=\textwidth]{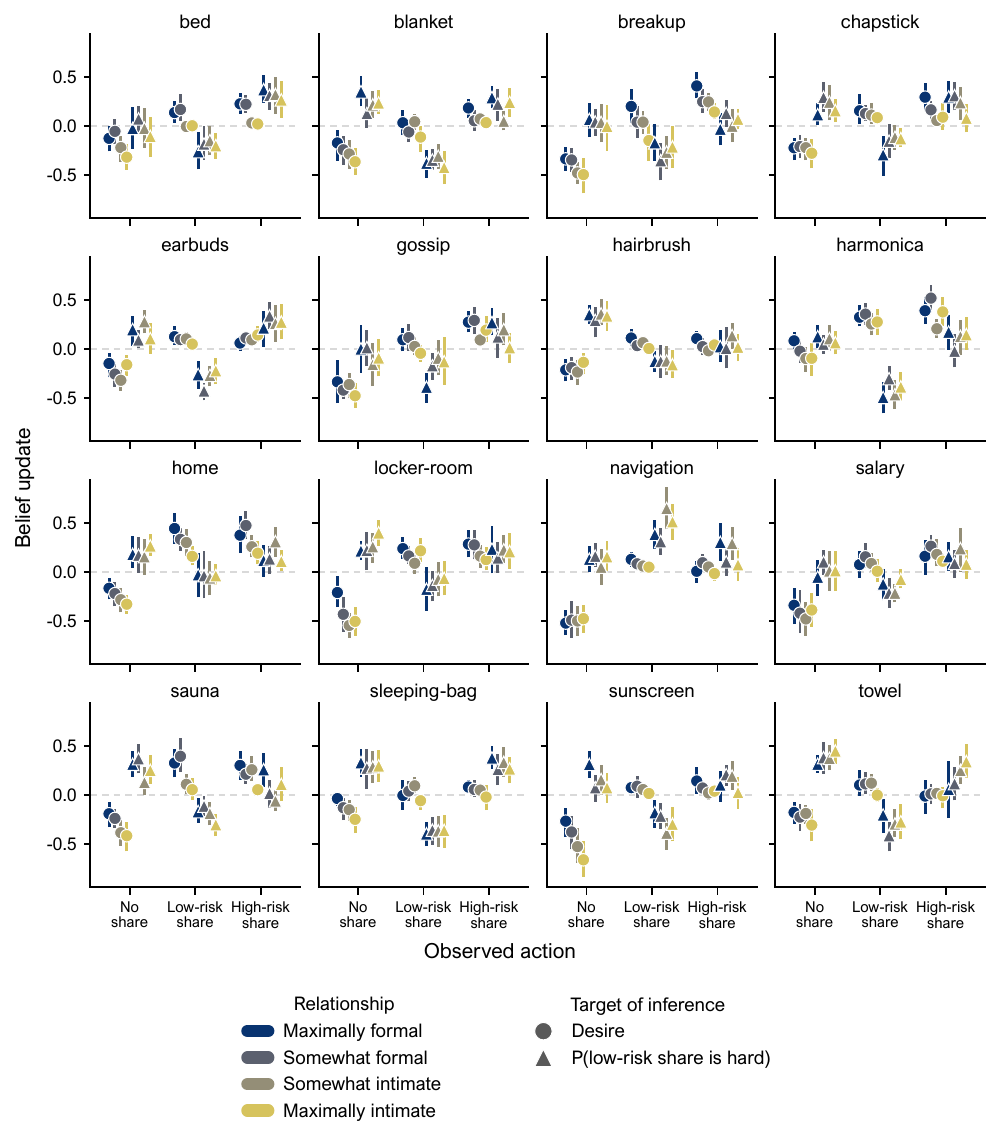}
  \caption{\textbf{Scenario-level results for Study~3a.} Error bars are 95\% confidence intervals from resampling participants.}
  \label{fig:si-scenarios-3a}
\end{figure}

\begin{figure}[p]
  \centering
  \includegraphics[width=\textwidth]{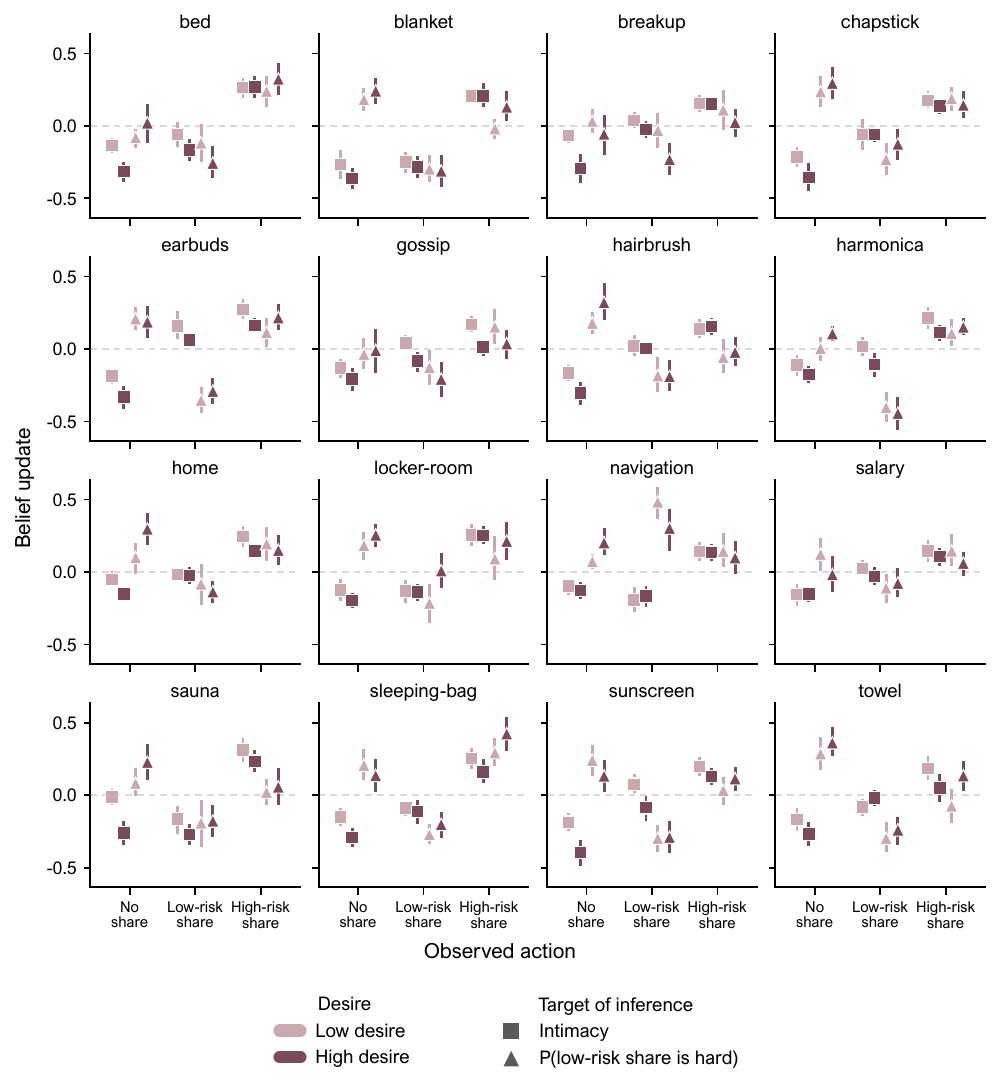}
  \caption{\textbf{Scenario-level results for Study 3b.} Error bars are 95\% confidence intervals from resampling participants.}
  \label{fig:si-scenarios-3b}
\end{figure}

\clearpage

\subsection{Prior and posterior rating levels}
\label{si:prior-posterior}

\begin{figure}[hp]
  \centering
  \includegraphics[width=\textwidth]{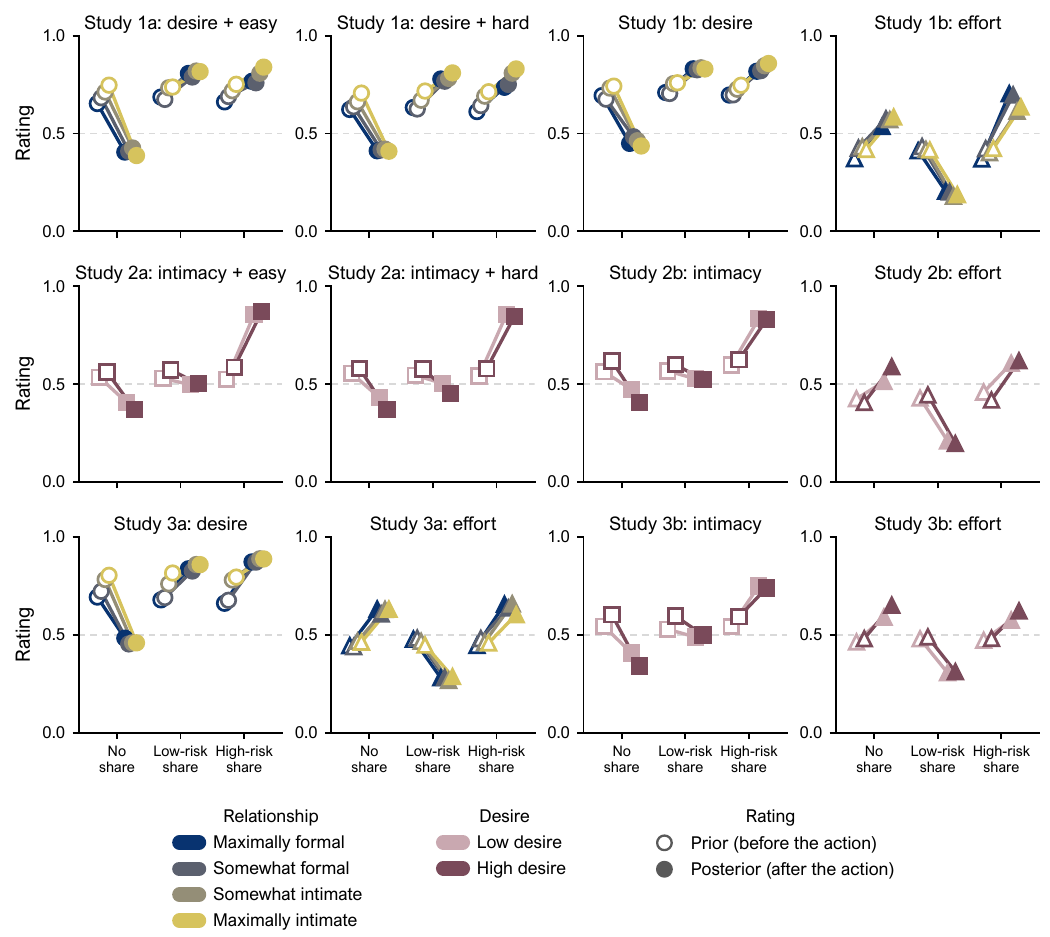}
  \caption{Participants' prior and posterior ratings.}
  \label{fig:si-prior-posterior-levels}
\end{figure}

\newpage

\section{Additional model evaluation}

\subsection{Held-out likelihood}
\label{si:modulation}

\begin{table}[h]
  \centering
  \small
  \setlength{\tabcolsep}{5pt}
\begin{tabular}{llccc}
\toprule
 & & \multicolumn{3}{c}{Held-out LL / participant} \\
\cmidrule(lr){3-5}
Study & Inferred target & Full & Full $-$ vanilla & Full $-$ discomfort-only \\
\midrule
1a & desire                     & \llFullOneA & \statBaseOneA & \statDiscOneA \\
1b & desire $+$ physical        & \llFullOneB & \statBaseOneB & \statDiscOneB \\
2a & relationship               & \llFullTwoA & \statBaseTwoA & \statDiscTwoA \\
2b & relationship $+$ physical  & \llFullTwoB & \statBaseTwoB & \statDiscTwoB \\
\midrule
\multicolumn{5}{l}{\itshape Non-food} \\
3a & desire $+$ physical        & \llFullThreeA & \statBaseThreeA & \statDiscThreeA \\
3b & relationship $+$ physical  & \llFullThreeB & \statBaseThreeB & \statDiscThreeB \\
\bottomrule
\end{tabular}

  \caption{\textbf{Model comparison using log-likelihood.} Values are held-out log-likelihoods per participant under leave-one-scenario-out cross-validation; the two columns on the right show the model's improvement over each ablation, with 95\% confidence intervals from resampling participants.}
  \label{tab:model-comparison}
\end{table}

\begin{table}[h]
  \centering \small
\begin{tabular}{llrrcr}
\toprule
Study & Inferred & Explainable & Manipulated & 95\% CI & Manipulated / explainable \\
\midrule
1a & desire & 37\% & 35\% & \ensuremath{[32,\ 37]}\% & 95\% \\
1b & desire & 42\% & 40\% & \ensuremath{[36,\ 44]}\% & 95\% \\
1b & physical & 34\% & 30\% & \ensuremath{[27,\ 33]}\% & 89\% \\
2a & relationship & 50\% & 48\% & \ensuremath{[45,\ 51]}\% & 96\% \\
2b & relationship & 42\% & 40\% & \ensuremath{[35,\ 44]}\% & 94\% \\
2b & physical & 28\% & 27\% & \ensuremath{[22,\ 32]}\% & 97\% \\
3a & desire & 44\% & 38\% & \ensuremath{[34,\ 41]}\% & 87\% \\
3a & physical & 27\% & 24\% & \ensuremath{[21,\ 27]}\% & 88\% \\
3b & relationship & 39\% & 38\% & \ensuremath{[35,\ 41]}\% & 96\% \\
3b & physical & 24\% & 21\% & \ensuremath{[19,\ 24]}\% & 88\% \\
\bottomrule
\end{tabular}

  \caption{\textbf{Trial-level variance by study and inferred variable.} ``Explainable'' is the share between experimental cells; the remainder arises within cells. ``Manipulated'' is the share attributable to the observed action and the manipulated relationship or desire condition taken together, holding scenario and any other given condition fixed; it includes their interactions. All estimates are corrected for sampling noise in the cell means.}
  \label{tab:variance-decomposition}
\end{table}

\clearpage

\subsection{Cross-experiment generalization}
\label{si:generalization}

The reported model estimates use separate utility weights across each experiment, so we tested whether a shared utility could predict judgments across experiments. 
We kept $\alpha_{\mathrm{obs}}$, $\sigma$, and $\eta$ free per experiment because they describe the observer and response scale, and because the scope of $\eta$ differs among experiments.

Across the two non-food experiments, the correlation across scenario-averaged condition means was $r=\rNonfoodOwnFit$ with separately fitted weights, $r=\rNonfoodFoodFit$ with weights fitted to the four food experiments, and $r=\rNonfoodPooledFit$ with one utility shared across all six experiments (\autoref{tab:generalization-primary}).
Sharing one utility across all six experiments changed held-out log-likelihood per participant by \statPoolAll{} across experiments, with most of the cost in Study~1 (\autoref{tab:generalization}).

\begin{table}[h!]
  \centering \small
\begin{tabular}{lr}
\toprule
Experiment & $\Delta$ held-out log-likelihood per participant \\
\midrule
1a & \statPoolOneA \\
1b & \statPoolOneB \\
2a & \statPoolTwoA \\
2b & \statPoolTwoB \\
\midrule
\multicolumn{2}{l}{\itshape Non-food} \\
3a & \statPoolThreeA \\
3b & \statPoolThreeB \\
\bottomrule
\end{tabular}

  \caption{\textbf{Held-out cost of sharing one utility across all six experiments.} Values compare shared weights with weights fitted separately to each experiment. Positive values mean the shared weights predicted held-out judgments better. Each paired difference uses the same trials and leave-one-scenario-out cross validation, with a 95\% confidence interval from resampling participants.}
  \label{tab:generalization}
\end{table}

\begin{table}[h!]
  \centering \small
  \setlength{\tabcolsep}{5pt}
\begin{tabular}{lccc}
\toprule
 & \multicolumn{3}{c}{Correlation $r$} \\
\cmidrule(lr){2-4}
Experiment & Own & Food & All six \\
\midrule
1a & \ensuremath{0.992} & \textit{n/a} & \ensuremath{0.992} \\
1b & \ensuremath{0.993} & \textit{n/a} & \ensuremath{0.978} \\
2a & \ensuremath{0.992} & \textit{n/a} & \ensuremath{0.992} \\
2b & \ensuremath{0.988} & \textit{n/a} & \ensuremath{0.987} \\
3a & \ensuremath{0.990} & \ensuremath{0.989} & \ensuremath{0.989} \\
3b & \ensuremath{0.982} & \ensuremath{0.942} & \ensuremath{0.958} \\
\bottomrule
\end{tabular}

  \caption{\textbf{Cross-experiment generalization of the correlation across scenario-averaged condition means.} ``Own'' uses each experiment's fitted utility, ``Food'' uses one utility fitted to the four food experiments to predict the non-food experiments, and ``All six'' uses one utility shared across every experiment.}
  \label{tab:generalization-primary}
\end{table}

\clearpage

\section{Validation of LM elicitations and likelihood}

\begin{figure}[h]
  \centering
  \includegraphics[width=0.85\textwidth]{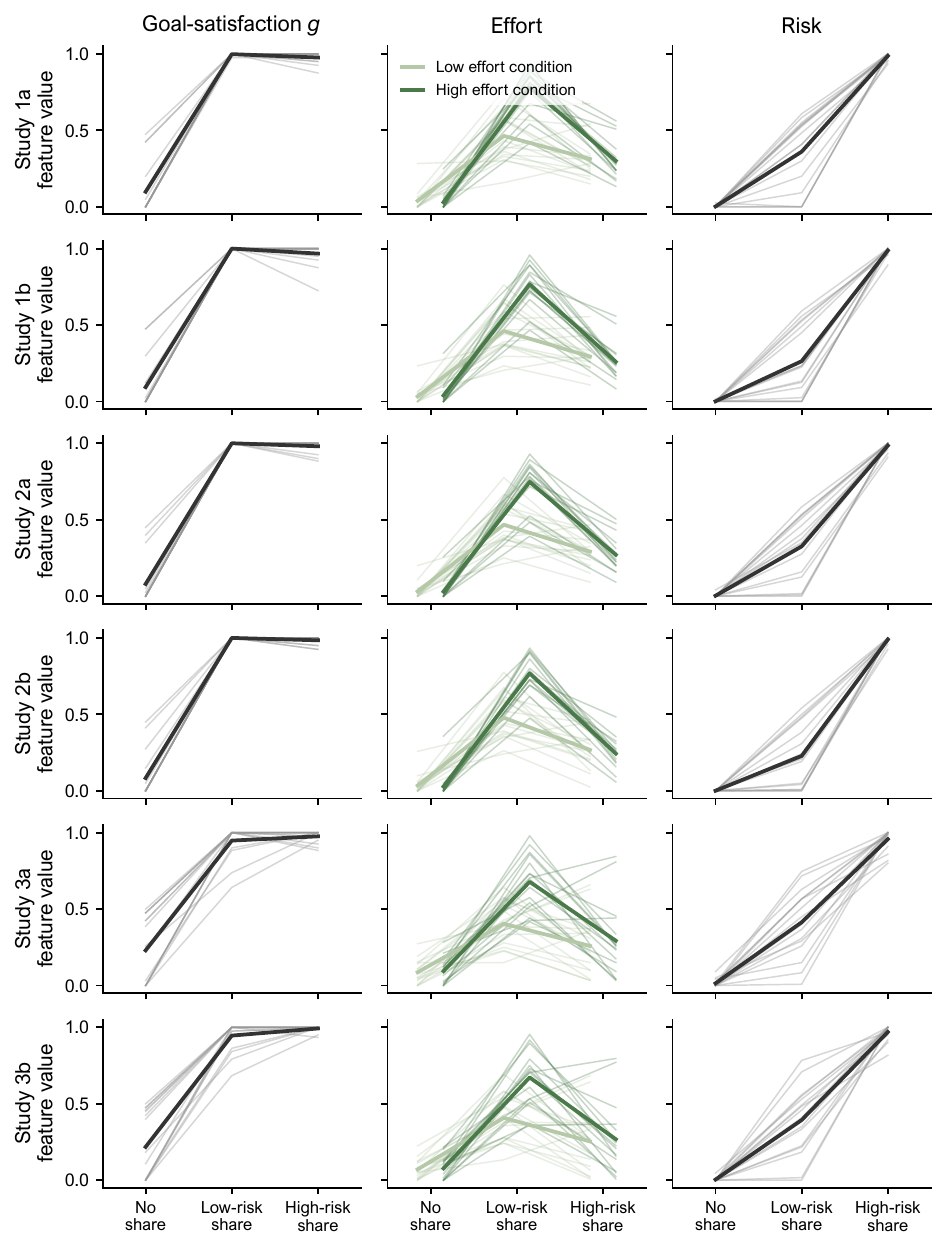}
  \caption{\textbf{LM-elicited features preserve the designed action structure.} Thin lines show individual scenarios, with features averaged over elicitation runs and the study's other conditions; thick lines show means across the \nScenarios{} scenarios.}
  \label{fig:si-feature-structure}
\end{figure}

\begin{figure}[p]
    \centering
    \includegraphics[width=0.72\textwidth]{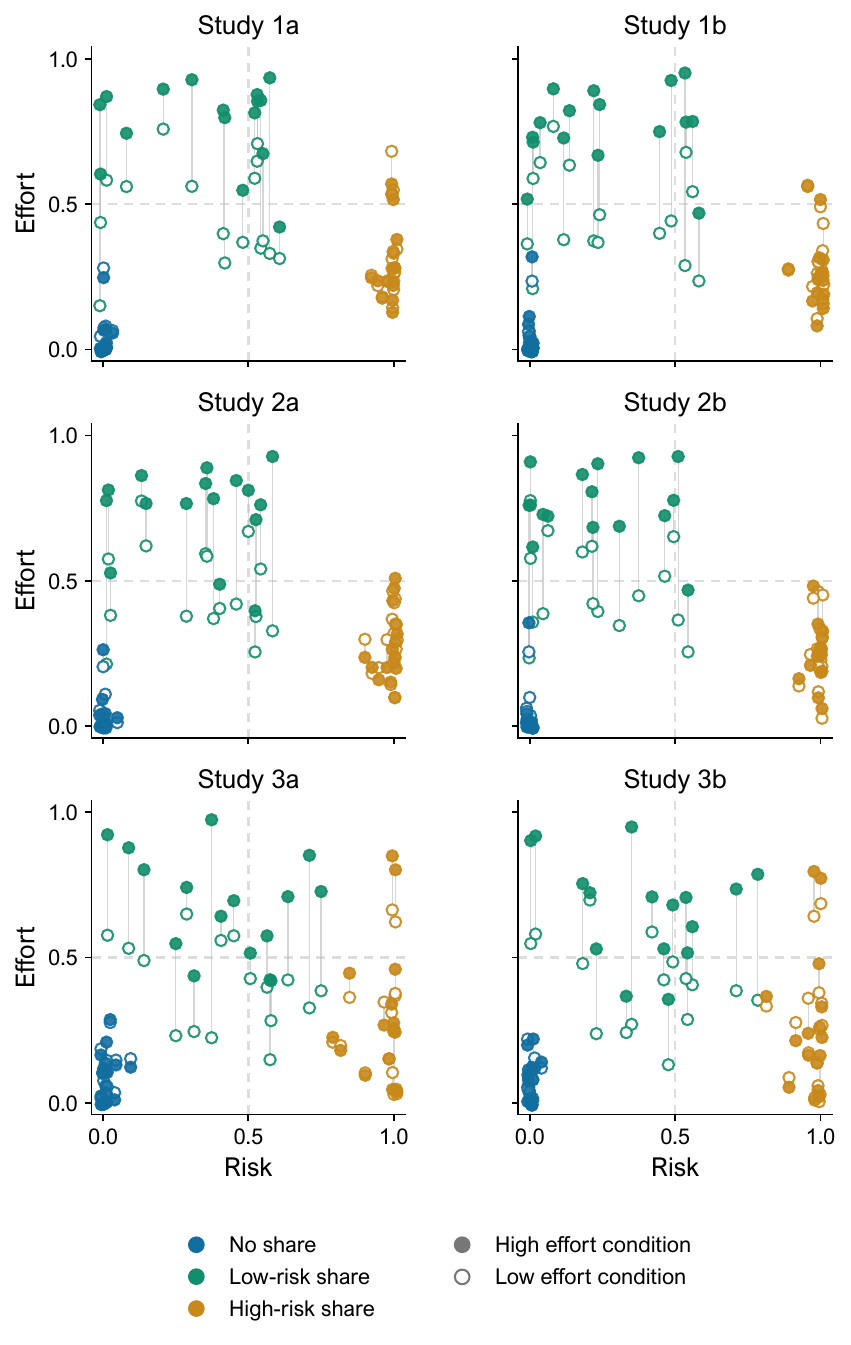}
    \caption{\textbf{Observed actions in the risk--effort feature space.} Each point is one scenario $\times$ action $\times$ effort condition, with features averaged over elicitation runs and the study's other conditions. Gray lines connect the two effort conditions for the same scenario and action.}
    \label{fig:si-observed-scatter}
  \end{figure}

\begin{figure}[p]
  \centering
  \includegraphics[width=0.95\textwidth]{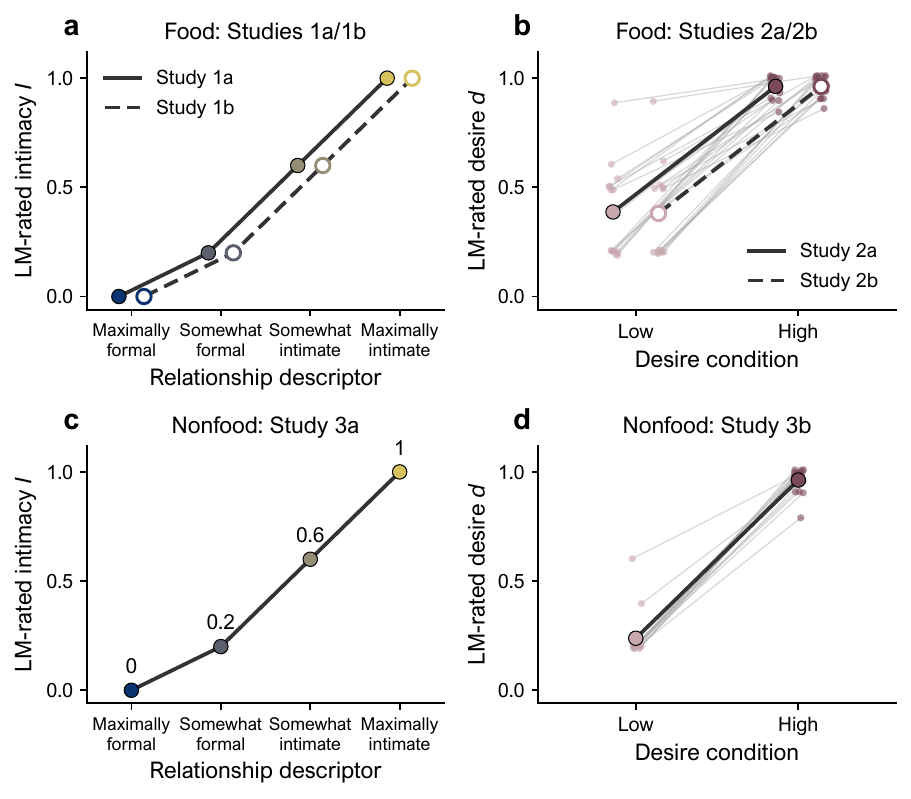}
  \caption{\textbf{LM ratings preserve the given relationship and desire manipulations.}
  \textbf{(a,c)} Rated intimacy $I$ for the four relationship descriptions in the food (a: Studies~1a/1b) and non-food studies (c: Study~3a).
  \textbf{(b,d)} Rated desire $d$ for the low- and high-desire descriptions in the food (b: Studies~2a/2b) and non-food studies (d: Study~3b).
  Each thin line is one scenario (ratings averaged over runs); the large markers show the mean across scenarios.}
  \label{fig:si-manipulation-checks}
\end{figure}

\begin{figure}[p]
  \centering
  \includegraphics[width=\textwidth]{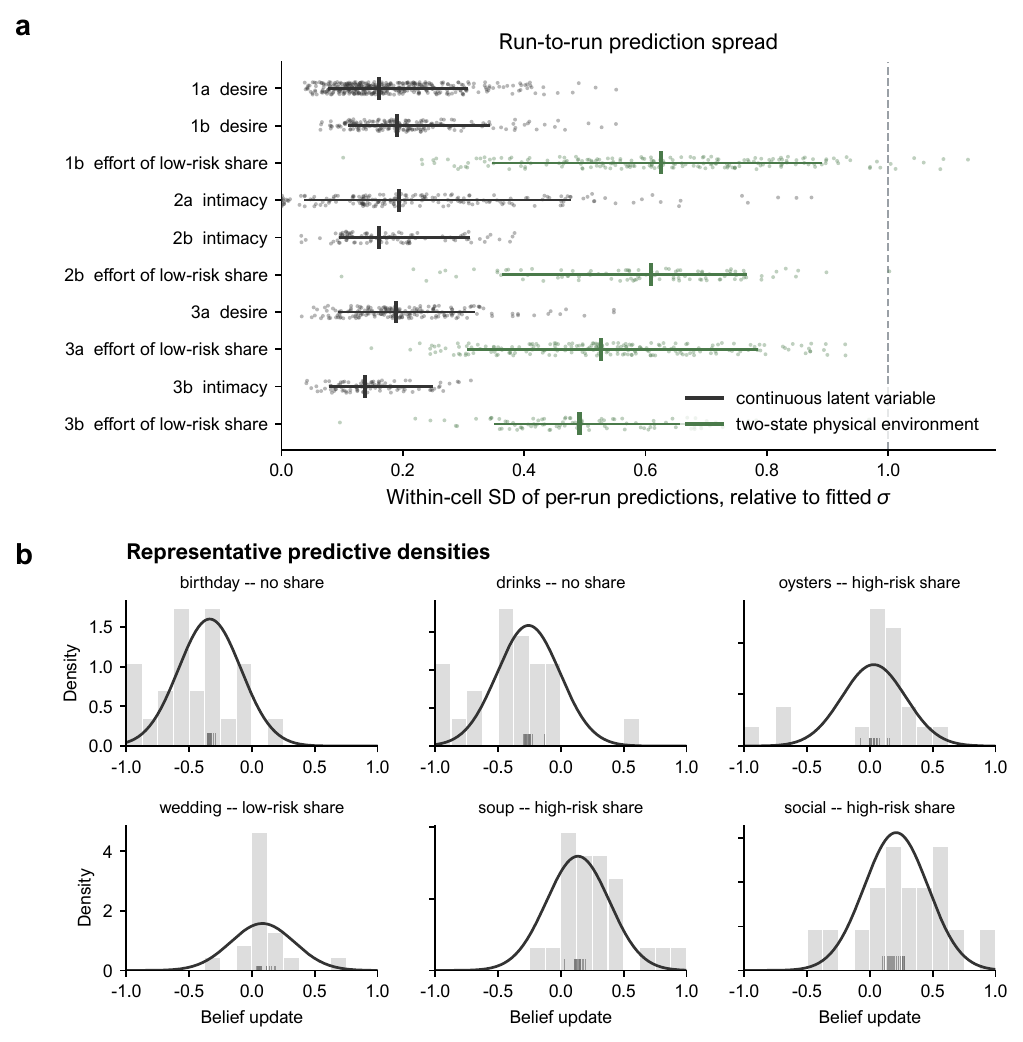}
  \caption{\textbf{Run-to-run variability and predictive likelihood checks.}
  \textbf{(a)} Within-cell standard deviation of the full social model's per-run held-out predictions, divided by the fitted response-noise scale $\sigma$. Points are held-out cells, vertical ticks are medians, horizontal bars span the 10th--90th percentiles, and the dashed line marks a run-to-run spread equal to $\sigma$. Model predictions are generally stable across LM elicitation runs, but predictions about the physical environment are more sensitive to LM sampling than predictions about desire or intimacy. 
  \textbf{(b)} Predictive densities for six Study~1a cells spanning the range of held-out desire updates. Lines show $\frac{1}{K}\sum_k\mathcal{N}(u\mid\delta_k,\sigma^2)$, histograms show participants' observed updates, and short ticks mark the $K$ per-run predictions $\delta_k$.}
  \label{fig:si-variability-checks}
\end{figure}

\clearpage

\section{Scenarios}
\label{si:scenarios}

\subsection{Food scenarios}
\label{si:food-scenarios}


{\small
\begin{longtable}{@{}>{\raggedright\arraybackslash}p{0.19\textwidth} >{\raggedright\arraybackslash}p{0.75\textwidth}@{}}
\toprule
\multicolumn{2}{@{}>{\raggedright\arraybackslash}p{0.94\textwidth}@{}}{\textbf{1.~Basketball}}\\
\midrule
Vignette & Carissa and Josh are attending a basketball game together. During halftime, they go to the hot-dog stand in the arena. When they get to the stand, they realize that it's cash only, and that between them they only have enough cash to get one hot dog.\\
Resource & the hot dog\\
Low-desire state & Neither of them is particularly hungry.\\
High-desire state & They are both very hungry.\\
Low-effort world & Right next to the hot-dog stand is a condiment station with a stack of clean plastic knives set out for customers to use.\\
High-effort world & There are no knives at the hot-dog stand or the nearby condiment area -- the nearest knives are at a sit-down restaurant on the far side of the arena, which would take several minutes to walk to and back.\\
Action: don't share & They leave the stand without a hot dog and go back to the game.\\
Action: low-risk share & They order a hot dog. They get a knife, cut the hot dog in half, and each eats from their own portion.\\
Action: high-risk share & They order a hot dog and trade off taking bites directly from it, until it is finished.\\
\addlinespace[10pt]
\multicolumn{2}{@{}>{\raggedright\arraybackslash}p{0.94\textwidth}@{}}{\textbf{2.~Birthday}}\\
\midrule
Vignette & Lio and Mitchell are at a birthday party. The cake served is a dark chocolate cake with a raspberry sauce. After everyone sings happy birthday, the birthday person cuts the cake into slices and puts the slices on plates with forks. The slices are really large, and Lio and Mitchell both don't think they can finish a full slice.\\
Resource & the cake\\
Low-desire state & Lio and Mitchell both feel neutral about dark chocolate raspberry cake.\\
High-desire state & Lio and Mitchell both really like dark chocolate raspberry cake.\\
Low-effort world & Extra forks and empty plates have been laid out on the cake table for anyone who wants to share a slice.\\
High-effort world & All the extra forks and plates have already been cleared back to the kitchen, and getting them would require walking to the kitchen and looking through the cabinets.\\
Action: don't share & Neither Lio nor Mitchell takes a slice of cake, since neither wants to take a whole large slice they can't finish.\\
Action: low-risk share & They decide to share one slice. They get an extra fork and plate, cut the slice in half, put each half on its own plate, and each eat their own half.\\
Action: high-risk share & They decide to share one slice, and share the single fork provided to eat the slice together from the same plate.\\
\addlinespace[10pt]
\multicolumn{2}{@{}>{\raggedright\arraybackslash}p{0.94\textwidth}@{}}{\textbf{3.~Brunch}}\\
\midrule
Vignette & Allison and Dana meet at a brunch restaurant together on Saturday morning. They each select their own dishes and then discuss what else to order. One intriguing choice is to also order a stack of the restaurant's famous pancakes, to eat for dessert after they eat their main dishes. They decide to wait until they have finished their main dishes, to decide whether to order the pancakes.\\
Resource & the pancakes\\
Low-desire state & After eating their main dishes, they are both pretty full.\\
High-desire state & After eating their main dishes, they are both still hungry.\\
Low-effort world & There is a box of extra utensils on their table.\\
High-effort world & To get extra utensils, they would need to walk to the utensils station on the opposite side of the restaurant.\\
Action: don't share & They decide not to order the pancakes.\\
Action: low-risk share & They order the pancakes. They get extra utensils and use them to cut portions of the stack onto their own plates and eat from their own plates.\\
Action: high-risk share & They order the pancakes. They use their own utensils, which they have already eaten with, to eat directly from the shared stack.\\
\addlinespace[10pt]
\multicolumn{2}{@{}>{\raggedright\arraybackslash}p{0.94\textwidth}@{}}{\textbf{4.~Takeout}}\\
\midrule
Vignette & Emily and Elizabeth are at a conference for work. After a long day of events, they get back to the conference hotel. It is late and all the nearby places are closed, so they decide to order chicken tenders delivered to the hotel. When they open the bag, they see that the restaurant has only included one container of honey mustard dipping sauce. They look in the bag and see if the restaurant has included anything else, and see that the restaurant has included several ketchup packets.\\
Resource & the honey mustard sauce\\
Low-desire state & Neither Emily nor Elizabeth minds whether they use ketchup or honey mustard.\\
High-desire state & Both Emily and Elizabeth prefer honey mustard sauce to ketchup.\\
Low-effort world & The takeout came with extra sauce containers that they can use to pour out and divide the sauce.\\
High-effort world & The takeout didn't come with extra sauce containers or a surface they can pour the sauce onto, so to get them they need to go to the hotel lobby downstairs.\\
Action: don't share & Emily and Elizabeth each use the ketchup packets and leave the single honey mustard container unopened.\\
Action: low-risk share & They get a sauce container and pour half of the honey mustard into it so that each person has their own dipping portion. Each person only dips from their own portion.\\
Action: high-risk share & They both dip their own chicken tenders into the single honey mustard container, double-dipping the same tenders back into the sauce after each bite.\\
\addlinespace[10pt]
\multicolumn{2}{@{}>{\raggedright\arraybackslash}p{0.94\textwidth}@{}}{\textbf{5.~Cooking}}\\
\midrule
Vignette & Liana and Serina are both major foodies who love cooking. They spend the evening cooking a big batch of a new pasta recipe to bring to an event the next day. While the pasta is still warm, they consider eating some for their own dinner tonight. There is a plate on the counter that they have used for placing their cooking utensils.\\
Resource & the pasta\\
Low-desire state & Neither of them especially feels like having any of the pasta tonight.\\
High-desire state & The pasta smells amazing and they're both tempted to have some tonight.\\
Low-effort world & There are more clean plates and utensils in the cabinets, ready to use.\\
High-effort world & All the other plates and utensils are mid-cycle in the dishwasher -- getting them out clean would require waiting for the cycle to finish.\\
Action: don't share & Liana and Serina don't have any pasta themselves tonight; they keep the whole batch for the party.\\
Action: low-risk share & They get clean plates and extra utensils, serve some pasta from the pot onto their individual plates, and eat from their own plates using their own utensils.\\
Action: high-risk share & They spoon a shared portion from the pot onto the single plate on the counter, and both eat from that plate together with the same fork.\\
\addlinespace[10pt]
\multicolumn{2}{@{}>{\raggedright\arraybackslash}p{0.94\textwidth}@{}}{\textbf{6.~Apples}}\\
\midrule
Vignette & Camille and Haoyu are at an orchard for a group apple-picking outing in the fall. There are many different apple varieties at the orchard, including Red Delicious, Yellow Delicious, Honeycrisp, and Jonagold. At the entrance, the employees tell them that visitors are welcome to taste the apples. When they enter the orchard, the apple variety in the first aisle is the Jonagold variety. Camille picks an apple from a tree.\\
Resource & the Jonagold variety\\
Low-desire state & Both Camille and Haoyu have tasted the Jonagold variety at this orchard before, so they are indifferent about tasting it again.\\
High-desire state & Neither Camille nor Haoyu has tasted the Jonagold variety at this orchard before, so they would both like to taste it.\\
Low-effort world & The Jonagold aisle is dense with ripe apples at eye level -- Haoyu could easily reach up and pick her own separate apple from the same tree.\\
High-effort world & This particular tree has only the one apple Camille picked within reach; all the other Jonagolds are on trees at the far end of the orchard, a long walk away.\\
Action: don't share & Camille takes bites out of the apple. Haoyu does not taste the apple.\\
Action: low-risk share & Haoyu picks a separate Jonagold apple for herself, and each of them eats their own apple.\\
Action: high-risk share & They share the one apple, passing it back and forth and each taking bites until it is finished.\\
\addlinespace[10pt]
\multicolumn{2}{@{}>{\raggedright\arraybackslash}p{0.94\textwidth}@{}}{\textbf{7.~Dip}}\\
\midrule
Vignette & Declan and Eric are preparing snacks for a house party. They get two kinds of chips, tortilla chips and pita chips, and prepare two kinds of dips: a buffalo chicken dip and a spinach and artichoke dip. Before putting the dips out for guests, they spoon a small tasting amount of each dip into a couple of small bowls to sample themselves.\\
Resource & the dips\\
Low-desire state & Neither of them thinks it's particularly important to taste the dips before the party.\\
High-desire state & They both really want to taste the dips before the party.\\
Low-effort world & There is a stack of clean small appetizer plates already unpacked and sitting on the snack table, ready to use.\\
High-effort world & The appetizer plates are still packed away in a box of party supplies that hasn't been unboxed yet, buried in the garage under other party boxes -- digging them out would take a while.\\
Action: don't share & Neither of them tastes the dips before the party; they put everything out for guests as-is.\\
Action: low-risk share & They get two small appetizer plates, divide the tasting portions between the two plates, and each tries the dips with chips from their own plate.\\
Action: high-risk share & They leave the tasting portions in the shared bowls and dip the same chips into them, each taking bites and double-dipping them back in.\\
\addlinespace[10pt]
\multicolumn{2}{@{}>{\raggedright\arraybackslash}p{0.94\textwidth}@{}}{\textbf{8.~Drinks}}\\
\midrule
Vignette & Noah and Ria are at a party. At the bar there is a one-off special cocktail the bartender invented -- tequila mixed with ice cream -- and only one serving is left.\\
Resource & the cocktail\\
Low-desire state & Neither of them is especially interested in the cocktail.\\
High-desire state & They are both very interested in the cocktail.\\
Low-effort world & There is a stack of clean empty cups right at the bar within arm's reach.\\
High-effort world & The bartender has stepped away to the back, and the extra clean cups are on a shelf behind the bar that's inaccessible without flagging someone down -- which would take a long time given how packed the party is.\\
Action: don't share & Neither of them drinks the special cocktail; they each get a regular drink from the bar instead.\\
Action: low-risk share & They get a second clean cup, pour half of the cocktail into it, and each sip from their own cup.\\
Action: high-risk share & They share the single cup of cocktail, passing it back and forth and sipping from the same rim.\\
\addlinespace[10pt]
\multicolumn{2}{@{}>{\raggedright\arraybackslash}p{0.94\textwidth}@{}}{\textbf{9.~Driving}}\\
\midrule
Vignette & Danielle and Katherine are driving together to an event a few hours away, leaving early in the morning. On their way out of town, they stop at a gas station to get coffee. The coffee machine is out of order partway through pouring, so only one cup gets filled.\\
Resource & the coffee\\
Low-desire state & Both of them got enough sleep last night, so they aren't particularly tired.\\
High-desire state & Both of them are extremely tired, and need the coffee to be able to stay awake.\\
Low-effort world & Right next to the coffee machine at this gas station is a self-serve stack of clean empty cups that customers can grab from.\\
High-effort world & There are no cups out at the coffee machine; getting a second cup would require asking at the counter, and the line to pay is long.\\
Action: don't share & They do not share the coffee. Danielle, who is driving the car, drinks the coffee.\\
Action: low-risk share & They get another cup, pour half of the coffee into the other cup, and each drink from their own cup.\\
Action: high-risk share & They share the coffee from the single cup, passing it back and forth.\\
\addlinespace[10pt]
\multicolumn{2}{@{}>{\raggedright\arraybackslash}p{0.94\textwidth}@{}}{\textbf{10.~Fair}}\\
\midrule
Vignette & Marianne and Lisa are at the county fair. They've spent a day walking around. Around the time the fair is closing, the food trucks are giving out their leftover food. They both go to a food truck that sells fresh corn on the cob. The food truck gives them a large skewer of corn. They then go to some other food trucks, and get some funnel cake and also some fried cheese. They put all the food on a plate, and need to decide how to share it.\\
Resource & the corn\\
Low-desire state & The corn looks okay, and they're more excited about the other food that they got.\\
High-desire state & The corn looks especially appealing to both of them, even more so than the other food.\\
Low-effort world & There is a cutlery station nearby with extra knives, plates, and napkins for visitors to use.\\
High-effort world & The food truck has packed up its cutlery station, and if they want cutlery, they need to go to the cutlery station on the other side of the fair.\\
Action: don't share & They eat the food on their sides of the plate, and don't share the individual items of food. The corn is on Marianne's side of the plate, so she eats the corn.\\
Action: low-risk share & They get cutlery and a plate, and use the knife to cut the corn off the cob onto the plate. They each eat the corn from the plate with their own forks.\\
Action: high-risk share & They switch off taking bites from the cob.\\
\addlinespace[10pt]
\multicolumn{2}{@{}>{\raggedright\arraybackslash}p{0.94\textwidth}@{}}{\textbf{11.~Gala}}\\
\midrule
Vignette & Elena and Todd are at a fancy gala. Elena is deciding what to order, and asks Todd what he ordered. Todd said that he ordered a pumpkin spice martini and that the bar offers many different types of interesting seasonal drinks. Intrigued, Elena decides to order an apple pie espresso martini.\\
Resource & each other's drinks\\
Low-desire state & When the drinks come, Elena and Todd do not particularly want to try each other's drinks.\\
High-desire state & When the drinks come, Elena and Todd want to try each other's drinks.\\
Low-effort world & There is a dispenser with extra clean straws sitting right on the bar that anyone can grab from.\\
High-effort world & The straw dispenser near them is out of straws, and the servers are deep in the banquet service -- flagging one down to bring extra straws would take a long time.\\
Action: don't share & They only drink from their own drinks, and do not try each other's drinks.\\
Action: low-risk share & They get two extra clean straws and use them to try each other's drinks, before drinking their own drinks directly from the rim of the glass.\\
Action: high-risk share & They try each other's drinks by drinking directly from the rims of the glass.\\
\addlinespace[10pt]
\multicolumn{2}{@{}>{\raggedright\arraybackslash}p{0.94\textwidth}@{}}{\textbf{12.~Hike}}\\
\midrule
Vignette & Tony and Alvin go on a day hike in New Hampshire. Alvin packs snacks and energy bars, while Tony brings peanut butter and jelly sandwiches. Halfway down the mountain, they take a snack break. Alvin realizes that he has run out of his food. Tony pulls out a sandwich from his pack.\\
Resource & the sandwich\\
Low-desire state & Neither of them is very hungry, and the hike is almost over.\\
High-desire state & They are both tired and hungry.\\
Low-effort world & Tony's backpacking knife is in the top lid of his pack and is easy to grab.\\
High-effort world & Tony's backpacking knife is buried at the very bottom of his pack, and grabbing it would require taking out all of his gear, looking for the knife, and re-packing everything.\\
Action: don't share & Tony and Alvin do not share the sandwich; Tony eats the sandwich himself.\\
Action: low-risk share & Tony gets his knife from the pack, uses it to cut the sandwich in half, and hands one half to Alvin.\\
Action: high-risk share & Tony and Alvin alternate taking bites directly from the sandwich.\\
\addlinespace[10pt]
\multicolumn{2}{@{}>{\raggedright\arraybackslash}p{0.94\textwidth}@{}}{\textbf{13.~Oysters}}\\
\midrule
Vignette & Will and Jay are at a seafood restaurant. They are interested in ordering the oyster platter. The server mentions that the restaurant has different types of oysters in stock, each type with unique flavors and notes. They decide to order one of each type of oyster.\\
Resource & all the oyster types\\
Low-desire state & They are both indifferent to tasting all the different oyster types.\\
High-desire state & They both would like to taste as many different oyster types as possible.\\
Low-effort world & The restaurant is quiet tonight and the server is attentive -- they expect that extra small plates and forks, for splitting the oysters, will come immediately whenever they ask.\\
High-effort world & The restaurant is packed and the server is slammed -- flagging him down for extra small plates and forks, for splitting the oysters, would take a long time.\\
Action: don't share & They only eat the oysters (the meat and the brine) on their own sides of the platter. They do not share the individual oysters.\\
Action: low-risk share & They ask the server for two extra small plates and cocktail forks. They use the forks to split each oyster's meat between the plates, and each eats half of each oyster from their own plate.\\
Action: high-risk share & For each oyster, they split it directly from the shell -- each biting off half of the meat with their teeth and drinking some of the brine from the same shell, before passing it to the other person.\\
\addlinespace[10pt]
\multicolumn{2}{@{}>{\raggedright\arraybackslash}p{0.94\textwidth}@{}}{\textbf{14.~Social}}\\
\midrule
Vignette & Sonia and Alan arrive late to an ice cream social hosted by their religious organization. Unfortunately, there is only one ice cream cone left. There are other kinds of desserts available, however, including many flavors of cookies.\\
Resource & the ice cream\\
Low-desire state & Neither of them particularly wants ice cream.\\
High-desire state & They both really want ice cream.\\
Low-effort world & There is a stack of clean spoons on the dessert table.\\
High-effort world & The dessert table has been mostly cleared and the spoons have been put away in the kitchen, so getting clean spoons would mean interrupting the hosts who are busy with other cleanup.\\
Action: don't share & Neither of them takes the ice cream cone. They instead go and eat cookies.\\
Action: low-risk share & They get two spoons and each uses their own spoon to eat the ice cream out of the cone.\\
Action: high-risk share & They pass the cone back and forth, taking turns licking the ice cream and biting into the cone.\\
\addlinespace[10pt]
\multicolumn{2}{@{}>{\raggedright\arraybackslash}p{0.94\textwidth}@{}}{\textbf{15.~Soup}}\\
\midrule
Vignette & Christina and Tanya are having dinner at a restaurant. They each order a soup as a starter. Christina orders the chicken noodle soup, and Tanya orders the lentil soup.\\
Resource & each other's soups\\
Low-desire state & Christina and Tanya are satisfied with their own soups.\\
High-desire state & Christina and Tanya want to try each other's soups.\\
Low-effort world & The restaurant is quiet and the server is attentive -- extra bowls would come to the table quickly if they asked.\\
High-effort world & The restaurant is packed and the server is slammed -- asking for extra bowls would mean a long wait.\\
Action: don't share & They eat only their own soups.\\
Action: low-risk share & They ask for two extra bowls, spoon some of their own soup into an extra bowl, and swap the extra bowls.\\
Action: high-risk share & They put both bowls in the middle of the table and eat directly from each other's bowls, going back and forth between the two bowls.\\
\addlinespace[10pt]
\multicolumn{2}{@{}>{\raggedright\arraybackslash}p{0.94\textwidth}@{}}{\textbf{16.~Wedding}}\\
\midrule
Vignette & Maxwell and Ralph are seated together at a wedding. After the appetizer, there are two dishes that guests can choose from for the main course. One dish is a mushroom risotto and the other dish is a coconut curry salmon. Both dishes look really good, so they both have a hard time selecting what to order. Maxwell ends up getting the mushroom risotto, and Ralph ends up getting the coconut curry salmon.\\
Resource & both dishes\\
Low-desire state & When the food comes, they are actually really happy with what they ended up selecting. Maxwell and Ralph each feel that their own dish looks better than the other one.\\
High-desire state & When the food comes, they realize how good both dishes look, and that they really want to taste both dishes.\\
Low-effort world & The catering staff has brought a tray of extra utensils to the table for guests.\\
High-effort world & The catering staff is deep in the next course of service, and there are no extra utensils on the table -- flagging a server down for new utensils would mean a long wait.\\
Action: don't share & Maxwell eats his mushroom risotto, and Ralph eats his coconut curry salmon. They do not share their food.\\
Action: low-risk share & They get an extra set of utensils, and use them to place some of each dish onto the other's plate. They then eat from their own plates using their original utensils.\\
Action: high-risk share & They eat from each other's plates throughout the meal, going back and forth between their own plate and the other's plate with their own forks.\\
\bottomrule
\end{longtable}
}

\newpage

\subsection{Non-food scenarios}
\label{si:nonfood-scenarios}


{\small
\begin{longtable}{@{}>{\raggedright\arraybackslash}p{0.19\textwidth} >{\raggedright\arraybackslash}p{0.75\textwidth}@{}}
\toprule
\multicolumn{2}{@{}>{\raggedright\arraybackslash}p{0.94\textwidth}@{}}{\textbf{1.~Chapstick} \hfill \textit{substance}}\\
\midrule
Vignette & Priya and Marcus are visiting a ski resort for the weekend. After a long day on the slopes in dry mountain air, they're back at the lodge. Priya pulls out a tube of lip ointment. Marcus realizes that he forgot to pack some for himself.\\
Resource & to use the lip ointment\\
Low-desire state & Both of their lips are only slightly dry, and neither of them is particularly bothered by it.\\
High-desire state & Both of their lips are cracked and painful, and will keep getting worse overnight without ointment.\\
Low-effort world & Getting the ointment on with a fingertip instead of the tube takes clean hands, and theirs are grimy from the day outside -- but there's a restroom with a sink just around the corner from where they're sitting.\\
High-effort world & Getting the ointment on with a fingertip instead of the tube takes clean hands, and theirs are grimy from the day outside -- and the only restroom with a sink is on the far side of the lodge, with a line out the door.\\
Action: don't share & Priya applies some of the ointment straight from the tube to her own lips; Marcus does not use any of it.\\
Action: low-risk share & Priya applies some of the ointment straight from the tube to her own lips. Marcus washes his hands, squeezes a dab of the ointment onto his own fingertip, and applies it to his lips with his finger.\\
Action: high-risk share & Priya applies some of the ointment straight from the tube to her own lips, and hands the tube to Marcus, who applies it straight from the tube to his own lips as well.\\
\addlinespace[10pt]
\multicolumn{2}{@{}>{\raggedright\arraybackslash}p{0.94\textwidth}@{}}{\textbf{2.~Towel} \hfill \textit{substance}}\\
\midrule
Vignette & Amira and Logan are at a public pool on a summer afternoon. When they get out of the water after swimming, Logan realizes he forgot to pack a towel in his bag.\\
Resource & to use the towel\\
Low-desire state & It's warm and sunny out, and neither of them minds air drying in the sun.\\
High-desire state & A cold wind has picked up, and they both want to get dry quickly to warm up.\\
Low-effort world & Amira's towel is an oversized beach towel, so each of them can easily dry off with separate ends of the towel.\\
High-effort world & Amira's towel is on the smaller side, so keeping to separate ends would mean each drying off with only a small patch of fabric, which would be slow and awkward.\\
Action: don't share & Amira dries herself off with her towel; Logan does not use it and air-dries on the pool deck.\\
Action: low-risk share & Amira uses one end of the towel to dry herself, and Logan uses the far end, each keeping to their own half so their drying surfaces don't overlap.\\
Action: high-risk share & Amira dries herself off using the whole towel, and then hands the damp towel to Logan, who dries himself with it.\\
\addlinespace[10pt]
\multicolumn{2}{@{}>{\raggedright\arraybackslash}p{0.94\textwidth}@{}}{\textbf{3.~Earbuds} \hfill \textit{substance}}\\
\midrule
Vignette & Isla and Theo are on a long train ride together. That morning, the newest episode of a podcast series they have both been following was released. Isla, but not Theo, has a pair of earbuds.\\
Resource & to listen to the new episode on the train\\
Low-desire state & Neither of them is in any rush to hear the new episode; they wouldn't mind waiting to listen to it another day.\\
High-desire state & Both of them really want to listen to the new episode on the train ride.\\
Low-effort world & Isla has a travel pack of sanitizing wipes in the outer pocket of her bag.\\
High-effort world & Neither of them has any sanitizing wipes with them, but the caf\'{e} car several cars away sells small travel packs of wipes.\\
Action: don't share & Isla listens to the episode on her earbuds by herself; Theo waits to listen until he's home.\\
Action: low-risk share & Isla wipes the earbuds clean with a sanitizing wipe. They split the pair -- each wearing one earbud -- and listen to the episode together.\\
Action: high-risk share & They split the pair as-is, without wiping them: Isla hands Theo one of the earbuds she has been wearing, he puts it straight into his ear, and they listen to the episode together.\\
\addlinespace[10pt]
\multicolumn{2}{@{}>{\raggedright\arraybackslash}p{0.94\textwidth}@{}}{\textbf{4.~Hairbrush} \hfill \textit{substance}}\\
\midrule
Vignette & Celeste and Tamsin are getting ready in a hotel room before an evening banquet. Celeste has been blow-drying and brushing her hair. Tamsin realizes that she forgot to pack her hairbrush.\\
Resource & to brush their hair\\
Low-desire state & Both of their hair will look fine regardless of whether they brush it.\\
High-desire state & Both of their hair got tangled and windblown on the trip over, and they each feel they really need to brush it to look presentable.\\
Low-effort world & Only a few loose strands are caught in the brush's bristles, so picking it clean and giving it a quick rinse and dry would take a minute at most.\\
High-effort world & The brush is densely matted with hair wound through the bristles, so picking it clean would mean working the strands out row by row before it could even be rinsed and dried.\\
Action: don't share & Celeste brushes her own hair; Tamsin does not use the brush, and smooths her hair with her fingers and a bit of water.\\
Action: low-risk share & Tamsin picks the hair out of the brush's bristles, rinses the brush under hot water, dries it off, and brushes her hair with the cleaned brush.\\
Action: high-risk share & Celeste passes the brush directly to Tamsin, who brushes her hair with it without cleaning it or picking out the hair.\\
\addlinespace[10pt]
\multicolumn{2}{@{}>{\raggedright\arraybackslash}p{0.94\textwidth}@{}}{\textbf{5.~Harmonica} \hfill \textit{substance}}\\
\midrule
Vignette & Jonas and Derek are at a jam session in a friend's backyard. Jonas has brought along a harmonica that he bought recently but has barely played. Neither he nor Derek has ever really played one before.\\
Resource & to play the harmonica\\
Low-desire state & Playing the harmonica isn't a big deal to either of them.\\
High-desire state & They are both very excited to try playing the harmonica.\\
Low-effort world & There's a pack of sanitizing wipes sitting out on the snack table a few steps away.\\
High-effort world & Nobody at the jam session has sanitizing wipes -- finding one would mean going into the house and digging through the host's bathroom cabinets.\\
Action: don't share & Jonas tries out the harmonica himself for a while; Derek does not play it.\\
Action: low-risk share & They take turns playing the harmonica, wiping the mouthpiece down thoroughly with a sanitizing wipe between turns.\\
Action: high-risk share & They take turns playing the harmonica, passing it directly back and forth without wiping it between turns.\\
\addlinespace[10pt]
\multicolumn{2}{@{}>{\raggedright\arraybackslash}p{0.94\textwidth}@{}}{\textbf{6.~Sunscreen} \hfill \textit{substance}}\\
\midrule
Vignette & Zara and Ben are at the beach. Neither of them can reach their own back to put sunscreen on it. Zara has brought a tube of sunscreen lotion.\\
Resource & to use sunscreen on their backs\\
Low-desire state & It's fairly overcast, and neither of them thinks their back is at much risk of burning.\\
High-desire state & The sun is strong, and they both know their backs will burn badly if they don't put sunscreen on it.\\
Low-effort world & The beach has a free sunscreen station stocked with cans of spray-on sunscreen, right next to where they've set up.\\
High-effort world & The beach has a free sunscreen station stocked with cans of spray-on sunscreen, but it's all the way down by the far parking lot -- a long walk across the hot sand.\\
Action: don't share & Neither of them puts sunscreen on their back.\\
Action: low-risk share & One of them goes to the beach's sunscreen station for a can of spray-on sunscreen, and they take turns spraying each other's backs.\\
Action: high-risk share & They take turns squeezing Zara's sunscreen lotion onto their hands and rubbing it into each other's backs.\\
\addlinespace[10pt]
\multicolumn{2}{@{}>{\raggedright\arraybackslash}p{0.94\textwidth}@{}}{\textbf{7.~Blanket} \hfill \textit{space}}\\
\midrule
Vignette & Anaya and Kai are at an outdoor movie screening in the park. About twenty minutes into the movie, the temperature drops sharply -- much more than they had expected. Anaya brought one large blanket.\\
Resource & to warm up under a blanket\\
Low-desire state & Neither of them feels particularly cold; their jackets are enough for now.\\
High-desire state & They are both shivering, and without some way to stay warm they won't be able to enjoy the rest of the movie.\\
Low-effort world & Kai has a blanket in his car, which is parked along the edge of the park just a short walk from where they're sitting.\\
High-effort world & Kai has a blanket in his car, but the parking near the park had filled up before the movie, so the car is a long walk away through the dark.\\
Action: don't share & Anaya wraps her blanket around herself; Kai pulls his jacket tighter and keeps his hands in his pockets.\\
Action: low-risk share & Kai goes to his car for his own blanket, and they each wrap up in a blanket of their own.\\
Action: high-risk share & They move closer to each other and wrap Anaya's blanket fully around both of them, sitting shoulder-to-shoulder.\\
\addlinespace[10pt]
\multicolumn{2}{@{}>{\raggedright\arraybackslash}p{0.94\textwidth}@{}}{\textbf{8.~Sleeping-bag} \hfill \textit{space}}\\
\midrule
Vignette & Gabriel and Henry are on a two-night backpacking trip in Yosemite. On the first day, Gabriel accidentally drops his sleeping bag in a stream, soaking it completely through.\\
Resource & to sleep in a sleeping bag\\
Low-desire state & The night is expected to be mild, and sleeping without a proper sleeping bag would be tolerable.\\
High-desire state & The night is expected to be cold, and sleeping without proper insulation would be miserable.\\
Low-effort world & There's a stack of dry firewood at the campsite, so keeping a fire going strong enough to dry the wet bag by bedtime would be easy.\\
High-effort world & There's almost no dry firewood at the campsite, so keeping a fire going strong enough to dry the wet bag would mean spending a long time looking through the woods for burnable wood.\\
Action: don't share & Henry sleeps in his bag, and Gabriel wraps himself in his spare layers on his own sleeping pad.\\
Action: low-risk share & They start a fire for drying Gabriel's bag. Once it is dry, they each sleep in their own bag.\\
Action: high-risk share & They push their two sleeping pads together, and try to sleep pressed side by side in Henry's bag, sharing body heat.\\
\addlinespace[10pt]
\multicolumn{2}{@{}>{\raggedright\arraybackslash}p{0.94\textwidth}@{}}{\textbf{9.~Bed} \hfill \textit{space}}\\
\midrule
Vignette & Ines and Sofia are staying at an Airbnb with a group of people over the weekend for a retreat. By the time they arrive, the other beds have been claimed, and only one full-size bed remains.\\
Resource & to sleep on the bed\\
Low-desire state & Neither of them particularly cares where they sleep; the couch in the living room would be fine too.\\
High-desire state & They are both exhausted from the long drive and really want a proper bed to sleep in.\\
Low-effort world & The bedroom closet has spare pillows and extra blankets stacked on the shelf.\\
High-effort world & There's no spare bedding in the room -- finding extra pillows and a second blanket would mean searching through the whole house's closets or messaging the host and waiting for a reply.\\
Action: don't share & One of them takes the couch, and one of them takes the bed.\\
Action: low-risk share & They both sleep on the bed, with a row of pillows between them and separate blankets for each of them.\\
Action: high-risk share & They both sleep on the bed under the same blanket.\\
\addlinespace[10pt]
\multicolumn{2}{@{}>{\raggedright\arraybackslash}p{0.94\textwidth}@{}}{\textbf{10.~Locker-room} \hfill \textit{space}}\\
\midrule
Vignette & Maya and Samira have just finished a workout class at their office building's gym. They step into the locker room to grab their bags. The changing area is a communal space with a long bench, and they consider whether to change out of their gym clothes here before the commute or wait until they're home.\\
Resource & to change before the commute home\\
Low-desire state & Neither of them minds staying in their workout clothes for the ride home; they can just change once they get home.\\
High-desire state & They both feel sweaty and gross, and commuting in damp workout clothes sounds miserable -- they would much rather change before heading out.\\
Low-effort world & A tall row of lockers juts into the middle of the bench area, so they can easily change on opposite sides of it without seeing each other.\\
High-effort world & The bench area is one open room with no dividers, so changing without seeing each other would mean carefully coordinating.\\
Action: don't share & Neither of them changes at the gym; they both change once they get home.\\
Action: low-risk share & They change at opposite ends of the bench area, positioned so that they don't see each other undressed.\\
Action: high-risk share & They change side by side at the bench area.\\
\addlinespace[10pt]
\multicolumn{2}{@{}>{\raggedright\arraybackslash}p{0.94\textwidth}@{}}{\textbf{11.~Sauna} \hfill \textit{space}}\\
\midrule
Vignette & Ronan and Aaron are at a wellness retreat that includes access to a traditional Finnish sauna, which is customarily used nude. Everyone in the group is invited to use it. When they arrive at the sauna hut, they find that they are the only two people there -- everyone else has gone to different activities.\\
Resource & to use the sauna\\
Low-desire state & Neither of them is especially in the mood for using the sauna right now -- they'd happily postpone and come back another time.\\
High-desire state & They were both really looking forward to using the sauna right now.\\
Low-effort world & The lounge next to the sauna hut is warm and comfortable, so waiting for a turn would be easy.\\
High-effort world & The lounge next to the sauna hut is locked up for the afternoon, so whoever waits would have to stand outside in the cold wind in just a robe.\\
Action: don't share & Neither of them uses the sauna now. They head to another activity and figure they'll try the sauna another time.\\
Action: low-risk share & They take turns in the sauna one at a time -- one goes in while the other waits, and then they switch off.\\
Action: high-risk share & They both go in at the same time following the traditional nude sauna practice, sitting near each other on the same bench.\\
\addlinespace[10pt]
\multicolumn{2}{@{}>{\raggedright\arraybackslash}p{0.94\textwidth}@{}}{\textbf{12.~Breakup} \hfill \textit{privacy}}\\
\midrule
Vignette & Tessa is visiting Naomi's city, and they meet up for coffee while she's in town. They haven't seen each other in person for a while. They each know, through mutual friends, that the other has gone through a breakup.\\
Resource & to talk about their breakups\\
Low-desire state & Both of their breakups feel like old news by now, and neither of them feels much need to talk about it.\\
High-desire state & Both of their breakups are recent, and each of them has been wanting to talk it through with someone going through the same thing.\\
Low-effort world & Both of their breakups were fairly self-contained, so a general version that gets the gist across would take only a few sentences to tell.\\
High-effort world & Both of their breakups were tangled up with other things -- moves, job changes, overlapping friend groups -- so telling even a general version that makes sense would take a lot of backstory.\\
Action: don't share & Neither of them brings up the breakups. They catch up on work and other lighter topics for the rest of the meet-up.\\
Action: low-risk share & They each give a careful, general account of their breakup.\\
Action: high-risk share & They each share exactly what is on their mind about their breakup, without thinking about how they are saying it.\\
\addlinespace[10pt]
\multicolumn{2}{@{}>{\raggedright\arraybackslash}p{0.94\textwidth}@{}}{\textbf{13.~Salary} \hfill \textit{privacy}}\\
\midrule
Vignette & Omar and June work in the same role in the same industry, at different companies. Over dinner after an industry event, the conversation turns to how work has been going, and eventually to pay in their field.\\
Resource & to know how their salaries compare\\
Low-desire state & They're each pretty content with their current pay, and only somewhat curious about how it compares.\\
High-desire state & They're each preparing to negotiate a raise this year, and knowing how their pay compares would genuinely help them.\\
Low-effort world & Their field's annual salary survey is widely known, and its pay bands line up cleanly with both of their jobs, so placing themselves against it takes only a sentence or two.\\
High-effort world & Their field's annual salary survey is widely known, but its categories don't line up well with either of their actual roles, so placing themselves against it takes a long chain of comparisons and qualifications to get the picture across.\\
Action: don't share & Neither of them shares anything about their own pay; they trade generalities about the job market and move on to other topics.\\
Action: low-risk share & Without either of them ever saying a number out loud, they each describe where their pay sits relative to their field's annual salary survey, in enough detail that each can work out roughly where the other stands.\\
Action: high-risk share & They each lay out exactly what they make -- base salary, bonus, and everything else -- down to the precise numbers.\\
\addlinespace[10pt]
\multicolumn{2}{@{}>{\raggedright\arraybackslash}p{0.94\textwidth}@{}}{\textbf{14.~Gossip} \hfill \textit{privacy}}\\
\midrule
Vignette & Wren and Oscar are meeting up for dinner. The conversation drifts to their mutual friends and what's been going on in their romantic and sexual lives.\\
Resource & to discuss the latest gossip in their social circle\\
Low-desire state & Neither of them is particularly interested in getting into other people's business tonight.\\
High-desire state & They both want to talk about what has happened with someone who knows the people involved.\\
Low-effort world & The recent developments are straightforward, so a light hint is enough for the other to know at once who and what is meant.\\
High-effort world & The recent developments are a set of long, tangled stories, so conveying them takes a lot of roundabout description before the other can piece together who and what is meant.\\
Action: don't share & They keep the conversation general and don't talk much about other people.\\
Action: low-risk share & They talk around the latest developments -- hinting at what has happened, and letting their own reactions show only implicitly -- but in enough detail that each can work out who and what the other means.\\
Action: high-risk share & They speak openly: naming names, spelling out exactly what has been going on, and saying what they really think about the people involved.\\
\addlinespace[10pt]
\multicolumn{2}{@{}>{\raggedright\arraybackslash}p{0.94\textwidth}@{}}{\textbf{15.~Home} \hfill \textit{privacy}}\\
\midrule
Vignette & Vera and Ivy know each other from college. Vera is in town for the day with a few hours between meetings, and she's near Ivy's apartment. They text hello and agree to meet. They need to decide where to meet. In its current state, Ivy's apartment has laundry on the floor, dishes from yesterday in the sink, books and personal items scattered around.\\
Resource & to meet at Ivy's apartment\\
Low-desire state & It's a nice day out for walking around the neighborhood or spending time at a cafe.\\
High-desire state & It's pouring rain outside, the coffee shops look full and chaotic, and it would be more comfortable to meet at Ivy's apartment.\\
Low-effort world & Vera is coming from across town, so Ivy has over an hour before she arrives -- plenty of time to tidy the living room properly.\\
High-effort world & Vera is only a few blocks away, so Ivy has just a few minutes before she arrives -- getting the living room presentable in time would be a frantic scramble.\\
Action: don't share & They meet at a coffee shop.\\
Action: low-risk share & Ivy tidies the living room before Vera arrives. They meet in the living room, and Ivy keeps the doors to the other rooms closed.\\
Action: high-risk share & Ivy does not tidy the apartment before Vera arrives. When Vera comes in, she makes herself at home and Ivy gives her a tour of the apartment in its current actual state, including the bedroom with the unmade bed and the bathroom with personal items out.\\
\addlinespace[10pt]
\multicolumn{2}{@{}>{\raggedright\arraybackslash}p{0.94\textwidth}@{}}{\textbf{16.~Navigation} \hfill \textit{privacy}}\\
\midrule
Vignette & Nadia and Ravi are driving together to an out-of-town event, along a rural highway with long stretches between exits. Nadia is driving, and they have been using her phone for directions since Ravi's phone battery died earlier in the drive. About halfway through the drive, they're considering whether to add a restaurant stop on the way, for a quick meal before continuing on.\\
Resource & to stop for a quick meal\\
Low-desire state & Neither of them is particularly hungry; they could also wait until they reach their destination to eat.\\
High-desire state & They are both starving and really want to eat.\\
Low-effort world & Nadia's phone is set to stay unlocked for a long time, so Ravi can work in the maps app without it locking on him.\\
High-effort world & Nadia's phone auto-locks after a few seconds, so Ravi would have to keep handing the phone back for Nadia to unlock, over and over.\\
Action: don't share & They keep driving straight to the event without stopping to eat.\\
Action: low-risk share & Nadia unlocks her phone, opens the maps app, and hands the phone to Ravi to find a restaurant and add it as a stop. Ravi works only within the maps app, and whenever the phone locks he hands it back to Nadia to unlock.\\
Action: high-risk share & Nadia tells Ravi her passcode. Ravi unlocks her phone with it and uses the phone freely.\\
\bottomrule
\end{longtable}
}

\newpage

\section{LM prompts}
\label{si:prompts}

%
%
%

\subsection{Alternative action generation}

\begin{promptbox}{System prompt --- alternative action generation ($G_{\mathrm{LM}}$)}
\begin{Verbatim}[breaklines=true,breakanywhere=false,breaksymbolleft={},breaksymbolright={},fontsize=\footnotesize]
You are a participant in a human study. Respond as if you were a regular adult from the United States.

In this survey, you will read a vignette about two people in a situation where some resource -- food, an object, a physical space, or a piece of information -- could be shared between them.

The vignette omits some information about the situation, which an observer will be asked to infer.

You will be told what action they took in the situation, and which question(s) an observer is asked to answer.

Your job is to list the actions that the two people were realistically choosing between, that an observer would compare with the action they actually took, to answer the question(s).

First, briefly explain step by step which actions the two people were realistically choosing between, and what an observer would need to compare with the action they actually took to answer the question(s). Then respond with a JSON array in this exact format, with no other text after the array:
[
  {"action": "description of alternative 1"},
  {"action": "description of alternative 2"}
]
\end{Verbatim}
\end{promptbox}

\begin{promptbox}{User prompt (Example) --- alternative action generation}
\begin{Verbatim}[breaklines=true,breakanywhere=false,breaksymbolleft={},breaksymbolright={},fontsize=\footnotesize]
The two people are in a relationship they would describe as <relationship descriptor>.
Scenario: <scenario vignette>
<physical-effort paragraph>
You do not know how much the two people would like <the resource at stake>.

The two people took the following action:
<observed action>

List the actions the two people were choosing between -- the comparison set you would use to interpret their choice and judge how much they would like <the resource at stake>. Do not include the action they actually took.
\end{Verbatim}
\end{promptbox}

\subsection{Utility-feature scoring}

\begin{promptbox}{System prompt --- goal-satisfaction $g_{\tau}(a)$}
\begin{Verbatim}[breaklines=true,breakanywhere=false,breaksymbolleft={},breaksymbolright={},fontsize=\footnotesize]
You are a participant in a human study. Respond as if you were a regular adult from the United States, just going off your intuition.

In this survey, you will read a vignette about two people in a situation where some resource -- food, an object, a physical space, or a piece of information -- could be shared between them. You will see a set of possible actions the two people could take.

For each action, evaluate how fully it results in the two people ending up with the thing at stake in the scenario -- the food they could eat, the object they could use, the space they could occupy, the information they could learn.

Judge only outcome attainment: whether, and how completely, the dyad ends up obtaining or consuming the thing. Judge each action by the outcome it leads to once it is carried through to completion. An action can deliver the outcome fully whether it is done together or separately, directly or via a safer indirect route. If an action involves extra steps along the way -- going to fetch a utensil, taking a longer route, acquiring something first -- rate it by the end state those steps arrive at, not by the fact that it is still unfinished partway through. How much work or time those steps take is a separate dimension (effort) that we are not asking about here.

An action that ends with both people getting and consuming the thing should be rated high; an action where only one person gets it, or where they end up with a reduced or incomplete version, should be rated in the middle; an action where they forgo or abandon it should be rated low.

Use this scale from 0 to 6 (continuous values allowed):
0 = The thing is not obtained (the action forgoes or abandons it)
3 = Partially obtained (a reduced portion, only one person, or an incomplete version)
6 = Fully obtained (both people end up getting or consuming the thing)

Respond with your numerical ratings as a JSON object whose keys are "action_0", "action_1", ... matching the number of actions given, no explanation needed. Example for 3 actions:
{"action_0": <number>, "action_1": <number>, "action_2": <number>}
\end{Verbatim}
\end{promptbox}

\begin{promptbox}{System prompt --- effort $\mathrm{effort}_{\tau}(a)$}
\begin{Verbatim}[breaklines=true,breakanywhere=false,breaksymbolleft={},breaksymbolright={},fontsize=\footnotesize]
You are a participant in a human study. Respond as if you were a regular adult from the United States, just going off your intuition.

In this survey, you will read a vignette about two people in a situation where some resource -- food, an object, a physical space, or a piece of information -- could be shared between them. You will see a set of possible actions the two people could take.

For each action, evaluate the total physical, executional, and temporal cost of completing the action. Count required work regardless of which person performs it, including work divided between them. The cost types below all count; integrate across them into a single rating:

- Physical motor cost: how much bodily work the action requires (preparing, serving, cutting, pouring, handing over, cleaning, wiping, drying, tidying, rearranging, applying).
- Equipment and preparation cost: whether the action needs extra items or setup (utensils, plates, containers, sanitizing supplies, barriers, separate furniture, separate spaces) that either person has to obtain, set up, or take care of.
- Executional and production cost: for actions that consist of speaking, telling, or disclosing, how much work goes into producing the utterance itself -- how long the account takes to deliver and how much context, backstory, or roundabout indirect phrasing the speaker must use, for it to land.
- Time cost: how long the action takes -- waiting for something to dry, sequential rather than simultaneous use, an extended preparation or telling.

Do NOT rate social awkwardness, relational discomfort, or how intimate, appropriate, or emotionally hard the action would feel -- those are separate dimensions that we are not asking about here. Here we want only the effort of carrying the action out.

Use this scale from 0 to 6 (continuous values allowed):
0 = No effort (neither person needs to do bodily work, obtain extra items, wait, compose, or explain)
3 = Moderate effort (a few bodily steps, such as setting out a clean utensil, dividing a portion, or briefly waiting; a small handful of extra items to obtain; or a short account that takes a little effort to produce)
6 = High effort (many bodily steps, substantial setup, or significant time -- for example, leaving to obtain something from far away and returning, waiting a long time, cleaning and assembling many separate items, or producing a long account that needs extensive backstory or roundabout phrasing to convey)

Respond with your numerical ratings as a JSON object whose keys are "action_0", "action_1", ... matching the number of actions given, no explanation needed. Example for 3 actions:
{"action_0": <number>, "action_1": <number>, "action_2": <number>}
\end{Verbatim}
\end{promptbox}

\begin{promptbox}{System prompt --- interpersonal risk $\mathrm{risk}_{\tau}(a)$}
\begin{Verbatim}[breaklines=true,breakanywhere=false,breaksymbolleft={},breaksymbolright={},fontsize=\footnotesize]
You are a participant in a human study. Respond as if you were a regular adult from the United States, just going off your intuition.

In this survey, you will read a vignette about two people in a situation where some resource -- food, an object, a physical space, or a piece of information -- could be shared between them. You will see a set of possible actions the two people could take.

For each action, evaluate how much it makes one person interpersonally vulnerable to the other -- how much it exposes them, opens them up, or lowers the boundary between them, letting something normally kept to oneself pass from one person's side to the other. This interpersonal vulnerability can take multiple forms, and a single action may involve more than one:

- Bodily / substance exposure: bodily substances (saliva, breath, skin oils, sweat) from one person reach the other, either directly or via a shared vessel or item that's been on the first person's body. Even brief contact counts -- the substance doesn't have to remain visible for the exposure to be real.
- Physical contact or shared space: the two people's bodies physically touch, or they share close physical space -- sustained proximity within a bounded space such as a bed, blanket, small room, or vehicle. The extent of contact or proximity and the body region involved both matter -- brief incidental touch or passing nearness is a small exposure; sustained skin contact, sharing a confined space, or contact with normally restricted body regions is a large one.
- Private or emotional disclosure: private, sensitive, or emotional information (personal details, or feelings one would not voice publicly), or access to personal resources (a private space, a personal item, a confidential record), from one person becomes accessible to the other.

Here we are asking what the action itself does -- the interpersonal vulnerability it creates -- independent of the relationship between the two people.

Co-presence without substance transfer, contact, close shared space, or disclosure does NOT by itself make one person vulnerable to the other -- for example, two people each handling their own separate utensils, standing close together in a public space like an elevator, or keeping a conversation to surface-level topics. These should be rated near zero.

Use this scale from 0 to 6 (continuous values allowed):
0 = No interpersonal vulnerability (the two people stay fully separate; no exchange of substance, no contact or shared interpersonal space, no disclosure)
3 = Limited or indirect vulnerability (e.g. bodily substances reaching the other person only indirectly, through an item that has touched one person's skin; deliberate but limited physical contact, such as a hand on the shoulder; being close to each other in an open or roomy space rather than a confined one; or disclosing somewhat personal but not deeply private information)
6 = Strong, direct vulnerability (e.g. direct bodily-substance transfer such as mouth-to-mouth contact or sharing a utensil that's been in one person's mouth, sustained skin-to-skin contact, sharing a bed or other close confined space, or disclosing private details)

Respond with your numerical ratings as a JSON object whose keys are "action_0", "action_1", ... matching the number of actions given, no explanation needed. Example for 3 actions:
{"action_0": <number>, "action_1": <number>, "action_2": <number>}
\end{Verbatim}
\end{promptbox}

\begin{promptbox}{User prompt (template) --- feature scoring (shared across $g$, effort, risk)}
\begin{Verbatim}[breaklines=true,breakanywhere=false,breaksymbolleft={},breaksymbolright={},fontsize=\footnotesize]
Scenario: <scenario vignette>

<feature-specific rating instruction (see below)>

Action 0: <observed action>
Action 1: <alternative action 1>
Action 2: <... alternative action k>
\end{Verbatim}
\end{promptbox}

\begin{promptbox}{Per-feature rating instruction (the one line that varies)}
\begin{Verbatim}[breaklines=true,breakanywhere=false,breaksymbolleft={},breaksymbolright={},fontsize=\footnotesize]
goal-satisfaction g:  Rate how much each action results in the two people actually getting or consuming <the resource at stake> (0-6 scale):
  (for desire objects phrased as an infinitive outcome in some
   non-food scenarios, "getting or consuming" becomes "getting",
   e.g.: "Rate how much each action results in the two people actually getting to try the harmonica (0-6 scale):")

effort:               Rate the total physical or executional cost of carrying out each action, counting work performed by either person -- how much physical work, preparation, or equipment it takes, or, for telling or disclosing, how much explaining and roundabout phrasing producing the account takes (0-6 scale):

interpersonal risk:   Rate how much each action makes one person interpersonally vulnerable to the other -- through bodily exposure, physical contact or shared space, or private disclosure (0-6 scale):
\end{Verbatim}
\end{promptbox}

\subsection{Given-magnitude ratings}

\begin{promptbox}{System prompt --- desire $d$}
\begin{Verbatim}[breaklines=true,breakanywhere=false,breaksymbolleft={},breaksymbolright={},fontsize=\footnotesize]
You are a participant in a human study. Respond as if you were a regular adult from the United States, just going off your intuition.

In this survey, you will read a vignette about two people in a situation where some resource -- food, an object, a physical space, or a piece of information -- could be shared between them, along with a short description of their current state. Judge how much the two people would like the thing at stake in the scenario, given that state, on a scale from 0 (would not like it at all) to 100 (would like it extremely). Rate only how much they would like it -- not what they end up doing, how much effort it takes, or how the two people are related.

Respond with a JSON object in this exact format, no explanation:
{"desire": <number>}
\end{Verbatim}
\end{promptbox}

\begin{promptbox}{User prompt (template) --- desire $d$}
\begin{Verbatim}[breaklines=true,breakanywhere=false,breaksymbolleft={},breaksymbolright={},fontsize=\footnotesize]
Scenario: <scenario vignette>

State: <motivational-state paragraph>

On a scale from 0 to 100, how much would the two people like <the resource at stake>, given their state? Respond with {"desire": <number>}.
\end{Verbatim}
\end{promptbox}

\begin{promptbox}{System prompt --- relationship intimacy $I$}
\begin{Verbatim}[breaklines=true,breakanywhere=false,breaksymbolleft={},breaksymbolright={},fontsize=\footnotesize]
You are a participant in a human study. Respond as if you were a regular adult from the United States, just going off your intuition.

You will read a short description of a relationship between two people. Judge how intimate the relationship is on a scale from 0 (maximally formal) to 100 (maximally intimate).

Respond with a JSON object in this exact format, no explanation:
{"intimacy": <number>}
\end{Verbatim}
\end{promptbox}

\begin{promptbox}{User prompt (template) --- relationship intimacy $I$}
\begin{Verbatim}[breaklines=true,breakanywhere=false,breaksymbolleft={},breaksymbolright={},fontsize=\footnotesize]
The two people are in a relationship they would describe as <relationship descriptor>.

On a scale from 0 to 100, how intimate is this relationship? Respond with {"intimacy": <number>}.
\end{Verbatim}
\end{promptbox}

\end{document}